\documentclass[usenatbib]{mn2e}
\usepackage{graphicx, txfonts, natbib}
\usepackage{epsf}
\usepackage{amssymb}
\usepackage{epstopdf}
\usepackage{longtable,lscape}
\DeclareMathAlphabet{\mathsc}{OT1}{cmr}{m}{sc}
\def\testbx{bx}%
\DeclareRobustCommand{\ion}[2]{%
\relax\ifmmode
\ifx\testbx\f@series
{\mathbf{#1\,\mathsc{#2}}}\else
{\mathrm{#1\,\mathsc{#2}}}\fi
\else\textup{#1\,{\mdseries\textsc{#2}}}%
\fi}

\newcommand{\gps}{\ensuremath{g_{\rm P1}}}
\newcommand{\rps}{\ensuremath{r_{\rm P1}}}
\newcommand{\ips}{\ensuremath{i_{\rm P1}}}
\newcommand{\zps}{\ensuremath{z_{\rm P1}}}
\newcommand{\yps}{\ensuremath{y_{\rm P1}}}

\newcommand{\grizy}{\emph{griz}\yps}
\newcommand{\PS}{\protect \hbox {PS1}}

\newcommand{\apjs}{ApJS}
\newcommand{\apj}{ApJ}
\newcommand{\apjl}{ApJL}
\newcommand{\nat}{Nature}
\newcommand{\scie}{Science}
\newcommand{\procspie}{Proc. SPIE}
\newcommand{\mnras}{MNRAS}
\newcommand{\aaa}{A\&A}

\newcommand{\ap}{PS1-11ap}
\newcommand{\dam}{PTF12dam}
\newcommand{\xk}{SN2011ke}
\newcommand{\bi}{SN2007bi}
\newcommand{\gx}{SN2010gx}
\newcommand{\f}{PS1-10ahf}
\newcommand{\m}{PS1-10pm}

\newcommand{\bilike}{SN2007bi-like}

\newcommand{\MDS}{MDS}

\begin{document}

\title[\m\ \& \f]
{Selecting superluminous supernovae in faint galaxies from the first year of the Pan-STARRS1 Medium Deep Survey}

\author[M. McCrum et al.]
  {M.~McCrum,$^1$\thanks{E-mail: mmccrum04@qub.ac.uk}
  S. J.~Smartt,$^1$
  A.~Rest,$^2$
  K.~Smith,$^1$  
  R.~Kotak,$^1$ 
 S. A.~Rodney,$^3$\thanks{Hubble Postdoctoral Fellow}
D. R.~Young,$^1$
    \newauthor
 R.~Chornock,$^4$
 E.~Berger,$^4$
 R. J.~Foley,$^4$
M.~Fraser,$^1$
D.~Wright,$^1$
 D.~Scolnic,$^3$
  J. L.~Tonry,$^5$ 
 \newauthor
 Y.~Urata,$^6$
 K.~Huang,$^7$
  A.~Pastorello,$^8$
 M. T.~Botticella,$^9$ 
S.~Valenti,$^{10,11}$
 S.~Mattila,$^{12}$
\newauthor
 E.~Kankare,$^{12}$
D. J.~Farrow,$^{13}$
 M. E.~Huber,$^5$
 C. W.~Stubbs,$^4$
 R. P.~Kirshner,$^4$
 F.~Bresolin,$^5$ 
\newauthor 
W. S.~Burgett,$^{14}$
K. C.~Chambers,$^5$
P. W.~Draper,$^{13}$
H.~Flewelling,$^5$
 R.~Jedicke,$^5$
 N.~Kaiser,$^5$ 
\newauthor
 E. A.~Magnier,$^5$
 N.~Metcalfe,$^{13}$
  J. S.~Morgan,$^5$ 
  P. A.~Price,$^5$
 W.~Sweeney,$^5$
 R. J.~Wainscoat,$^5$
 \newauthor
  C.~Waters$^5$ \\
  $^1$Astrophysics Research Centre, School of Maths and Physics, Queen's University Belfast, Belfast BT7 1NN, UK\\
  $^2$Space Telescope Science Institute, 3700 San Martin Drive, Baltimore, MD 21218, USA\\
  $^3$Department of Physics and Astronomy, Johns Hopkins University, 3400 North Charles Street, Baltimore, MD 21218, USA\\
  $^4$Department of Physics, Harvard University, Cambridge, MA 02138, USA\\
  $^5$Institute for Astronomy, University of Hawaii at Manoa, Honolulu, HI 96822, USA\\
  $^6$Institute of Astronomy, National Central University, Chung-Li 32054, Taiwan\\
  $^7$Department of Mathematics and Science, National Taiwan Normal University, Lin-kou District, New Taipei City 24449, Taiwan\\
  $^8$INAF - Osservatorio Astronomico di Padova, Vicolo dell`Osservatorio 5, 35122 Padova, Italy\\
  $^9$INAF - Osservatorio astronomico di Capodimonte, Salita Moiariello 16, I- 80131 Napoli, Italy\\  
  $^{10}$Las Cumbres Observatory Global Telescope Network, 6740 Cortona Dr., Suite 102, Goleta, California 93117, USA \\
  $^{11}$Department of Physics, University of California Santa Barbara, Santa Barbara, CA 93106, USA \\
  $^{12}$Finnish Centre for Astronomy with ESO (FINCA), University of Turku, V{\"a}is{\"a}l{\"a}ntie 20, FI-21500 Piikki{\"o}, Finland \\
  $^{13}$Department of Physics, Durham University, South Road, Durham DH1 3LE, UK \\
  $^{14}$GMTO Corporation, 251 S. Lake Ave., Suite 300, Pasadena, CA  91101, USA \\
}
\maketitle

\begin{abstract} 
The Pan-STARRS1 (\PS) survey has obtained imaging in 5 bands (\grizy) over 10 Medium Deep Survey (MDS) fields covering a total of 70 square degrees. This paper describes the search for apparently hostless supernovae (SNe) within the first year of PS1 MDS data with an aim of discovering  superluminous supernovae (SLSNe). A total of 249 
hostless transients were discovered down to a limiting magnitude of $M_{\rm AB}\sim23.5$, of which 76 were classified as type Ia SNe. There were 57 SNe with complete light curves that are likely core-collapse SNe (CCSNe) or type Ic SLSNe and 12 of these have had spectra taken. Of these 12 hostless, non-type Ia SNe, 7 were SLSNe of type Ic at redshifts between 0.5-1.4. 
This illustrates that the discovery rate of type Ic SLSNe  can be maximised by concentrating on hostless transients and removing normal SNe\,Ia. We  present data for two possible SLSNe; PS1-10pm ($z=1.206$) and PS1-10ahf ($z=1.1$), and estimate the rate of type Ic SLSNe to be between 
$3^{+3}_{-2}\times10^{-5}$ and $8^{+2}_{-1}\times10^{-5}$ that of the CCSN rate within $0.3\leq z\leq1.4$ 
by applying a Monte-Carlo technique. The rate of slowly evolving, type Ic SLSNe (such as SN2007bi) is estimated as a factor of 10 lower than this range. 
\end{abstract}

\begin{keywords}
supernovae: general -- supernovae: individual: PS1-10pm, PS1-10ahf
\end{keywords}

\section{Introduction}
\label{intro}

Optical transients which are spatially coincident or associated with elongated and extended sources have a high probability of being supernovae (SNe).  Thus when searching for SNe, concentrating on high mass or intrinsically high luminosity galaxies can be a fruitful endeavour to optimise the yield of recorded events as it will maximise the number of stars observed that can potentially explode as SNe.  To date the majority of such searches at low redshift have adopted this approach to good effect, for example the Lick Observatory Supernova Search \citep{loss}.  The unbiased nature of  surveys like the Palomar Transient Factory \citep[PTF,][]{dwarfrates} and Pan-STARRS1 \citep[PS1,][]{PS1_system} mean that transients are being discovered without a bright galaxy bias and the neighbourhoods in which SNe are being found are not restricted to large, star forming galaxies.  \cite{dwarfrates} found that the core-collapse SNe (CCSNe) population in dwarf galaxies is different to that found in giant galaxies, in the sense that there are many more broad-lined, Type Ic SNe in the former population. They link this to the metallicity of the underlying stellar population and its effect on stellar evolution. 

Another class of SNe that have so far been discovered almost exclusively hosted by smaller, fainter galaxies is the relatively rare breed of superluminous SNe (SLSNe).
\cite{bluedeath} unravelled the mysteries of the luminous SN2005ap and SCP 06F6 by grouping them with a number of PTF discoveries to suggest these SLSNe as the death throes of at least some of the most massive of stars.  
Detailed studies of some of these events, such as SN2010gx
\citep[also known as PTF10cwr,][]{10gx,bluedeath} and PTF12dam \citep{12dam},
have increased the knowledge base on these highly energetic events and the recent \PS\ discoveries of  PS1-10ky, PS1-10awh \citep{10kyawh},  PS1-10bzj \citep{lun} and PS1-11ap \citep{11ap} are supporting and expanding progenitor and physical explosion mechanism theories.
The highest redshift discoveries ($z>1.5$) from PS1 of \cite{berger} and from the Supernova Legacy Survey \citep[SNLS,][]{2013arXiv1310.0470H} illustrate that SLSNe can be spectroscopically followed at significantly higher redshifts than Type Ia SNe (SNe\,Ia).

\cite{bluedeath} used the identification of narrow Mg\,{\sc ii} $\lambda\lambda$2796,2803 absorption lines from foreground gas to place robust lower limits on the redshift of these SLSNe, which immediately provided an estimate of the enormous luminosities. In a few cases the redshifts of the Mg\,{\sc ii} absorption exactly matches the emission lines of the host galaxy, confirming the reasonable assumption that the Mg\,{\sc ii} absorption arises in the host galaxy itself.
The redshifts derived by \cite{bluedeath} and then by \cite{10kyawh} using this method find peak absolute SLSNe magnitudes of $M_{u}\simeq-22\pm0.5$\,mag and total radiated energies $\gtrsim10^{51}$\,erg, making them substantially more luminous than any other SN-type events.

\begin{table*}
\caption{\PS\ Medium Deep Field Centres. }
\label{table:fields}
\begin{tabular}{lrr}
  \hline
  \hline
{\bf Field} & {\bf RA (degrees, J2000)} & {\bf Dec (degrees, J2000)} \\
    \hline
MD00  &  10.675 & $ 41.267$ \\
MD01  &  35.875 & $ -4.250$ \\
MD02  &  53.100 & $-27.800$ \\
MD03  & 130.592 & $ 44.317$ \\
MD04  & 150.000 & $  2.200$ \\
MD05  & 161.917 & $ 58.083$ \\
MD06  & 185.000 & $ 47.117$ \\
MD07  & 213.704 & $ 53.083$ \\
MD08  & 242.787 & $ 54.950$ \\
MD09  & 334.188 & $  0.283$ \\
MD10  & 352.312 & $ -0.433$ \\
MD11  & 270.000 & $ 66.561$ \\
\hline
 \end{tabular}
 \medskip
\end{table*}

\begin{table*}
\caption{\PS\ Medium Deep Survey, typical cadence. FM$\pm$3 designates 3 nights on either side of Full Moon.}
\label{table:cadence}
\begin{tabular}{ccc}
  \hline
  \hline
{\bf Night} & {\bf Filter} & {\bf Exposure Time} \\
    \hline
1         & \gps \& \rps & 8$\times$113s each\\
2         & \ips         & 8$\times$240s \\
3         & \zps         & 8$\times$240s \\
repeats... & . . . & . . . \\
FM$\pm3$  & \yps         & 8$\times$240s \\
\hline
 \end{tabular}
 \medskip
\end{table*}

Despite this common lower limit, differences in the photometric and spectroscopic evolution of the observed SLSNe suggests a number of progenitor possibilities.
\cite{10gx} find iron and other features normally associated with Type Ic SNe in the spectra of \gx\ at $30-50$ days after peak showing that the transient evolved to resemble an energetic Type Ic SN but on a much slower timescale.  \cite{11xk} present data on another five such objects at redshifts 0.1 - 0.2 with detailed modelling suggesting that the explosions are simply `normal' Type Ic SNe with an additional power source providing a boost to the luminosity.  The best fitting models presented by \cite{11xk} are those where a fast rotating neutron star (magnetar) provides the extra energy required, an idea proposed and developed by \cite{pulsara}, \cite{pulsarb} and \cite{magn1}.  Other power sources, such as the radioactive decay of $^{56}$Ni \citep{arnett}, are also explored but could not be reconciled with the light curves of SLSNe.
These kind of models were used by \cite{12dam} and \cite{11ap} to explain the slower photometric evolution of the SLSNe \dam\ and \ap, where again the magnetar models gave the most satisfactory fits to the data.  Prior to the discovery of these two SLSNe however, the only well studied object of this slowly evolving type was \bi\ \citep{07bi,DY} which was believed to have been the result of a pair-instability supernova \citep[PISN,][]{1stpisn, 2ndpisn, 3rdpisn, 4thpisn}.
\cite{snsh} suggest the interaction of the SN shock with a dense circumstellar material of H-poor material as a possible mechanism for the production of the observed features of the light curves and spectra.  This theory also accounts for the Ic-like features and the unusually high magnitudes associated with the aforementioned objects.  
Thus we will refer to these two subclasses of SLSNe, those sharing the properties of the \cite{bluedeath} sample and those that display the prolonged light curve evolution of SN2007bi, as SLSNe-Ic and slowly evolving SLSNe-Ic respectively.

A number of discoveries of a completely separate class of SLSNe that display strong H emission similar to the Type II classes of SNe have also been made.  SN2006gy \citep{06gya,06gyb} and SN2003ma \citep{03ma} are examples of this class, where the observed features are likely produced by the interaction between an  energetic SN explosion and a very dense circumstellar medium.  
Any objects of this class will be referred to here as SLSNe-II \citep{gal-yam}.
In contrast to the SLSNe-Ic,  SN2006gy and SN2003ma occured in bright host galaxies, suggesting that
 SLSNe-II do not follow the apparent trend visible for SLSNe Ic which have, almost exclusively, dwarf hosts.

An important feature of the known SLSNe-Ic sample is the noticeable preference for them appearing in faint host galaxies. This trend has been apparent right from the early discoveries with
 a faint host galaxy of no brighter than $M_{r}\sim-18$  found for SN2005ap \citep{05ap} and an upper  limit of $-18.1$ mag set at the pre-explosion location of SCP06F6 \citep{scp06f6}.  \cite{10gxgal} and \cite{bluedeath} found a faint dwarf host for SN2010gx in archive SDSS images (SDSS J112546.72-084942.0), with an absolute magnitude of $M_r\sim-18$ and a more recent study  further refined this value to $M_g=-17.42\pm0.17$ \citep{janet}. 
  \cite{DY} also calculated an absolute magnitude for the host of SN2007bi using SDSS archive images (SDSS J131920.14+085543.7) and found an $M_{B}\sim-16.4$.
A similar trend has been noted with virtually all subsequent discoveries \citep{bluedeath,10kyawh,10gx,06oz,scp06f6, janet,lun,2014arXiv1405.1325N}, which 
led \cite{2013arXiv1311.0026L} to study a large sample of host galaxies and suggest that they are similar hosts to those of Gamma-Ray Bursts (GRBs).  An exception to this supposed trend is the event so far reported with one of the highest redshifts.
PS1-11bam \citep{berger}, which has $z=1.55$ derived from strong absorption of both Fe\,{\sc ii} and Mg\,{\sc ii} and the detection of [OII] 3727\AA\ from the host galaxy in emission, has the most luminous host discovered so far for a SLSN-Ic,
with a near UV absolute magnitude of M$_{\rm NUV} \simeq -20.3$. 
However the host is still some 2 magnitudes fainter than the SN itself.

In summary, the currently known SLSNe-Ic sample typically appear to be $>2-4$ mag brighter than their hosts
\citep[see][for a compilaton of host magnitudes]{2013arXiv1311.0026L}
indicating that a simple way of isolating them in higher redshift 
searches could be to target transients with either no host detected or with a significant difference between total host luminosity and peak SLSN-Ic magnitude. It is striking that 
no SLSNe-Ic erupting in bright galaxies have  been uncovered by any previous wide-field survey or galaxy targeted low redshift searches. Although one might argue there is always a preference for observers to take spectra of isolated transients to avoid galaxy contamination effects, the very large
sample of low redshift supernovae now classified do not contain any examples of SLSNe Ic 
in galaxies close to $L^{\star}$  \citep[the characteristic luminosity in the Schechter luminosity function for galaxies, approximately corresponding to $M_{B}\simeq-21$][]{1976ApJ...203..297S}. This is 
somewhat surprising since the bulk of stellar mass, and star formation, occurs in galaxies close
to $L^{\star}$. 

The focus of this paper is to use this apparent preference for dwarf galaxy hosts as a  method of finding SLSNe-Ic in the first year of the \PS\ Medium Deep Survey (\MDS),  to quantify their numbers in a magnitude limited survey and to approximately estimate the volumetric rates between the redshift range of $z = 0.3 - 1.4$. \cite{quimbrate} previously estimated the SLSN-Ic rate to be $32^{+77}_{-26}$ events Gpc$^{-3}$yr$^{-1}h^{3}_{71}$ at a redshift of $z\sim0.2$ (although based on 1 event) and the SLSN-II rate to be $151^{+151}_{-82}$ events Gpc$^{-3}$yr$^{-1}h^{3}_{71}$ at $z \sim 0.15$.

At the beginning of the  PS1 survey we used reference images which were made
of a small number  ($\sim8$) of single input images from a 
good night. As the survey has progressed we have been able to 
stack the images to build much deeper template and stack images. 
However at the beginning of the PS1 survey and during the first year of operations, we used reference stacks which reached a limiting apparent magnitudes of $\sim23.5$ in the \gps\rps\ips-bands. 
Hence SNe which are brighter than 22, but have no host visible brighter than 23.5 were immediately candidates for hunting down these exciting phenomena. This paper focuses on all hostless transients discovered during the first year of the \PS\ survey operations.

\section{The \PS\ Medium Deep Survey}
\label{sec:observations}

\begin{figure*}
\begin{center}$
\begin{array}{cc}
\includegraphics[scale=0.3,angle=270]{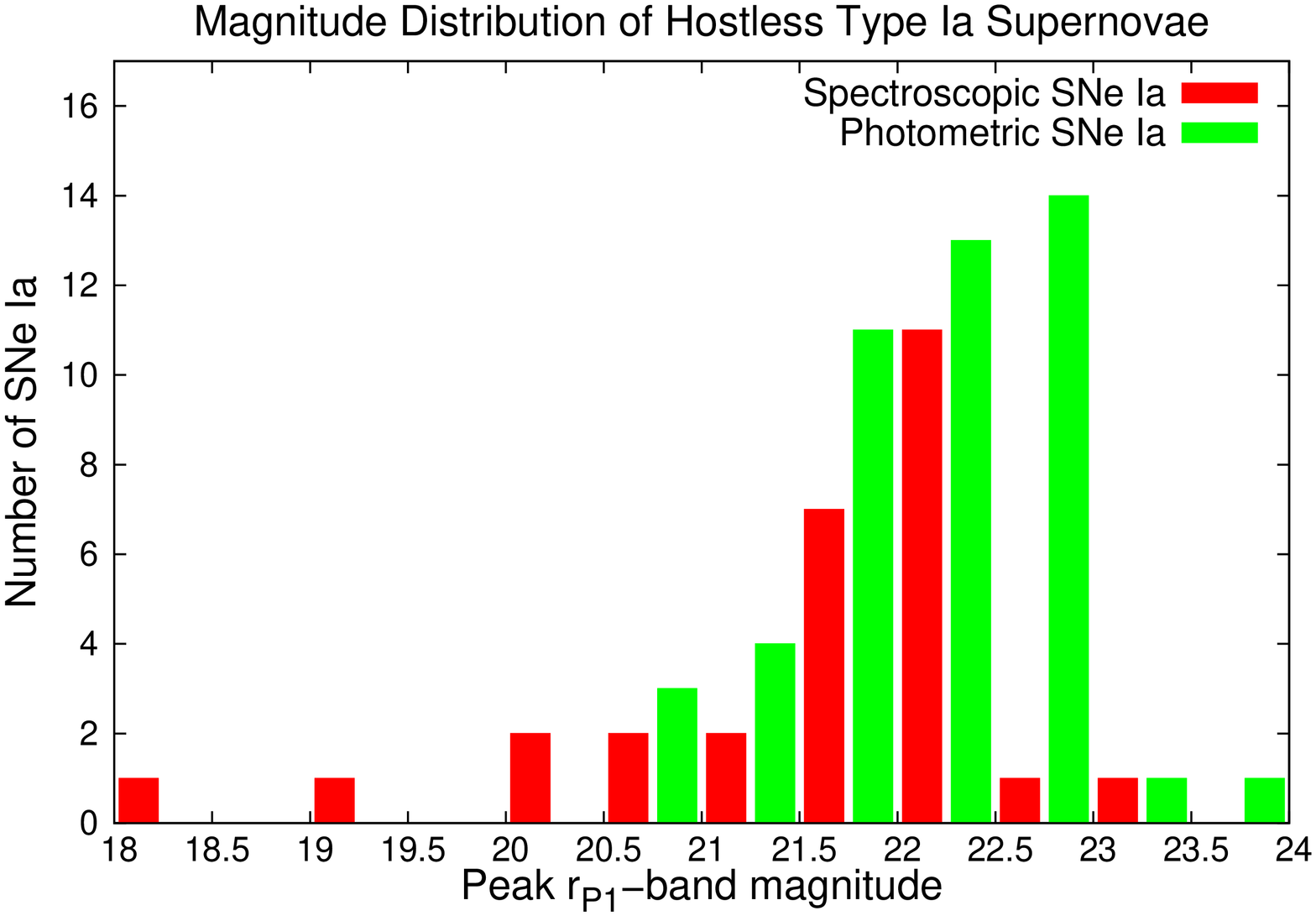} &
\includegraphics[scale=0.3,angle=270]{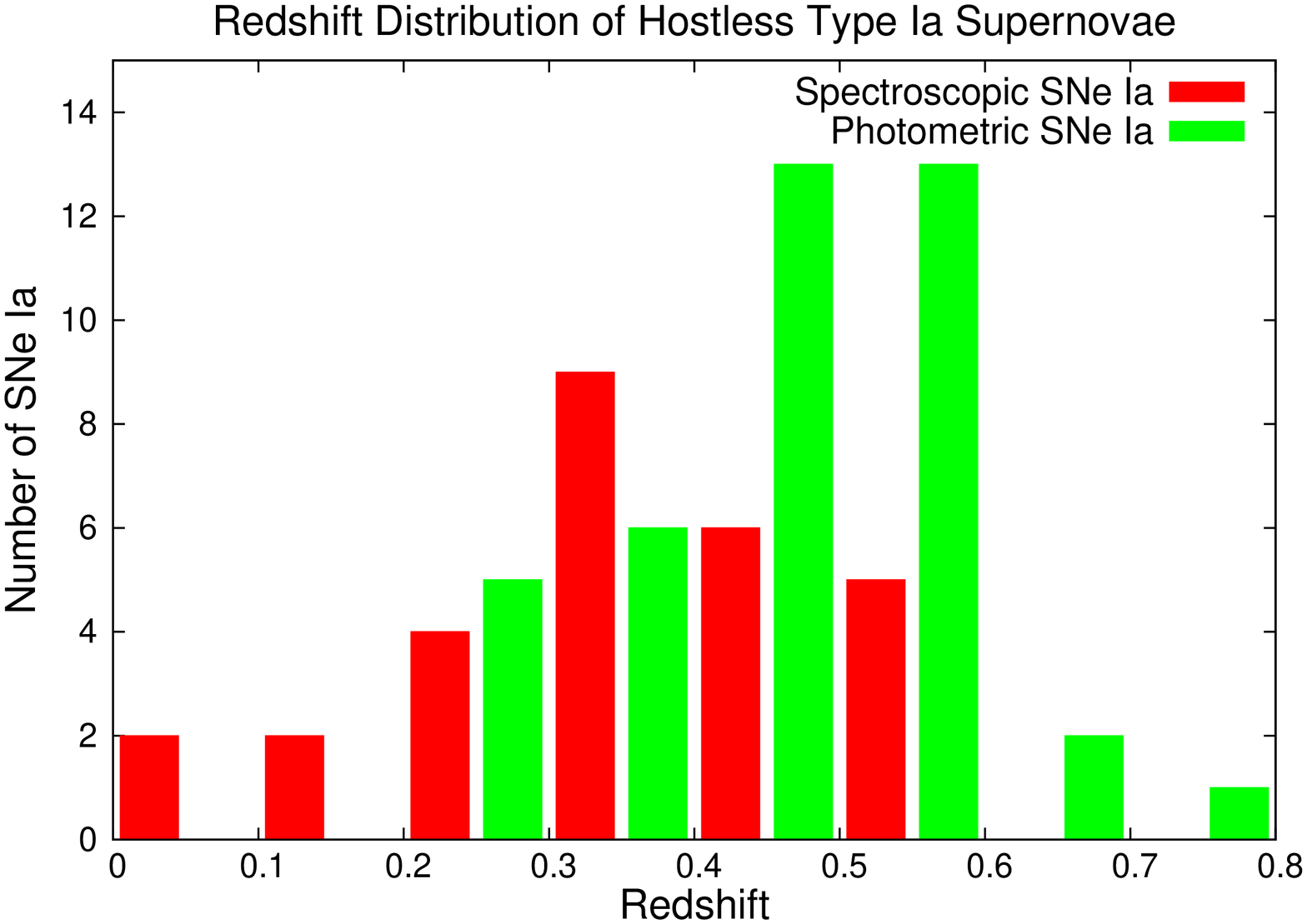} \\
\includegraphics[scale=0.3,angle=270]{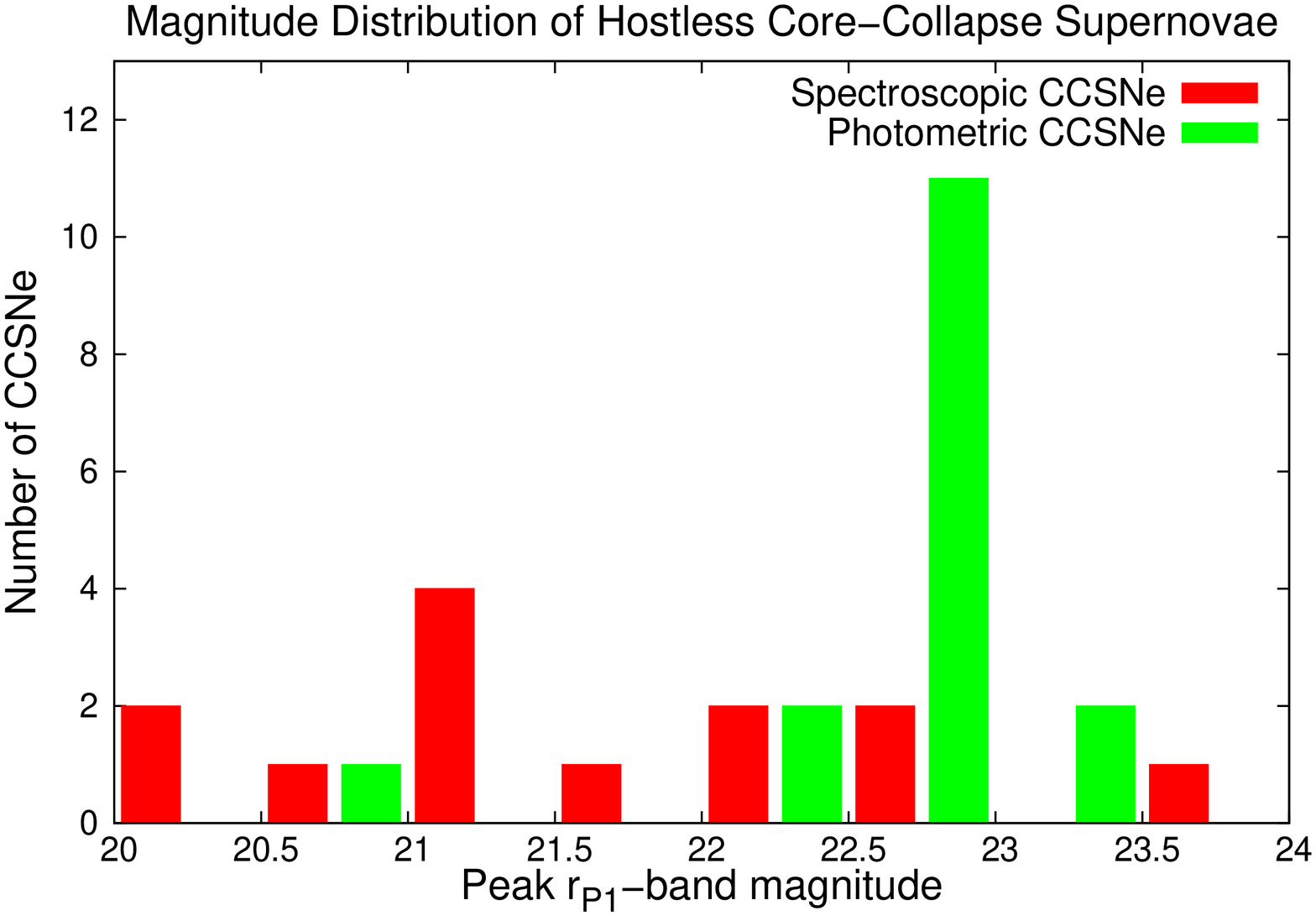} &
\includegraphics[scale=0.3,angle=270]{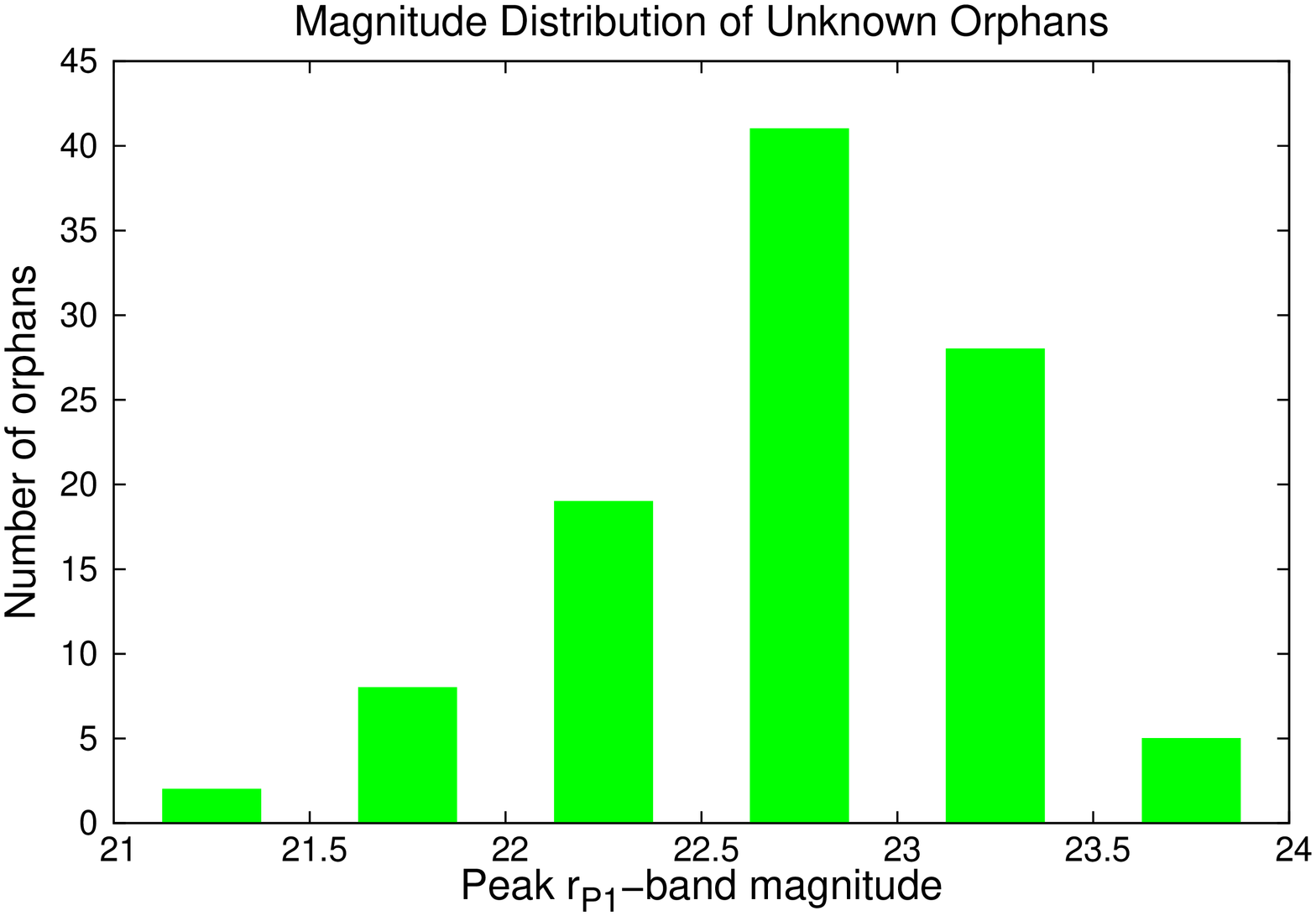}
\end{array}$
\end{center}
\caption{Magnitude (\rps band) and redshift distributions of the Type Ia SNe orphans and magnitude distributions of the core-collapse orphans and the miscellaneous orphans.  The classifications are split into two mutually exclusive classes, spectroscopic (red) classifications from a variety of telescopes and photometric (green) classifications using the \textsc{soft} and \textsc{psnid} algorithms from Rodney \& Tonry (2009) and Sako et al. (2008, 2011). Note that in the bottom left panel there are 16 Photometric CCSNe, whereas Table\,\ref{table:CC?} has 17. This is due to PS1-11ag not having \rps-band points at peak, hence we do not plot it.}
\label{fig:Ia}
\end{figure*}

The \PS\ system is a high-\'{e}tendue wide-field imaging system, designed for dedicated survey observations. The system is installed on the peak of Haleakala on the island of Maui in the Hawaiian island chain. 
 The telescope has a 1.8-m diameter primary mirror and the gigapixel camera (GPC1) located at the $f/4.4$ cassegrain focus consists of sixty 4800$\times$4800 pixel detectors (pixel scale 0.26$''$) giving a field of view of 3.3$^{\circ}$ diameter.
Routine observations are conducted remotely, from the Waiakoa Laboratory in Pukalani.  A more complete description of the \PS\ system, both hardware and software, is provided by \cite{PS1_system}. The survey philosophy and execution strategy are described in Chambers et al. (in preparation).

The \PS\ observations are obtained through a set of five broadband filters, which we have designated as \gps, \rps, \ips, \zps, and \yps.  Although the filter system for \PS\ has much in common with that used in previous surveys, such as SDSS \citep{SDSS}, there are important differences. The \gps\ filter extends 20~nm redward of $g_{SDSS}$, paying the price of 5577\AA\ sky emission for greater sensitivity and lower systematics for photometric redshifts, and the \zps\ filter is cut off at 930~nm, giving it a different response than the detector response which defined $z_{SDSS}$.  SDSS has no corresponding \yps\ filter.  Further information on the passband shapes is described in \cite{PS_lasercal}.  
The \PS\ photometric system and its response is covered in detailed in \cite{tonry12}.  Photometry is in the ``natural'' \PS\  system, $m=-2.5\mathrm{log}(flux)+m'$, with a single zeropoint adjustment $m'$ made in each band to conform to the AB magnitude scale. 
This paper uses images and photometry from the \PS\ \MDS, the observations of which are described in more detail in \cite{JTwds}, \cite{PS1a} and \cite{scol}.
 \PS\ has observed 12 \MDS\ fields, but we will only describe data from ten of them (MD01 to MD10). MD00 is centered on M31 and we have not been searching systematically for SNe in this field behind Andromeda. MD11 was observed for a short period in 2010-11, but has since been dropped from the \MDS\ schedule. The MD field centres and the exposure times in the 5 filters are listed in Tables  \ref{table:fields} and  \ref{table:cadence}.  Observations of between 3-5 MD fields are taken each night and the filters are cycled through in the following pattern : \gps\ and \rps\ in the same night (dark time), followed by \ips\ and \zps\ on the subsequent second and third night respectively. Around full moon only \yps\ data are taken. Any one epoch consists of 8 dithered exposures of either $8\times113$s for \gps\ and \rps\ or $8\times240$s for the other three, giving nightly stacked images of 904s and 1920s duration. 

Images obtained by the \PS\ system are processed through the Image Processing Pipeline (IPP) \citep{PS1_IPP}, on a computer cluster at the Maui High Performance Computer Center (MHPCC). The pipeline runs the images through a succession of stages including device ``de-trending'', a flux-conserving warping to a sky-based image plane, masking and artifact location. De-trending involves bias and dark correction and flatfielding using white light flatfield images from a dome screen, in combination with an illumination correction obtained by rastering sources across the field of view.   After determining 
an initial astrometric solution the flat-fielded images were then warped onto the tangent plane of the sky using a flux conserving algorithm. The plate scale for the warped images was originally set at 0.200 arcsec/pixel, 
but has since been changed to 0.25 arcsec/pixel in what is known internally as the V3 tesselation for the MD fields. 
Bad pixel masks are applied to the individual images and carried through the stacking stage to give the ``nightly stacks'' of 904s and 1920s total duration.  

\begin{table*}
\begin{center}
\caption{28 spectroscopically confirmed \PS\ SNIa.  For the objects marked with an `*', the spectral classifications for these as SNe Ia can be found in Rest et al. (2014). The MD06 object PS1-11acn is designated as PTF11dws in the cited source of classification.}
\label{table:Ia!}
\begin{tabular}{lcccccccc}
  \hline
  \hline
{\bf Field} & {\bf PS1 ID} & {\bf RA (deg, J2000)} & {\bf Dec (deg, J2000)} & {\bf \emph{z}$_{\bf SPEC}$} & {\bf Peak \emph{r}$_{\bf P1}$} & {\bf Telescope} & {\bf Date}  & {\bf Ref.} \\
    \hline
MD01 & PS1-10nu & 36.8001 & -4.5347 & 0.065 & 18.276 (0.004) & Magellan & 14/08/2010  & * \\
MD01 & PS1-11cn & 35.7730 & -3.6101 & 0.250 & 20.866 (0.028) & GN & 31/01/2011 & * \\
MD03 & PS1-1000026 & 129.0524 & 44.0071 & 0.140 & 20.370 (0.025) & WHT & 23/02/2010 & \cite{val1a} \\
MD03 & PS1-10bka & 131.6802 & 44.0035 & 0.247 & 22.179 (0.096) & GN & 31/01/2011 & * \\
MD03 & PS1-10bzt & 132.2495 & 44.8656 & 0.420 & 22.291 (0.224) & MMT & 28/12/2010 & * \\
MD04 & PS1-10iv & 150.5018 & 2.0604 & 0.369 & 21.478 (0.055) & GN & 14/05/2010 & * \\
MD04 & PS1-10l & 151.2314 & 2.5908 & 0.370 & 23.435 (0.201) & Magellan & 20/01/2010 & * \\
MD04 & PS1-11s & 150.7889 & 2.143 & 0.400 & 22.251 (0.091) & Magellan & 12/01/2011 & * \\
MD04 & PS1-11t & 150.5261 & 2.0877 & 0.450 & 21.863 (0.093) & Magellan & 12/01/2011 & * \\
MD04 & PS1-11p & 148.7919 & 1.7302 & 0.480 & 22.372 (0.132) & Magellan & 12/01/2011 & * \\
MD04 & PS1-11bh & 149.6284 & 2.8717 & 0.350 & 21.830 (0.072) & Magellan & 12/01/2011 & * \\
MD05 & PS1-10ix & 162.0978 & 57.1481 & 0.381 & 21.950 (0.283) & GN & 14/05/2010 & * \\
MD06 & PS1-11acn & 184.7782 & 47.3557 & 0.150 & 20.079 (0.010) & Keck & 13/06/2011 & \cite{galatel} \\   
MD06 & PS1-10kj & 183.5271 & 46.9923 & 0.350 & 22.136 (0.106) & MMT & 18/06/2010  & * \\
MD06 & PS1-11jo & 184.0016 & 47.9204 & 0.330 & 21.632 (0.045) & MMT & 23/02/2011 & * \\
MD06 & PS1-11xw & 184.9750 & 48.1402 & 0.270 & 21.318 (0.042) & MMT & 11/06/2011 & * \\
MD06 & PS1-11yr & 186.5881 & 46.5959 & 0.530 & 22.573 (0.131) & GN & 11/06/2011 & * \\
MD07 & PS1-10ig & 211.8623 & 53.3429 & 0.260 & 20.891 (0.035) & MMT & 04/04/2010 & * \\
MD07 & PS1-10iy & 214.1196 & 54.0535 & 0.443 & 22.171 (0.124) & GN & 15/05/2010 & * \\
MD07 & PS1-10iw & 214.4605 & 52.8010 & 0.447 & 21.845 (0.092) & GN & 15/05/2010 & * \\
MD07 & PS1-10kf & 212.9905 & 52.0718 & 0.450 & 22.188 (0.133) & MMT & 18/06/2010 & * \\  
MD07 & PS1-10kv & 212.6639 & 53.9895 & 0.530 & 22.242 (0.148) & GN & 07/07/2010 & * \\
MD07 & PS1-11zd & 214.6670 & 54.1830 & 0.100 & 19.214 (0.006) & WHT & 08/06/2011 & * \\
MD08 & PS1-10jz & 241.7032 & 54.9809 & 0.550 & 22.283 (0.145) & MMT & 18/06/2010 & * \\
MD08 & PS1-10jv & 244.4487 & 55.3022 & 0.360 & 21.586 (0.057) & MMT & 17/06/2010 & * \\
MD10 & PS1-10bjz & 353.2949 & -1.2353 & 0.310 & 21.503 (0.175) & MMT & 10/12/2010 & * \\
MD10 & PS1-10byj & 353.3821 & 0.1340 & 0.511 & 22.136 (0.118) & GN & 17/12/2010 & * \\
MD10 & PS1-10axm & 353.3285 & -0.9505 & 0.510 & 22.274 (0.224) & GN & 16/10/2010 & * \\
\hline
 \end{tabular}
 \medskip
\end{center}
\end{table*}

\begin{table*}
\begin{center}
\caption{48 plausible \PS\ SNIa, classified using the \textsc{soft} and \textsc{psnid} photometric classification codes (Rodney \& Tonry 2009; Sako et al. 2008, 2011).  The values in the \textsc{soft} and \textsc{psnid} columns represent the probability that the object is classified as a SNIa.}
\label{table:Ia?}
\begin{tabular}{lcccccccc}
  \hline
  \hline
{\bf Field} & {\bf PS1 ID} & {\bf RA (deg, J2000)} & {\bf Dec (deg, J2000)} & {\bf SOFT} & {\bf PSNID} & {\bf \emph{z}$_{\bf PHOT}$} & {\bf \emph{dz}} & {\bf Peak \emph{r}$_{\bf P1}$} \\
    \hline
MD01 & PS1-10aat & 34.6005 & -4.2033 & - & 0.999 & - & - & 22.532 (0.172) \\
MD01 & PS1-10bcd & 36.4632 & -5.4049 & 0.996 & 0.974 & 0.480 & 0.028 & 22.289 (0.078) \\
MD01 & PS1-10blx & 35.2726 & -3.8241 & 0.970 & 0.987 & 0.280 & 0.028 & 22.678  (0.181) \\
MD01 & PS1-10zv & 36.5108 & -4.1125 & - & 1.000 & - & - & 21.421 (0.041) \\
MD01 & PS1-10zz & 36.8222 & -3.2903 & 0.981 & 0.946 & 0.340 & 0.045 & 22.594 (0.104) \\
MD02 & PS1-10afj & 52.6562 & -28.3717 & - & 0.987 & 0.260 & 0.028 & 22.948 (0.192) \\  
MD02 & PS1-10bxr & 54.1715 & -28.3673 & 0.890 & 0.998 & 0.460 & 0.117 & 22.483 (0.227) \\  
MD03 & PS1-10ayn & 131.2068 & 43.8823 & 0.999 & 0.997 & 0.520 & 0.057 & 22.738 (0.132) \\
MD03 & PS1-10bkm & 129.6447 & 44.8684 & - & 1.000 & - & - & 21.510 (0.051) \\
MD03 & PS1-10cbs & 131.8921 & 44.5374 & 0.997 & 1.000 & 0.480 & 0.028 & 22.292 (0.116) \\
MD03 & PS1-11bw & 129.8787 & 43.9868 & 0.998 & 0.996 & 0.560 & 0.117 & 22.648 (0.178) \\
MD03 & PS1-11ex & 130.1915 & 43.8002 & 0.941 & 0.977 & 0.580 & 0.117 & 23.223 (0.165) \\
MD03 & PS1-11gs & 131.3379 & 44.6013 & 1.000 & 0.923 & 0.760 & 0.028 & -  \\  
MD04 & PS1-11du & 149.6446 & 1.2582 & 0.876 & 0.819 & 0.620 & 0.117 & 23.599 (0.295) \\
MD04 & PS1-11r & 150.2982 & 1.5754 & 0.954 & 1.000 & 0.400 & 0.045 & 22.357 (0.105) \\
MD05 & PS1-10uu & 162.7694 & 58.4253 & 0.963 & 0.865 & 0.400 & 0.126 & 22.093 (0.122) \\
MD05 & PS1-10wb & 159.8594 & 57.2051 & 0.998 & 1.000 & 0.480 & 0.146 & 21.748 (0.296) \\
MD05 & PS1-10jx & 160.7420 & 56.9001 & - & 0.955 & - & - & 21.568 (0.055) \\
MD05 & PS1-11bp & 161.9585 & 57.2893 & 0.998 & 0.998 & 0.480 & 0.128 & 21.911 (0.112) \\
MD05 & PS1-11oh & 162.7667 & 59.1037 & - & 0.981 & - & - & 21.465 (0.067) \\  
MD06 & PS1-11ql & 183.4322 & 46.2902 & - & 0.97 & - & - & 21.357 (0.054) \\
MD06 & PS1-11tc & 186.4050 & 46.7988 & - & 0.844 & - & - & 22.802 (0.316) \\
MD06 & PS1-10qu & 186.4676 & 47.8554 & 0.999 & 1.000 & 0.380 & 0.028 & 22.129 (0.103) \\
MD06 & PS1-10qv & 184.0243 & 47.3677 & 1.000 & 1.000 & 0.280 & 0.028 & 20.977 (0.049) \\
MD06 & PS1-10rj & 184.9626 & 47.7476 & 1.000 & 1.000 & 0.300 & 0.028 & 21.624 (0.065) \\
MD06 & PS1-10tb & 185.7344 & 46.9296 & 0.877 & 0.986 & 0.480 & 0.172 & 22.434 (0.134) \\
MD06 & PS1-10wn & 186.7645 & 46.8898 & 0.949 & 0.905 & 0.460 & 0.057 & 22.582 (0.219)\\ 
MD06 & PS1-10xc & 185.3965 & 46.0963 & 0.992 & 0.914 & 0.520 & 0.117 & 22.671 (0.120)\\ 
MD06 & PS1-10xe & 186.4073 & 46.7506 & 1.000 & 1.000 & 0.500 & 0.057 & 22.678 (0.127) \\
MD07 & PS1-11nb & 214.9577 & 53.3471 & - & 1.000 & - & - & 20.721 (0.030) \\
MD07 & PS1-10lb & 212.4660 & 53.7906 & 0.866 & 0.998 & 0.480 & 0.122 & 21.827 (0.081) \\
MD08 & PS1-10acd & 243.1638 & 55.0691 & 1.000 & 1.000 & 0.500 & 0.108 & 22.410 (0.181) \\
MD08 & PS1-10aex & 244.6181 & 55.2436 & 0.942 & 0.997 & 0.420 & 0.082 & 22.157 (0.131) \\
MD08 & PS1-10afb & 243.1847 & 56.0324 & 1.000 & 0.999 & 0.500 & 0.045 & 21.903 (0.077) \\
MD08 & PS1-10afq & 242.3236 & 56.4400 & 1.000 & 1.000 & 0.200 & 0.045 & 21.137 (0.054) \\
MD08 & PS1-10lh & 243.0192 & 56.1721 & 1.000 & 0.998 & 0.500 & 0.082 & 22.441 (0.168) \\
MD08 & PS1-10nf & 243.4779 & 54.1900 & 0.989 & 1.000 & 0.380 & 0.028 & 22.064 (0.149) \\
MD08 & PS1-10np & 240.4345 & 55.0052 & 1.000 & 0.954 & 0.540 & 0.100 & 22.521 (0.195) \\
MD08 & PS1-10zo & 240.3358 & 55.1399 & 0.988 & 0.997 & 0.520 & 0.146 & 21.927 (0.091) \\
MD09 & PS1-10aac & 334.6540 & 0.6184 & 1.000 & 1.000 & 0.600 & 0.028 & 21.970 (0.106) \\
MD09 & PS1-10afi & 333.5490 & 0.7323 & 0.964 & 0.957 & 0.500 & 0.072 & 22.868 (0.170) \\
MD09 & PS1-10axb & 332.6959 & 0.1951 & 0.947 & 0.966 & 0.520 & 0.141 & 22.663 (0.218) \\
MD09 & PS1-10ayl & 333.8780 & -0.8293 & 0.983 & 0.985 & 0.460 & 0.072 & 22.956 (0.210) \\
MD09 & PS1-10ls & 333.9379 & 0.4928 & 0.999 & 0.990 & 0.420 & 0.057 & 22.408 (0.236) \\
MD09 & PS1-10lw & 333.2522 & 0.8601 & 0.979 & 0.959 & 0.500 & 0.028 & 21.869 (0.058) \\
MD09 & PS1-10mi & 334.6715 & 0.4075 & 1.000 & 0.997 & 0.360 & 0.161 & 22.440 (0.211) \\
MD10 & PS1-10act & 353.0132 & 0.6024 & 1.000 & 1.000 & 0.240 & 0.028 & 20.857 (0.033) \\
MD10 & PS1-10lp & 351.3700 & -0.1235 & 1.000 & 1.000 & 0.360 & 0.045 & 21.646 (0.059) \\
\hline
 \end{tabular}
 \medskip
\end{center}
\end{table*}

\subsection{Image Subtraction Pipelines}
We have had two parallel difference image pipelines running since the start of full PS1 science operations in  May 2010. The \textsc{photpipe} pipeline \citep{Rest} is hosted at Harvard/CfA and this is the primary source of the final photometry 
presented in this paper.  We briefly outline the process below, but the reader is referred to \cite{PS1a}
for a full description of this pipeline.  
 This pipeline produces difference images from the MD nightly stacks compared to a 
deep, good image quality reference made from pre-season data.  Forced-centroid, point-spread function-fitting photometry is applied
on its difference images, with a point-spread function (PSF) derived from reference stars in each nightly stack. The zeropoints were
measured for the AB system from comparison with field stars in the SDSS catalog. We propagate
the Poisson error on the pixel values through the resampling and difference imaging. Since this does not take the covariance between neighbouring pixels
into account, we also do forced photometry in apertures at random positions and calculate
the standard deviation of the ratio between the flux and the error. We then multiply all errors by
the standard deviation to correct for the covariance. Nightly difference images typically yield 3$\sigma$
 limiting magnitudes of $\sim$23.5 mag in \gps, \rps, \ips, and \zps. 

In parallel, the \PS\ system has developed the Transient Science Server
(TSS) which uses the difference imaging and photometric data from the IPP running in 
Hawaii. This process was described initially in \cite{2012Natur.485..217G} and is repeated here, expanded
upon for completeness. 
The TSS automatically takes the nightly stacks created by the IPP
in the MHPCC, creates difference images with manually created reference
images, carries out PSF fitting photometry on the difference images
and returns catalogues of variables and transient candidates. In the current version 
forced photometry is not implemented. Mask
and variance arrays are carried forward at each stage of the IPP
processing. Photometric and astrometric measurements performed by the
IPP system are described in \cite{PS1_photometry} and
\cite{PS1_astrometry} respectively. Individual detections made on the
difference images are copied nightly from the MHPCC and ingested into a MySQL database (located at Queen's University) 
after an initial culling of objects based on the detection of saturated, masked or
suspected defective pixels within the PSF area. Sources detected on
the nightly difference images are assimilated into potential real
astrophysical transients based on a set of quality tests. The TSS
requires more than 3 quality detections within the last 7 observations
of the field, including detections in more than one filter, and an RMS
scatter in the positions of $\leq0.5"$.  Each of these quality
detections must be of $5\sigma$ significance (defined as an instrumental
magnitude error $<0.2^{m}$) {\em and} have a Gaussian morphology
($XY_{moments} < 1.2$). Transient candidates which pass this automated 
filtering system are promoted for human screening, which currently
runs at around 10\% efficiency (i.e. 10\% of the transients promoted
automatically are judged to be real after human screening). 

The overlap with the \textsc{photpipe} system is good, with most high significance, real transients found by both pipelines. Each pipeline has been used to inform the other of small numbers missed and the reasons.  Within the PS1 TSS, real transients are crossmatched with all available catalogues of astronomical sources in the MDS fields (e.g. SDSS, GSC, 2MASS, APM, Veron AGN, X-ray catalogues) in order to have a first pass classification of supernovae, variable star, active galactic nuclei (AGN) and nuclear transients.  While the difference imaging runs within the PS1 IPP system in the MHPCC, the TSS database is hosted at Queen's University Belfast. The long term goal is to fully integrate the TSS into the PS1 system in Hawaii.

\begin{table*}
\begin{center}
\caption{12 confirmed \PS\ CCSNe.  Of note is the high percentage of SLSNe-Ic confirmed.  The two objects marked with an `*' are explored further in this paper.  The RA and Dec values are in the deg, J2000 format.  The values in the \textsc{soft} and \textsc{psnid} columns represent the probability that the object is classified as the SNe type given in brackets.  The numbered references refer to the following papers; [1] McCrum et al. (2014), [2] This paper, [3] Lunnan et al. (2014), [4] Chomiuk  et al. (2011), [5] Chornock et al. (2013), [6] Quimby et al. (2013), [7] Quimby et al. (2014). PS1-10afx
has been shown to be a lensed SN Ia, and is included here for completeness of objects, although
we do not use it in any of our rate estimates.}
\label{table:CC!}
\begin{tabular}{lccccccccccc}
  \hline
  \hline
{\bf Field} & {\bf PS1 ID} & {\bf RA} & {\bf Dec} & {\bf Type} & {\bf \emph{z}$_{\bf SPEC}$} & {\bf SOFT} & {\bf PSNID} & {\bf Peak \emph{r}$_{\bf P1}$} & {\bf Telescope} & {\bf Date} & {\bf Reference} \\
    \hline
MD05 & PS1-11ap & 162.1155 & 57.1526 & SLSN Ic & 0.524 & - & 1 (II) & 20.217 (0.017) & NOT & 07/02/2011 & [1] \\
MD05 & PS1-11ad & 164.0876 & 57.6654 & IIn & 0.422 & 1 (II) & 0.992 (II) & 20.935 (0.053) & GN & 15/02/2011 & [2] \\
MD06 & PS1-10pm* & 183.1758 & 46.9915 & SLSN Ic & 1.206 & 1 (II) & 1 (II) & 22.090 (0.141) & GN & 03/06/2010 & [2] \\
MD06 & PS1-11afv & 183.9074 & 48.1801 & SLSN Ic & 1.407 & - & - & 22.211 (0.186) & GN & 09/07/2011 & [3] \\
MD07 & PS1-11yh & 212.7977 & 51.9868 & II & 0.146 & - & - & 21.189 (0.048) & MMT & 05/06/2011 & [2] \\
MD08 & PS1-11tt & 243.1907 & 54.0713 & SLSN Ic & 1.283 & - & - & 22.654 (0.167) & GN & 07/06/2011 & [3] \\
MD09 & PS1-10ky & 333.4076 & 1.2398 & SLSN Ic & 0.956 & - & - & 21.190 (0.077) & GN & 17/07/2010 & [4]\\
MD09 & PS1-10afx & 332.8507 & 0.1621 & Lensed SNIa & 1.388 & 0.999 (Ibc) & - & 23.730 (0.200) & GS & 06/09/2010 & [5,6,7] \\   
MD09 & PS1-10ahq & 333.5172 & 1.1084 &  Ic  & 0.283 & 1 (Ibc) & 1 (Ibc) & 21.474 (0.046) & MMT & 18/10/2010 & [2] \\
MD09 & PS1-10awh & 333.6242 & -0.0676 & SLSN Ic & 0.908 & 1 (II) & - & 21.607 (0.075) & GN & 12/10/2010 & [4] \\
MD10 & PS1-10acl & 352.4529 & -0.2916 & IIn & 0.260 & - & - & 21.313 (0.054) & MMT & 08/10/2010 & [2] \\
MD10 & PS1-10ahf* & 353.1180 & -0.3621 & SLSN Ic & 1.1 & 1(II) & 1 (II) & 22.680 (0.158) & GS & 11/06/2010 & [2] \\
\hline
 \end{tabular}
 \medskip
\end{center}
\end{table*}

\begin{table*}
\begin{center}
\caption{17 plausible \PS\ CCSNe.  The values in the \textsc{soft} and \textsc{psnid} columns represent the probability that the object is classified as the Type II sub-class listed.}
\label{table:CC?}
\begin{tabular}{lccccccc}
  \hline
  \hline
{\bf Field} & {\bf PS1 ID} & {\bf RA (deg, J2000)} & {\bf Dec (deg, J2000)} & {\bf Type} & {\bf SOFT} & {\bf PSNID} & {\bf Peak \emph{r}$_{\bf P1}$}\\
    \hline
MD01 & PS1-10acp & 35.4861 & -3.8369 & IIL & 1 & 1 & 22.699 (0.147) \\
MD01 & PS1-10add & 35.0764 & -4.1370 & IIP & 1 & 1 & 22.170 (0.080) \\
MD03 & PS1-10axq & 130.5385 & 44.5480 & IIL & 1 & 1 & 22.929 (0.154) \\
MD04 & PS1-10dq & 150.2418 & 2.2475 & IIP & 1 & 1 & 22.568 (0.128) \\  
MD04 & PS1-11ag & 149.2396 & 3.2529 & IIL & 1 & 1 & - \\  
MD04 & PS1-11er & 149.1978 & 2.4177 & IIL & 0.959 & 0.999 & 23.427 (0.340) \\
MD06 & PS1-10sq & 186.3077 & 46.6273 & IIL & 1 & 0.996 & 22.624 (0.133) \\
MD06 & PS1-10vu & 184.9374 & 46.1197 & IIL & 0.999 & 0.804 & 22.642 (0.107) \\
MD07 & PS1-10wk & 211.6645 & 52.4217 & IIP & 1 & 1 & 22.515 (0.140) \\
MD08 & PS1-10acq & 244.6603 & 55.1885 & IIP & 1 & 0.999 & 22.880 (0.331) \\  
MD09 & PS1-10aal & 334.1762 & -0.5736 & IIL & 1 & 1 & 22.358 (0.139) \\
MD09 & PS1-10abf & 334.0138 & 1.0291 & IIL & 1 & 1 & 23.431 (0.250) \\
MD09 & PS1-10agf & 334.7398 & -0.1656 & IIP & 0.999 & 0.911 & 22.573 (0.328) \\
MD09 & PS1-10aht & 334.9031 & 1.2136 & IIP & 0.889 & 0.928 & 22.935 (0.201) \\  
MD10 & PS1-10acn & 351.7261 & -0.2600 & IIL & 1 & 1 & 22.985 (0.271) \\  
MD10 & PS1-10kz & 351.2487 & -0.3529 & II & - & 1 & 20.970 (0.032) \\
MD10 & PS1-10ayg & 350.9872 & -0.3808 & IIL & 1 & 0.837 & 22.649 (0.118) \\
\hline
 \end{tabular}
 \medskip
\end{center}
\end{table*}

\section{The Transient Sample}
\label{sec:results}

From the period starting February 25$^{th}$ 2010 and ending July 9$^{th}$ 2011, 249 hostless transients or `orphans'  were discovered in the \PS\ Medium Deep fields.  For the practical purposes of this paper, which will become clear for our scientific motivations, an orphan is defined as an object that is $>3.4''$ away from the centre of a catalogued galaxy or point source brighter than approximately 23.5$^{m}$ (in any of the \gps\rps\ips filters that the transient was detected in).   This magnitude limit was chosen for two reasons. At the beginning of the search period in 2010 the limit of a reference stack was not significantly deeper than the nightly stack, hence the transients which were observed as hostless by definition had no host brighter than $23.5^{m}$ in the specific band that they were detected in. Although the reference stacks now reach around 1 magnitude deeper, this limit is still useful as the transients we discuss in this paper are typically brighter than $22 - 22.5 ^{m}$, hence significantly brighter than the their hosts.   In many cases deeper images \citep{2013arXiv1311.0026L}, or deep \PS\ stacks \citep{JTwds} do indeed reveal a host or stellar counterpart.
Deep imaging in the \emph{z}-band with the Subaru telescope revealed hosts for 5 of our orphans with magnitudes listed in the Appendix in Table \ref{table:subaru}.
As can be seen in the table, two of these transients do have hosts brighter than $23.5^{m}$ in the $z$-band. Whilst our definition of `hostless' here is somewhat arbitrary, it is reasonably well defined and serves the science motivations of this paper well as we shall see. 

We set out with a goal of spectroscopically classifying as many of this hostless sample as possible, which peaked brighter than approximately $22 - 22.5$ in any of the \gps\rps\ips\ filters. This magnitude limit was chosen as a practical limit of the largest aperture telescopes (8 metre) that we had significant access to. The primary source was the
Gemini observatory, for which we had UK, US and UH time access although we also used the UK resources of 
the 4.2m William Herschel Telescope and the CfA Harvard access to Magellan and the Multi-mirror Telescope. 
A combination of the finite time resources available for these spectroscopic programme, ambient weather conditions, field visibility and scheduling, meant that spectra could not be obtained for every transient brighter than our chosen limit. The general PS1 spectroscopic follow-up of transients is also described in \cite{PS1a} and we emphasise that during the period described here, there were several multi-purpose spectroscopic classification and follow-up programmes running at these facilities which combined extensive classification with follow-up of scientifically interesting targets. Due to the ease of observing hostless transients (because of a lack of host galaxy contamination), a significant effort was invested to classify as many as possible. The results of the spectroscopic classification programmes are listed in Tables\,\ref{table:Ia!} and \ref{table:CC!} \citep[virtually all of the Type Ia sample are discussed in the cosmological analysis of][]{PS1a}. 

This sample is of course not spectroscopically complete, as illustrated in Fig.\,\ref{fig:Ia}. Our next step was to attempt photometric classification of those SNe for those which we did not manage to get spectra, but for which we had well sampled and relatively complete light curves. This sample included all those candidates for which we weren't able to take spectra and also those which had peak magnitudes too faint for inclusion in the spectroscopic typing programmes.  
We did not use this photometric fitting method for selecting targets for spectroscopy, since the lightcurve fitting methods (PSNID and SOFT) require a well sampled and complete lightcurve which obviously is not available when a spectrum needs to be triggered around peak brightness.
As expected, 
the distribution for the photometrically classified objects peaks at about a magnitude fainter than that of the spectroscopic sample. 
The light curve fitting is described below in Sections\,\ref{Ia} and \ref{ccnse-slsne}

At a canonical redshift of $z\sim0.2$, the $3.4$ arcsec separation corresponds to a minimum separation of $\sim11$\,kpc.
\cite{sullivan} used a host-association algorithm to ensure that the SNe\,Ia under study were being associated with appropriate host galaxies.
This was important for conclusions within the paper involving specific host galaxy properties, such as the star formation rate (SFR), however
for the purposes of this study the $3.4$ arcsec separation parameter we defined was sufficient.  
Our motivation is simply that a transient has no apparent host, we are not concerned with finding the most likely offset galaxy. The two possibilities for these orphans are either that the host is fainter than $\rps\sim23.5$ or that the transient has been expelled by a nearby galaxy and has a long enough lifetime that it can travel $>10$\,kpc before some energetic event generates sufficient luminosity to be captured by \PS.  Due to the shorter life cycle of CCSNe, runaway transients are likely SNe\,Ia or at least some type of thermonuclear event involving a WD \citep[for example see the transients described in][which are significantly offset from their likely host galaxies]{faintsne}. 
The alternative, where the host galaxy is simply too faint to be imaged by \PS\ suggests that the transient could be intrinsically bright or the galaxy 
intrinsically faint. For example, the host galaxy could be undetected simply because it is at a high redshift which places it beyond the PS1 detection limit. If the host were a typical $L^{\ast}$ galaxy (but undetected due to distance) this would imply the transient has an AB magnitude $\lesssim-22$.  Alternatively the transient could have a typical 
core-collapse magnitude of $M_{\rm AB}\sim-18$ and if the host galaxy is undetected then it is likely an intrinsically 
faint dwarf, possibly of low metallicity, but either way this combination has previously been found to be associated with SLSNe-Ic \citep{bluedeath,10kyawh}. They are not always dwarf galaxies as indicated by the high-\emph{z} discovery of \cite{berger} but they are typically, with very few exceptions in the current SLSNe-Ic sample,
at least 2\,mag brighter than their hosts.  The small subset of events which do not meet this criterion are all at higher redshifts, possibly indicating trends of metallicity and luminosity in these younger galaxies.  This is explored in greater detail in Section\,\ref{sec:discussion}.
As we show below, this search method allows a fairly straightforward way of identifying high-\emph{z}, superluminous transients. 

Through the spectroscopic classification programmes discussed above, optical spectra were taken for  40 orphans in total. While we would have liked to be spectroscopically complete to a defined 
magnitude limit (around $22^{m}$), spectroscopic facility access and weather constraints 
did not allow it. As with all spectroscopic programmes, there is some human preference that 
plays a role in target selection. In PS1, we have been particularly looking for transients that 
might be a high-z which could mean red colors and/or slow rise times.  Later in the paper we 
discuss an estimate of the volumetric rates of SLSNe and the spectroscopic completeness
plays a role in this. The number of SLSNe in the transient set without spectra (Tables\,\ref{table:CC?} and \ref{table:CC??}) is then the important question which we discuss in Section\,\ref{sec:MC-rates}.

\subsection{Hostless Type Ia supernovae}
\label{Ia}

Of these 40 transients which were spectroscopically confirmed, 28 turned out to be Type Ia SNe at redshifts between approximately $0.2-0.7$ and these are listed in Table \ref{table:Ia!}.  All but two of these are already presented in \citep{PS1a}, with one extra from \cite{val1a} and one more which is 
the same object as discovered and reported by PTF \citep{galatel} \footnote{We thank Avishay Gal-Yam and Peter Nugent for access to the spectrum to confirm classification.}.

 All transients for which no spectra were available (and which had relatively complete light curves) were passed through the  \textsc{soft} and \textsc{psnid} photoclassification algorithms 
\citep{SOFT,SAKO,SAKO2}.  We  identified 48 SNe as having light curves matching SNe\,Ia between redshifts 0.2 and 0.7. In order to make these confident Type Ia classifications we initially demanded that {\em both} the \textsc{soft} and \textsc{psnid} algorithms gave a SN\,Ia classification with a probability of $>$ 80$\%$. This resulted in 40 SNe\,Ia, however we found a further 8 that failed to get a secure classification in \textsc{soft} but \textsc{psnid} returned a high probability of being a SN\,Ia. A visual inspection of these light curves suggests to us that they are plausible SNe Ia and all 48 photometrically classified SNe Ia are listed in Table \ref{table:Ia?}. 

The redshift values and their associated errors that are listed in this table are output from the \textsc{soft} photoclassification code.  The method for generating these numbers is described in \cite{SOFT2}, including a figure comparing the photo-\emph{z} of a test sample of SNe\,Ia against \emph{z} values obtained from spectroscopic data.  As can be seen in the paper, the RMS scatter about $z_{SOFT}=z_{SPEC}$ is very small ($\sim0.05$) indicating that the photo-\emph{z} values offer a fair approximation of the actual redshift values of the probable SNe in question.

In summary, through spectroscopy and light curve fitting we find that 76 of the orphan transient sample 
are confidently identified as SNe\,Ia.

\subsection{Core-collapse and superluminous supernove}
\label{ccnse-slsne}

The other 12 transients for which we gathered spectra are listed in Table\,\ref{table:CC!} with their
redshifts and classifications.  Of these 12 spectroscopically confirmed SNe, there are four 
normal CCSNe, of types II, IIn and Ic. 
 \cite{10afx} presented the discovery of PS1-10afx at a redshift of $z=1.388$, suggesting it to be a SLSN which is different to the currently known population. 
But this has now been shown to be more likely a normal SN\,Ia lensed by a foreground galaxy 
\cite{2013ApJ...768L..20Q, quimblens}.  We include it in Table\,\ref{table:CC!} for completeness, 
as we originally had identified is  a transient which was not obviously a normal type Ia supernova, although we do not use it any  further in SLSN rate calculations. This leaves  seven which are confirmed as SLSNe-Ic, lying at redshifts beyond $z\sim0.5$.  These events include PS1-10pm and PS1-10ahf, the nature of which are discussed further in this paper.  Detailed analysis of PS1-10ky, PS1-10awh, PS1-11ap can be found in \cite{10kyawh} and \cite{11ap}.  The 
classifications of PS1-11tt and PS1-11afv are presented in \cite{2013arXiv1311.0026L} (and more details will be given in Lunnan et al. in prep.). This immediately suggests quite a high fraction of 
SLSNe, if one could remove the SNe\,Ia from the sample efficiently and early enough.

There are another 45 transients for which we were unable to get spectroscopic confirmation, but have 
well sampled and complete light curves which resemble CCSNe rather than SNe\,Ia. 
We passed these light curves through the  \textsc{psnid} and \textsc{soft} photoclassification algorithms, 
finding that  17 were classified as Type II by both algorithms with greater than 80\% confidence. These
are listed in Table\,\ref{table:CC?}, and we propose that they are likely to be core-collapse, Type II SNe given the high confidence light curve fits by both light curve fitters.  The other 28 transients had light curves which appeared SN-like but the fitting algorithms gave lower confidence levels for specific fits of SNe\,Ia, Type II or Type Ibc. 
These 28 transients are listed in Table \ref{table:CC??}, along with the lower confidence results from 
 \textsc{psnid} and \textsc{soft}. Additionally, 
a selection of these light curves are shown in Fig.\,\ref{fig:CCLC} and show typical trends such as a 
single asymmetric peak or an extended, declining plateau. We propose that these are high confidence
SNe but the classification of the light curve as SNe Ia, Ibc or II is uncertain. 

Fig.\,\ref{fig:Ia} shows the magnitude distribution of the confirmed and plausible CCSNe, which again illustrates the spectroscopic and photometric limits of our sample. A simple conclusion from this is that if one selects orphan candidates from a wide-field, magnitude limited survey (such as the PS1 MDS)  and one photometrically selects objects from the data stream with light curves which are unlike SNe\,Ia, then a large fraction of the brighter objects that remain are actually high redshift SLSNe-Ic. Of course there is still an outstanding question about how to identify the SNe\,Ia early enough in their light curve that one can securely identify them from photometry alone. This issue still remains open, but we show here that if it can be done then a large fraction of the transients
brighter than about 22$^{m}$ are SLSNe-Ic candidates. 

One major caveat to this is that there may be more SLSN-Ic candidates in the photometrically classified 
samples (either the Type Ia or core-collapse samples, or both), since the light curve fitters do not contain SLSNe-Ic (or SLSNe-II) template light curves.  As a check, we report the values output from \textsc{soft} and \textsc{psnid} for the light curves of the 
spectroscopically confirmed, hostless CCSNe sample (see Table \ref{table:CC!}).
None of the CCSNe sample were photometrically misclassified as SNe Ia, which supports our proposal that
the sample in Table\,\ref{table:Ia?} is relatively pure.

However two of the SLSNe-Ic would be misclassified as Type II SNe. Hence it is possible that there are
further SLSNe-Ic masquerading as normal CCSNe in the objects in Tables \ref{table:CC?}
and \ref{table:CC??}.  The lower left hand histogram in Fig.\,\ref{fig:Ia} again highlights the magnitude
limit of approximately $22-22.5$ in the spectroscopic observations. The spectroscopically confirmed
sample peaks at a significantly brighter \rps-band magnitude than the photometrically classified sample.  It is therefore possible that some of the photometrically classified CCSNe presented in Table \ref{table:CC?} could be unclassified SLSNe-Ic. When we discuss the rates of SLSNe-Ic, we note that they may be lower limits. 

In summary, our spectroscopic follow-up programmes took spectra of 12 transients which were not 
Type Ia SNe. A large fraction of these (7) were confirmed to be SLSNe-Ic at redshifts greater than 
$z\sim0.5$, and another is an unusually luminous transient at $z=1.388$. We photometrically 
classified 17 transients as plausible CCSNe and a further 28 had lower confidence photometric classifications.

\subsection{Miscellaneous orphans}

The remaining 116 hostless transients discovered by the \PS\ survey have insufficient data for any reliable classification to be made due to light curve data either being incomplete, variable or the object being too faint and only sneaking into the limiting magnitude the survey can reach.  As can be seen in the lower right hand plot in Fig.\,\ref{fig:Ia} these objects represent the fainter end of the detected orphans and so only limited data is currently available concerning them.  A typical trend seen in some of these orphans is that of a discrete peak suggesting that some of the objects are at a high redshift and peak just above our magnitude limit giving us an incomplete light curve. It is probable that these
are SNe or SNe-like transients. It is interesting to note that the
spectroscopically and photometrically classified samples do not contain any obvious AGN or 
QSO type variable sources, suggesting that the likelihood of finding such black-hole driven 
events is low if an underlying galaxy is not detected. We find many AGN and QSO variables in the 
full PS1 MDS transient search, but imposing the requirement for a host brighter than $\sim 23.5$
does seem to reduce their detected frequency.

\section{SLSNe-Ic analysis}
\label{sec:analysis}

\begin{figure}
\begin{center}
\includegraphics[angle=270,scale=0.36]{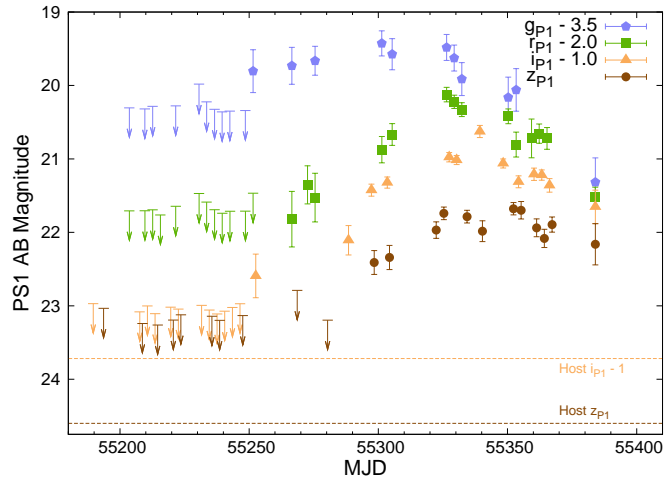}
\caption{Observed \emph{gri}\zps\ light curves of PS1-10pm.  The non \PS, WHT detections are the \emph{griz} points shown at MJD 55384 and any upper limits are marked with arrows.  Measurements of the host galaxy from late-time, deep GN \emph{i}- and \emph{z}-band images are shown as horizontal lines.  During the 2011 season, between MJD 55594.56 and 55736.33, there have been 10 non-detections in \gps\ down to a 3$\sigma$ indicative magnitude limit of 23.89, 12 non-detections in \rps\ down to 23.90, 12 non-detections in \ips\ down to 24.10 and 10 non-detections in \zps\ down to a limit of 23.24.  Over the entire course of the observations there are also 24 non-detections in \yps, however given that the limiting magnitude for this filter only reached a maximum magnitude of 21.86, observations of an object at the redshift of PS1-10pm were unfeasible.}
\label{fig:10pmoblc}
\end{center}
\end{figure}

\begin{table*}
\tiny
\caption{Observed photometry for PS1-10pm.  No \emph{K}-corrections have been applied and the phase values have not been corrected to the restframe.  Note that the \PS\ observations have had any flux from a previous reference image removed through image subtraction (although note that no host object can be seen at the location of PS1-10pm) whereas the late time WHT and GN observations have not.}
\label{table:10pm}
\begin{tabular}{c c c c c c c c c c c c}
  \hline
  \hline
{\bf Date} & {\bf MJD} & {\bf Phase (days)} & {\bf g$_{P1}$} & {\bf r$_{P1}$} & {\bf i$_{P1}$} & {\bf z$_{P1}$} & {\bf \emph{g}} & {\bf \emph{r}} & {\bf \emph{i}} & {\bf \emph{z}} & {\bf Telescope} \\
    \hline
24/12/2009 & 55189.65 & -136.35 & - & - & $>$23.97 & - & - & - & - & - & PS1 \\
28/12/2009 & 55193.65 & -132.35 & - & - & - & $>$23.03 & - & - & - & - & PS1 \\
07/01/2010 & 55203.61 & -122.39 & $>$23.80 & $>$23.71 & - & - & - & - & - & - & PS1 \\
11/01/2010 & 55207.65 & -118.35 & - & - & $>$24.08 & - & - & - & - & - & PS1 \\
12/01/2010 & 55208.65 & -117.35 & - & - & - & $>$23.24 & - & - & - & - & PS1 \\
13/01/2010 & 55209.64 & -116.36 & $>$23.82 & - & - & - & - & - & - & - & PS1 \\
13/01/2010 & 55209.65 & -116.35 & - & $>$23.71 & - & - & - & - & - & - & PS1 \\
14/01/2010 & 55210.61 & -115.39 & - & - & $>$24.00 & - & - & - & - & - & PS1 \\
16/01/2010 & 55212.64 & -113.36 & $>$23.78 & $>$23.69 & - & - & - & - & - & - & PS1 \\
17/01/2010 & 55213.56 & -112.44 & - & - & $>$24.11 & - & - & - & - & - & PS1 \\
18/01/2010 & 55214.56 & -111.44 & - & - & - & $>$23.26 & - & - & - & - & PS1 \\
19/01/2010 & 55215.63 & -110.37 & - & $>$23.76 & - & - & - & - & - & - & PS1 \\
23/01/2010 & 55219.65 & -106.35 & - & - & $>$24.02 & - & - & - & - & - & PS1 \\
24/01/2010 & 55220.63 & -105.37 & - & - & - & $>$23.19 & - & - & - & - & PS1 \\
25/01/2010 & 55221.56 & -104.44 & $>$23.78 & $>$23.64 & - & - & - & - & - & - & PS1 \\
26/01/2010 & 55222.59 & -103.41 & - & - & $>$24.04 & - & - & - & - & - & PS1 \\
27/01/2010 & 55223.57 & -102.43 & - & - & - & $>$23.12 & - & - & - & - & PS1 \\
03/02/2010 & 55230.60 & -95.40 & $>$23.48 & $>$23.47 & - & - & - & - & - & - & PS1 \\
04/02/2010 & 55231.58 & -94.42 & - & - & $>$23.99 & - & - & - & - & - & PS1 \\
06/02/2010 & 55233.53 & -92.47 & $>$23.72 & $>$23.59  & - & - & - & - & - & - & PS1 \\
07/02/2010 & 55234.56 & -91.44 & - & - & $>$24.06 & - & - & - & - & - & PS1 \\
08/02/2010 & 55235.59 & -90.41 & - & - & - & $>$23.14 & - & - & - & - & PS1 \\
09/02/2010 & 55236.60 & -89.40 & $>$23.82 & - & - & - & - & - & - & - & PS1 \\
09/02/2010 & 55236.61 & -89.39 & - & $>$23.69 & - & - & - & - & - & - & PS1 \\
10/02/2010 & 55237.57 & -88.43 & - & - & $>$24.11 & - & - & - & - & - & PS1 \\
11/02/2010 & 55238.57 & -87.43 & - & - & - & $>$23.20 & - & - & - & - & PS1 \\
12/02/2010 & 55239.54 & -86.46 & $>$23.86 & $>$23.74 & - & - & - & - & - & - & PS1 \\
13/02/2010 & 55240.53 & -85.47 & - & - & $>$24.07 & - & - & - & - & - & PS1 \\
15/02/2010 & 55242.54 & -83.46 & $>$23.85 & $>$23.71 & - & - & - & - & - & - & PS1 \\
16/02/2010 & 55243.53 & -82.47 & - & - & $>$24.02 & - & - & - & - & - & PS1 \\
19/02/2010 & 55246.46 & -79.54 & - & - & $>$23.97 & - & - & - & - & - & PS1 \\
20/02/2010 & 55247.54 & -78.46 & - & - & - & $>$23.13 & - & - & - & - & PS1 \\
21/02/2010 & 55248.53 & -77.47 & $>$23.84 & $>$23.71 & - & - & - & - & - & - & PS1 \\
24/02/2010 & 55251.51 & -74.49 & 23.31 (0.29) & $>$23.47 & - & - & - & - & - & - & PS1 \\
25/02/2010 & 55252.55 & -73.45 & - & - & 23.59 (0.30) & - & - & - & - & - & PS1 \\
11/03/2010 & 55266.54 & -59.46 & 23.23 (0.25) & 23.68 (0.38) & - & - & - & - & - & - & PS1 \\
13/03/2010 & 55268.58 & -57.42 & - & - & - & $>$22.78 & - & - & - & - & PS1 \\
17/03/2010 & 55272.58 & -53.42 & - & 23.35 (0.26) & - & - & - & - & - & - & PS1\\ 
20/03/2010 & 55275.52 & -50.48 & 23.16 (0.20) & 23.53 (0.33) & - & - & - & - & - & - & PS1\\ 
25/03/2010 & 55280.35 & -45.65 & - & - & - & $>$23.20 & - & - & - & - & PS1\\  
02/04/2010 & 55288.44 & -37.56 & - & - & 23.11 (0.20) & - & - & - & - & - & PS1\\ 
11/04/2010 & 55297.34 & -28.66 & - & - & 22.43 (0.08) & - & - & - & - & - & PS1\\ 
12/04/2010 & 55298.39 & -27.61 & - & - & - & 22.41 (0.16) & - & - & - & - & PS1\\ 
15/04/2010 & 55301.33 & -24.67 & 22.93 (0.17) & 22.87 (0.18) & - & - & - & - & - & - & PS1\\ 
17/04/2010 & 55303.49 & -22.51 & - & - & 22.32 (0.08) & - & - & - & - & - & PS1\\ 
18/04/2010 & 55304.32 & -21.68 & - & - & - & 22.34 (0.16) & - & - & - & - & PS1\\ 
19/04/2010 & 55305.34 & -20.66 & 23.07 (0.21) & 22.67 (0.15) & - & - & - & - & - & - & PS1\\ 
06/05/2010 & 55322.39 & -3.61 & - & - & - & 21.97 (0.11) & - & - & - & - & PS1\\ 
09/05/2010 & 55325.37 & -0.63 & - & - & - & 21.74 (0.09) & - & - & - & - & PS1\\ 
10/05/2010 & 55326.41 & 0.41 & 22.98 (0.18) & 22.12 (0.10) & - & - & - & - & - & - & PS1\\
11/05/2010 & 55327.43 & 1.43 & - & - & 21.98 (0.06) & - & - & - & - & - & PS1\\ 
13/05/2010 & 55329.37 & 3.37 & 23.13 (0.18) & 22.22 (0.09) & - & - & - & - & - & - & PS1\\ 
14/05/2010 & 55330.30 & 4.30 & - & - & 22.02 (0.06) & - & - & - & - & - & PS1\\ 
16/05/2010 & 55332.31 & 6.31 & 23.41 (0.22) & 22.33 (0.09) & - & - & - & - & - & - & PS1\\ 
18/05/2010 & 55334.33 & 8.33 & - & - & - & 21.79 (0.09) & - & - & - & - & PS1\\ 
23/05/2010 & 55339.28 & 13.28 & - & - & 21.63 (0.08) & - & - & - & - & - & PS1\\ 
24/05/2010 & 55340.28 & 14.28 & - & - & - & 21.98 (0.14) & - & - & - & - & PS1\\ 
01/06/2010 & 55348.34 & 22.34 & - & - & 22.06 (0.06) & - & - & - & - & - & PS1\\ 
03/06/2010 & 55350.30 & 24.30 & 23.66 (0.28) & 22.42 (0.10) & - & - & - & - & - & - & PS1\\ 
05/06/2010 & 55352.33 & 26.33 & - & - & - & 21.68 (0.08) & - & - & - & - & PS1\\ 
06/06/2010 & 55353.35 & 27.35 & 23.56 (0.29) & 22.80 (0.17) & - & - & - & - & - & - & PS1\\ 
07/06/2010 & 55354.26 & 28.26 & - & - & 22.31 (0.08) & - & - & - & - & - & PS1\\ 
08/06/2010 & 55355.26 & 29.26 & - & - & - & 21.70 (0.12) & - & - & - & - & PS1\\ 
12/06/2010 & 55359.31 & 33.31 & - & 22.72 (0.26) & - & - & - & - & - & - & PS1\\ 
13/06/2010 & 55360.26 & 34.26 & - & - & 22.21 (0.08) & - & - & - & - & - & PS1\\ 
14/06/2010 & 55361.26 & 35.26 & - & - & - & 21.94 (0.12) & - & - & - & - & PS1\\ 
15/06/2010 & 55362.28 & 36.28 & - & 22.66 (0.13) & -  & - & - & - & - & - & PS1\\ 
16/06/2010 & 55363.27 & 37.27 & - & - & 22.22 (0.07) & - & - & - & - & - & PS1\\ 
17/06/2010 & 55364.26 & 38.26 & - & - & - & 22.08 (0.12) & - & - & - & - & PS1\\ 
18/06/2010 & 55365.28 & 39.28 & - & 22.72 (0.15) & - & -  & - & - & - & - & PS1\\ 
19/06/2010 & 55366.27 & 40.27 & - & - & 22.36 (0.09) & - & - & - & - & - & PS1\\ 
20/06/2010 & 55367.26 & 41.26 & - & - & - & 21.89 (0.10) & - & - & - & - & PS1\\
06/07/2010 & 55383.97 & 57.97 & - & - & - & - & - & 23.52 (0.13) &  22.65 (0.23) & - & WHT\\
07/07/2010 & 55384.00 & 58.00 & - & - & - & - & 24.82 (0.33) & - & - & 22.16 (0.28) & WHT\\
\hline
30/01/2011 & 55591.62 & 265.62 & - & - & - & - & - & - & 24.72 (0.30) & 24.60 (0.30) & GN\\ 
\hline
 \end{tabular}
 \medskip
\end{table*}

\subsection{PS1-10pm}
\label{sec:pm}

PS1-10pm was first detected with \PS\ in the \rps-band on MJD 55248 (24$^{th}$ February 2010) in MD06 at a location of RA\,=\,12$^h$12$^m$42$^s$.18, DEC\,=\,46$^\circ$59$'$29$''$.5 (J2000).  Detections in \emph{gri}\zps\ continued until a final \zps-band point on MJD 55367 (20$^{th}$ June 2010) when the MD06 season ended and \PS\ no longer continued to observe the object.  Further data were taken on the 7$^{th}$ July 2010 with the ACAM instrument at the WHT (\emph{griz}) and deep imaging in \emph{i} and \emph{z} was performed with the GMOS instrument on GN on the 30$^{th}$ January 2011.  The details of the photometry performed can be found in Table \ref{table:10pm}.  The transient was caught as it began to rise and the \PS\ coverage captures a reasonably well-defined peak.  Fig.\,\ref{fig:10pmoblc} shows observed \emph{gri}\zps\ light curves for PS1-10pm.

Spectra of PS1-10pm were obtained with GMOS on GN on the 3$^{rd}$ June and the 2$^{nd}$ July 2010\footnote{Gemini Program ID: GN-2010A-Q-45}.  The first spectrum was taken using the R400 grating with a GG455 filter and a single, 2400s exposure gave a signal-to-noise ratio (SNR) in the detected continuum of $\sim10$ per pixel (at approximately 6500\AA).  With the R400 grating, a 1$''$ slit provides a resolution of 7.9\AA\ and the actual useful wavelength range of the obtained spectrum was from $\sim5000-9000$\AA.
The second spectrum consists of $4\times1800$s exposures taken using the R150 grating (G5306).  The 1$''$ slit provided a resolution of 22.7\AA\  with this grating but the useful wavelength range of the obtained spectrum increased to $\sim4000-9500$\AA.
The SNR in the continuum was similar to the first spectrum (around 10 per pixel), albeit at lower spectral resolution. 

\begin{figure*}
\begin{center}$
\begin{array}{cc}
\includegraphics[angle=270,scale=0.3]{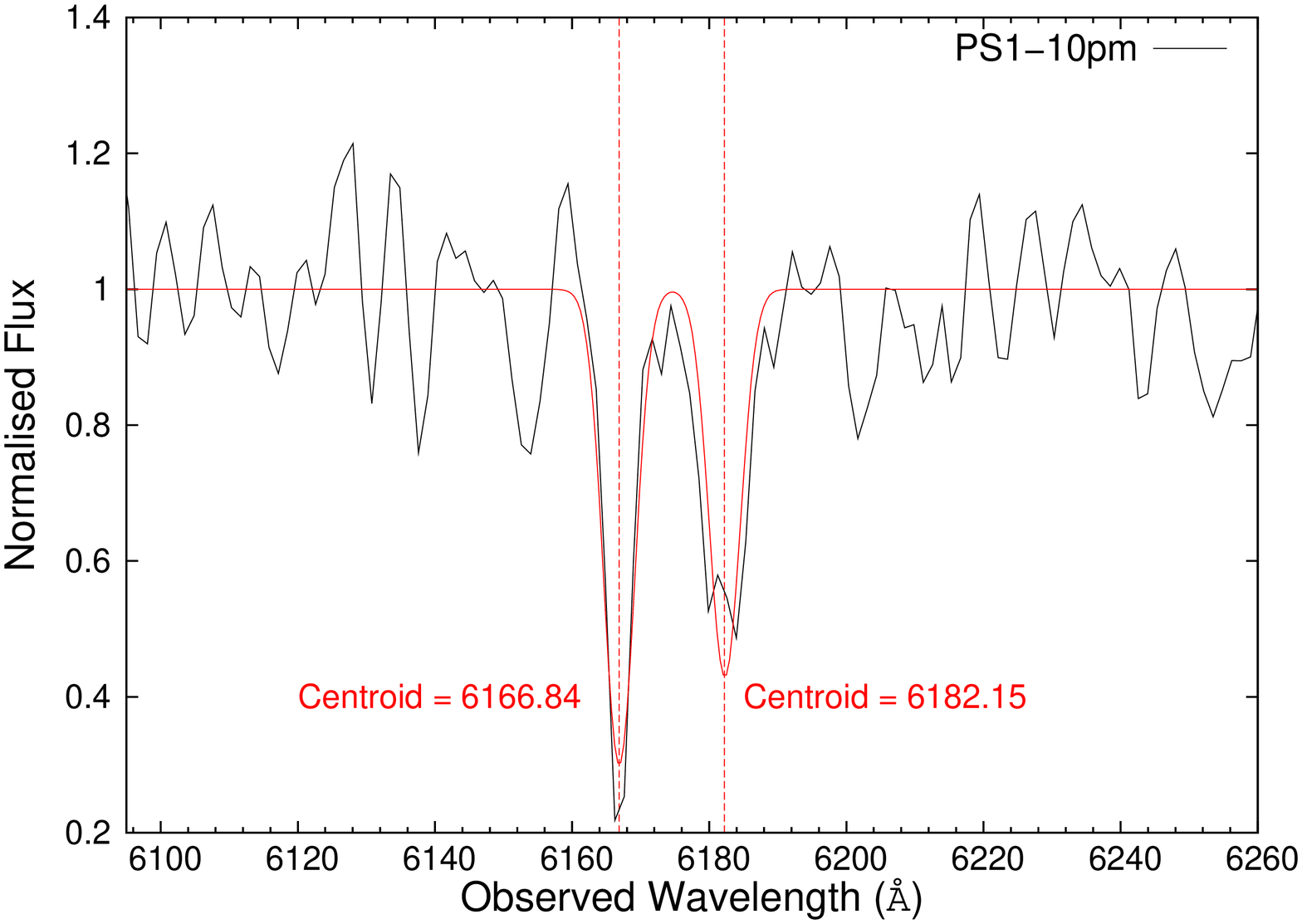} &
\includegraphics[angle=270,scale=0.3]{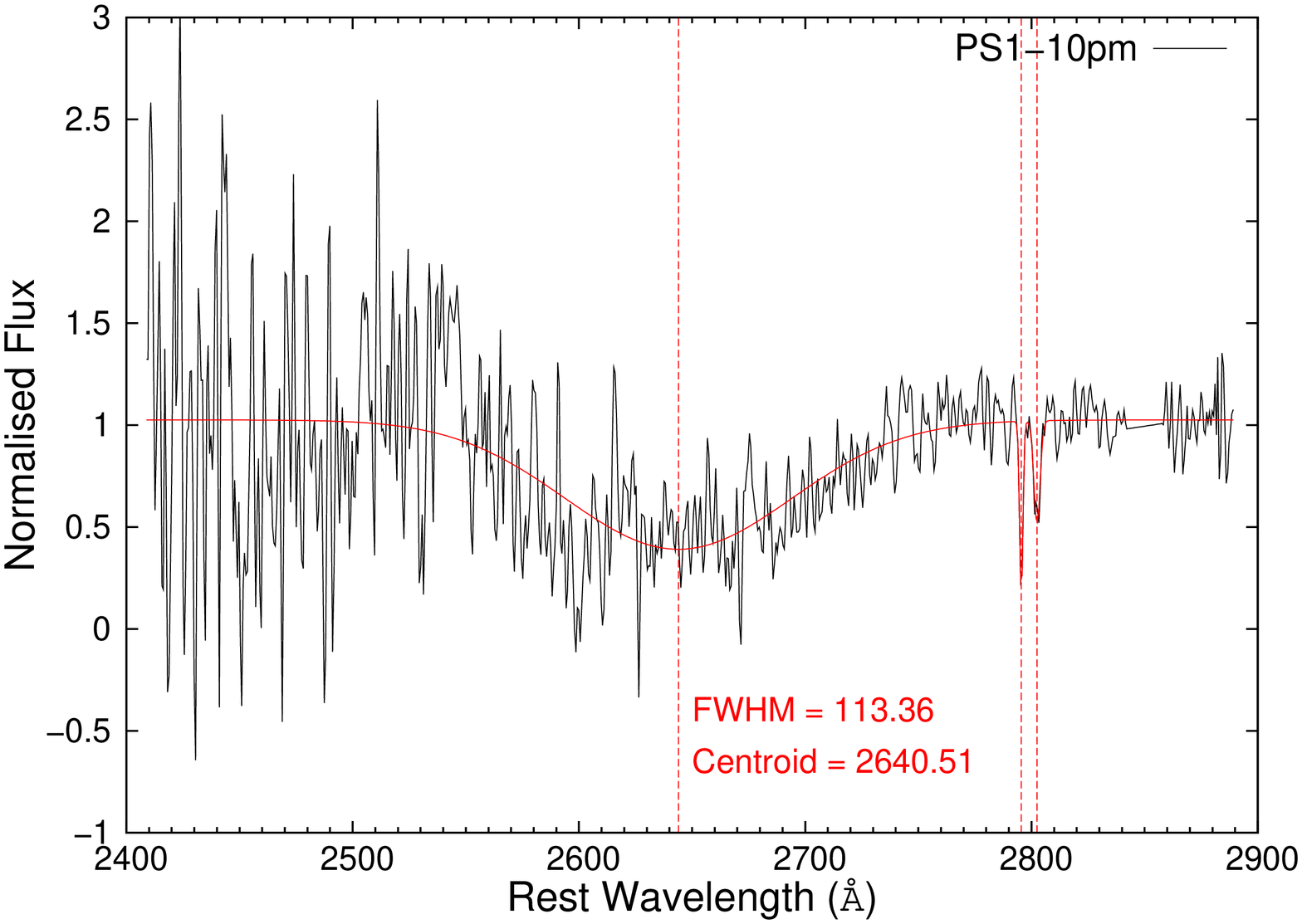}
\end{array}$
\end{center}
\caption{Detail of the GMOS, R400 PS1-10pm spectrum showing the observed wavelength of the two absorption features thought to be the Mg\,{\sc ii}  $\lambda\lambda$2796,2803 doublet used to determine a redshift of 1.206.  By taking the narrow doublet (seen here at the observed wavelength in the left hand figure and at the implied rest wavelength of $\sim2800\AA$ in the right hand figure) to be Mg\,{\sc ii} in the host galaxy and thus using it as a rest frame for the wider, bluer profile from the supernova, simple Gaussian profiles could be fitted to the absorption profiles and an expansion velocity of $\sim17,000$\,kms$^{-1}$ determined for PS1-10pm.}
\label{fig:10pmprofile}
\end{figure*}

The centroids and widths of the two strong absorption lines at around 6170\AA\ were measured by fitting simultaneous Gaussian profiles (see Fig.\,\ref{fig:10pmprofile}). This was done using our custom built \textsc{idl} spectral analysis package {\em procspec}, and checked with the \textsc{starlink} spectral analysis package \textsc{dipso}. 
The FWHM was allowed to vary in tandem, and the centroids were measured at 6166.84\AA\ and 6182.15\AA.  
We found a best fit with FWHM=5.4\AA\, which is slightly lower than the expected instrument resolution of 
6.4\AA\  at this wavelength, for a 1$''$ slit width. The image quality at the time of observations was 
lower than
the slit width (around 0.7-0.8$''$) and the source did not completely fill the slit. The lines widths are hence 
effectively unresolved.  If these were the 
Mg\,{\sc ii}  $\lambda\lambda$2795.528,2802.704\footnote{http://www.nist.gov/pml/data/asd.cfm}
 doublet, then the centroids both imply redshifts of z = 1.206. Hence this is a robust identification of the absorption components, likely in the interstellar medium (ISM) of the host galaxy of the transient. The Mg\,{\sc ii} absorber could conceivably be foreground, which would then imply an even higher redshift for the transient. 
 We do detect a probable host galaxy, coincident with PS1-10pm in deep Gemini images after the transient has faded (see below). Although this could be a foreground or background source, the simplest explanation is that the Mg\,{\sc ii} absorption is associated with the host galaxy and the redshift of the transient PS1-10pm is the same
as the Mg\,{\sc ii} absorption. In all reported cases in the literature where SLSNe-Ic have detections of both Mg\,{\sc ii} absorption and host galaxy emission lines, the redshifts are the same.

\begin{figure*}
\begin{center}
\includegraphics[angle=270,scale=0.6]{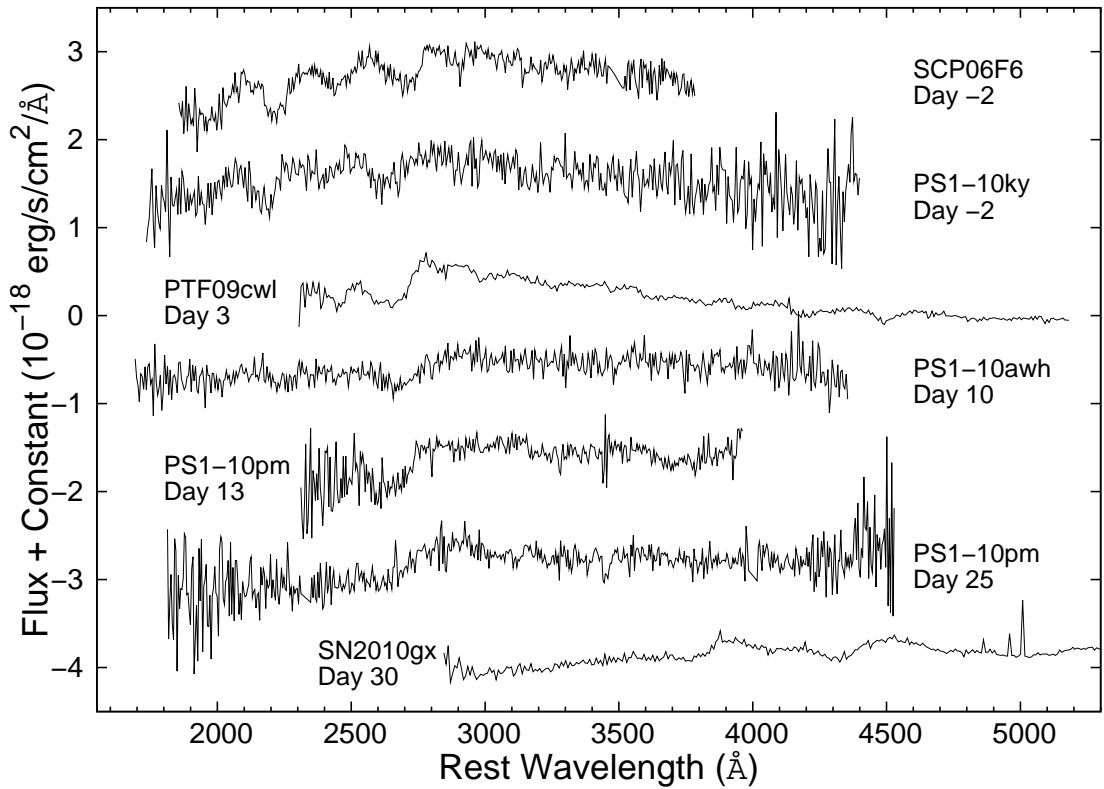}
\caption{Two GMOS spectra of PS1-10pm at $z=1.206$ compared with PS1-10awh at $z=0.908$, PS1-10ky at $z=0.956$, PTF09cwl at $z=0.349$, SCP06F6 at $z=1.189$ and SN2010gx at $z=0.23$ (see references in text).  All of the spectra have been corrected to restframe and re-binned to 10\AA\ and some chip gaps have been smoothed over.}
\label{fig:10pmspec}
\end{center}
\end{figure*}

Fig.\,\ref{fig:10pmspec} shows the two PS1-10pm spectra compared with spectra of SCP06F6, SN2010gx, PTF09cwl, PS1-10awh and PS1-10ky 
\citep{scp06f6,10gx,bluedeath,10kyawh}.  The Mg\,{\sc ii} $\lambda\lambda$2796,2803 absorption doublet can be seen in most of the spectra, corrected for each 
respective redshift, which allowed \cite{bluedeath} and \cite{10kyawh} to determine the redshifts and luminosities of these transients.  The
broad absorption in PS1-10pm is almost certainly due to the same Mg\,{\sc ii} resonance transition, but in the expanding photosphere of 
the transient. The similarity of the depth and strength of the feature immediately suggests that PS1-10pm could be a SLSN, similar to the SLSNe-Ic class illustrated here. 

\begin{figure*}
\begin{center}
\includegraphics[angle=270,scale=0.5]{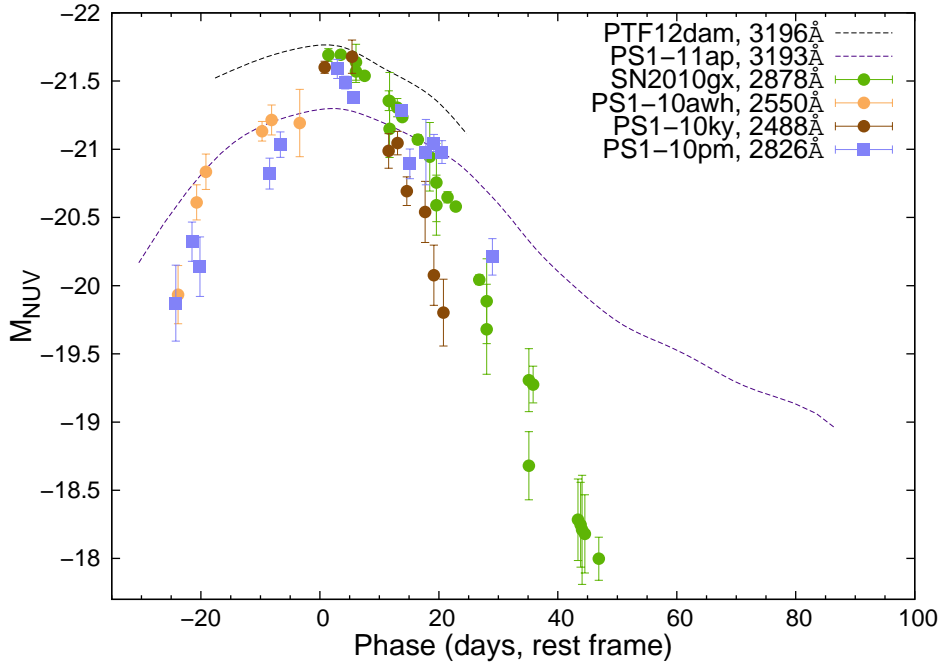}
\caption{A comparison of the PS1-10pm \rps-band absolute magnitude light curve with a \emph{u}-band SN2010gx light curve and \gps-band PS1-10awh and PS1-10ky light curves.  Light curve shapes obtained from \emph{u}-band PTF12dam and \gps-band PS1-11ap data are also presented here, highlighting the difference in the evolution of the SLSNe-Ic dataset.  The rest wavelengths of these bands for each object are given in Table \ref{table:filters} and the comparison here approximately represents the NUV, with a range of $\sim2500-3200$\AA.}
\label{fig:10pmlc}
\end{center}
\end{figure*}

\begin{table*}
\caption{Central rest wavelengths ({\AA}) of optical passbands for each of the SLSNe-Ic used in the photometric comparisons in this paper, where values in \emph{italic} represent filters used for comparison purposes.  The redshift of each object is given in the top row and the central wavelength of each filter in the second column.}
\label{table:filters}
\begin{tabular}{lcccccccc}
  \hline
  \hline
{\bf Filter} && {\bf PS1-10pm} & {\bf PS1-10ahf} & {\bf PS1-10awh} & {\bf PS1-10ky} & {\bf SN 2010gx} & {\bf PS1-11ap} & {\bf PTF12dam} \\ 
    \hline
&& 1.206 & 1.1 & 0.908 & 0.956 & 0.230 & 0.524 & 0.107\\
\hline
u & 3540 & - & - & - & - & \emph{2878} & - & \emph{3196}\\
g & 4860 & 2206 & 2256 & \emph{2550} & \emph{2488} & 3878 & \emph{3193} & \emph{4396} \\
r & 6230 & \emph{2826} & \emph{2890} & 3267 & 3188 & 5065 & \emph{4091} & 5632\\
i & 7525 & 3412 & 3489 & 3944 & 3848 & 6199 & 4939 & 6799\\
z & 8660 & 3924 & \emph{4012} & 4536 & 4426 & 7426 & 5680 & 7820\\
y & 9720 & 4409 & 4508 & 5097 & 4974 & - & 6382 & 8786\\
\hline
 \end{tabular}
 \medskip
\end{table*}

Fig.\,\ref{fig:10pmprofile} shows a 500\AA\ segment of the first, R400 GMOS spectrum, corrected for the redshift obtained above.  If we assume that the broad absorption line is Mg\,{\sc ii} then the centroid is at a blue shifted velocity of $\sim17,000$\,kms$^{-1}$ in comparison to the Mg\,{\sc ii} ISM doublet.  \cite{10kyawh} find similar velocities of  $\sim19,000$\,kms$^{-1}$ and  $\sim12,000$\,kms$^{-1}$ for PS1-10ky and PS1-10awh respectively.  Line widths ranging from $9000-12,000$\,kms$^{-1}$ were found by \cite{10kyawh}, which are slightly less than the $\sim13,000$\,kms$^{-1}$ found for PS1-10pm. 
However differences could arise due to the simplistic approach in fitting a FHWM to such a broad feature and ignoring possible blends. Also, we have used only the Mg\,{\sc ii} line whereas \cite{10kyawh} used multiple features in the analysis of PS1-10ky and PS1-10awh. 

To compare the respective absolute AB magnitude of each supernova we used the following:

\begin{equation}
M=m-5\mathrm{log}(\frac{d_{L}}{10(pc)})+2.5\mathrm{log}(1+z)
\label{eq:abmag}
\end{equation}

\citep{Hogg}, where \emph{m} is the apparent AB magnitude.  As can be seen, the measured magnitudes were corrected for cosmological expansion but not given a full $K-$correction.  However, suitable filters were chosen to make the comparison valid (see Table \ref{table:filters} for central wavelengths in the rest frame) with the resulting wavelength range ($\sim2500-4000$\AA) falling approximately in the near-ultraviolet (NUV).  We applied a correction for foreground reddening due to the Galactic line of sight only \citep{extinct}, as we have no information on the extinction in the host. The foreground extinction and the \cite{exlaw} extinction law implies $A_{r} \simeq 0.05$.
A standard cosmology with $H_0=72$ kms$^{-1}$, $\Omega_M=0.27$ and $\Omega_\lambda=0.73$ is used throughout.

Fig.\,\ref{fig:10pmlc} shows an absolute magnitude light curve of the PS1-10pm \rps-band (assuming a redshift of 1.206), along with \emph{u}-band data for SN2010gx \citep{10gx} and \gps-band data for PS1-10awh and PS1-10ky \citep{10kyawh}.  As can be seen from Table \ref{table:filters}, these are restframe light curves in the NUV ($\sim2500-2900$\AA) and they have been corrected for time dilation for the figure.  PS1-10pm reached a peak absolute magnitude M$_{NUV}=-21.59\pm0.07$ after a rise time lasting $\sim35$d.   The 
similarities in the light curve shapes further supports the classification of PS1-10pm as a SLSN-Ic,  but at one of the highest known redshifts of $z = 1.206$.  Two further light curves, obtained from \emph{u}-band PTF12dam data \citep{12dam} and \gps-band PS1-11ap data \citep{11ap}, are also presented in this figure, showing a clear distinction between these slowly evolving, SLSNe-Ic and the normal SLSNe-Ic class. The light curves for the former are much broader and easily distinguishable. 

\begin{figure}
\begin{center}
\includegraphics[angle=270,scale=0.35]{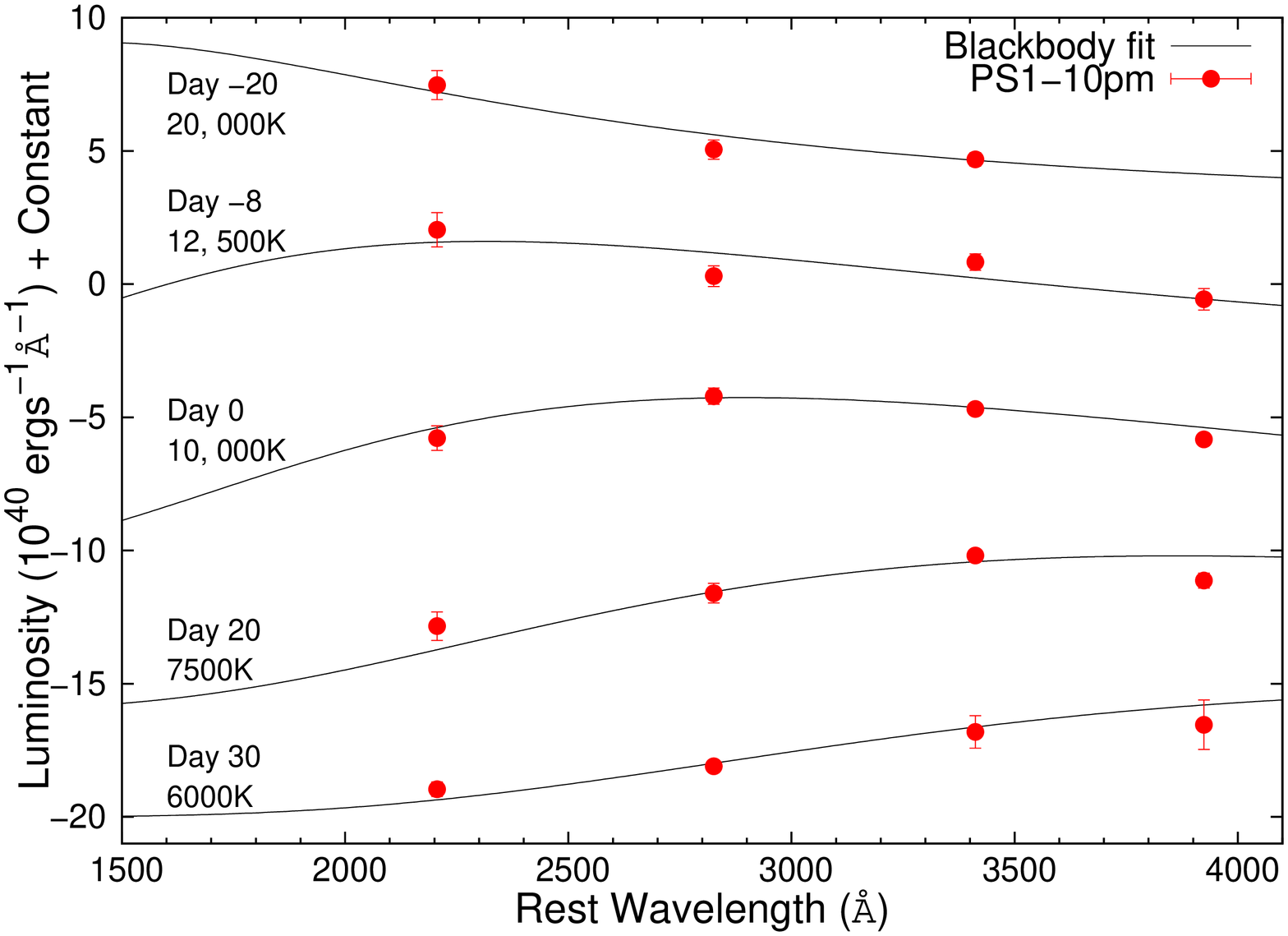}
\caption{Blackbody fitting of PS1-10pm across 5 epochs.}
\label{fig:10pmbb}
\end{center}
\end{figure}

\cite{10kyawh} and \cite{11xk} showed that the colour and luminosity evolution of these SLSNe-Ic are physically consistent
with hot blackbody temperatures ranging from $T_{\rm eff}\sim20000$\,K at 20 days before peak through 
 $T_{\rm eff}\sim15000$\,K at peak luminosity. Hence we can use the well sampled PS1 multi-colour light curve to trace the 
blackbody temperature of PS1-10pm to check for consistency with the known population of these transients. 
The PS1 bandpasses probe the restframe NUV wavelengths 2200-4000\AA\ for PS1-10pm and although this is 
a rather narrow window for a spectral energy distribution (SED) it is well suited to the high photospheric temperatures.     
We chose to determine a blackbody fit at five epochs which had approximately simultaneous and consistent $griz$ coverage 
(see Table \ref{table:bbfits}), giving a range between -20d and +30d with respect to peak. The fits and temperatures 
are shown in Fig.\,\ref{fig:10pmbb}, which illustrate  a physically consistent evolution of temperature which is similar to the 
lower redshift SLSNe-Ic as shown in Fig.\,8 of \cite{10kyawh}. At peak luminosity, a temperature of $T_{\rm eff}\sim10000$
provides a blackbody spectrum fit to the 
flux of PS1-10pm. This gives an integrated luminosity (between 1000 and 10000\AA) of $\sim3\times10^{44}$erg\,s$^{-1}$, or  
$7.3\times10^{10}$\,L$_{\odot}$. This is again very similar to the SLSNe-Ic in \cite{bluedeath}, \cite{10gx}  
and \cite{10kyawh}. The radius of the emitting surface must then be of order 6$\times10^{15}$\,cm, which is 
a factor of two larger than previously determined by \cite{10kyawh}  for PS1-10awh and PS1-10aky, due to the lower peak
$T_{\rm eff}$ that we determine. However within the intrinsic uncertainties of  the assumptions of blackbody radiation, the 
narrow spectral energy range and flux measurements,  we cannot say if this is real diversity or limitations of the fairly simple physics we employ. 
In conclusion, the PS1 measured multi-colour light curve is physically consistent with PS1-10pm being a SLSN-Ic at 
$z=1.206$. 
The spectrum in Fig,\,\ref{fig:10pmspec} illustrates the difficulties in classifying high$-z$ SN candidates
from optical spectra. At $z>1$, one typically gets a region of the rest frame UV that is 
a factor of two smaller in wavelength coverage than the observer frame spectrum. The lack 
of large numbers  of SNe (particularly unusually luminous SNe which will be 
preferentially detected at high-z) with restframe UV spectra  often 
makes the classification and redshift determination difficult.

\subsection{Host galaxy of PS1-10pm}
\label{sec:host}

\begin{table}
\begin{center}
\caption{PS1-10pm estimated temperatures from blackbody fitting.}
\label{table:bbfits}
\begin{tabular}{c c c}
  \hline
  \hline
{\bf MJD} & {\bf Phase (days, rest)} & {\bf T$_{BB}$ (K)} \\
    \hline
$\sim$55283 & -20 & 20000 $\pm$ 5000 \\
$\sim$55305 & -8 & 12500 $\pm$ 2500 \\
$\sim$55325 & 0 & 10000 $\pm$ 1000 \\
$\sim$55367 & 20 & 7500 $\pm$ 1000 \\
$\sim$55384 & 30 & 6000 $\pm$ 1000 \\
\hline
 \end{tabular}
 \medskip
\end{center}
\end{table}

\begin{figure*}
\begin{center}
\includegraphics[scale=0.5]{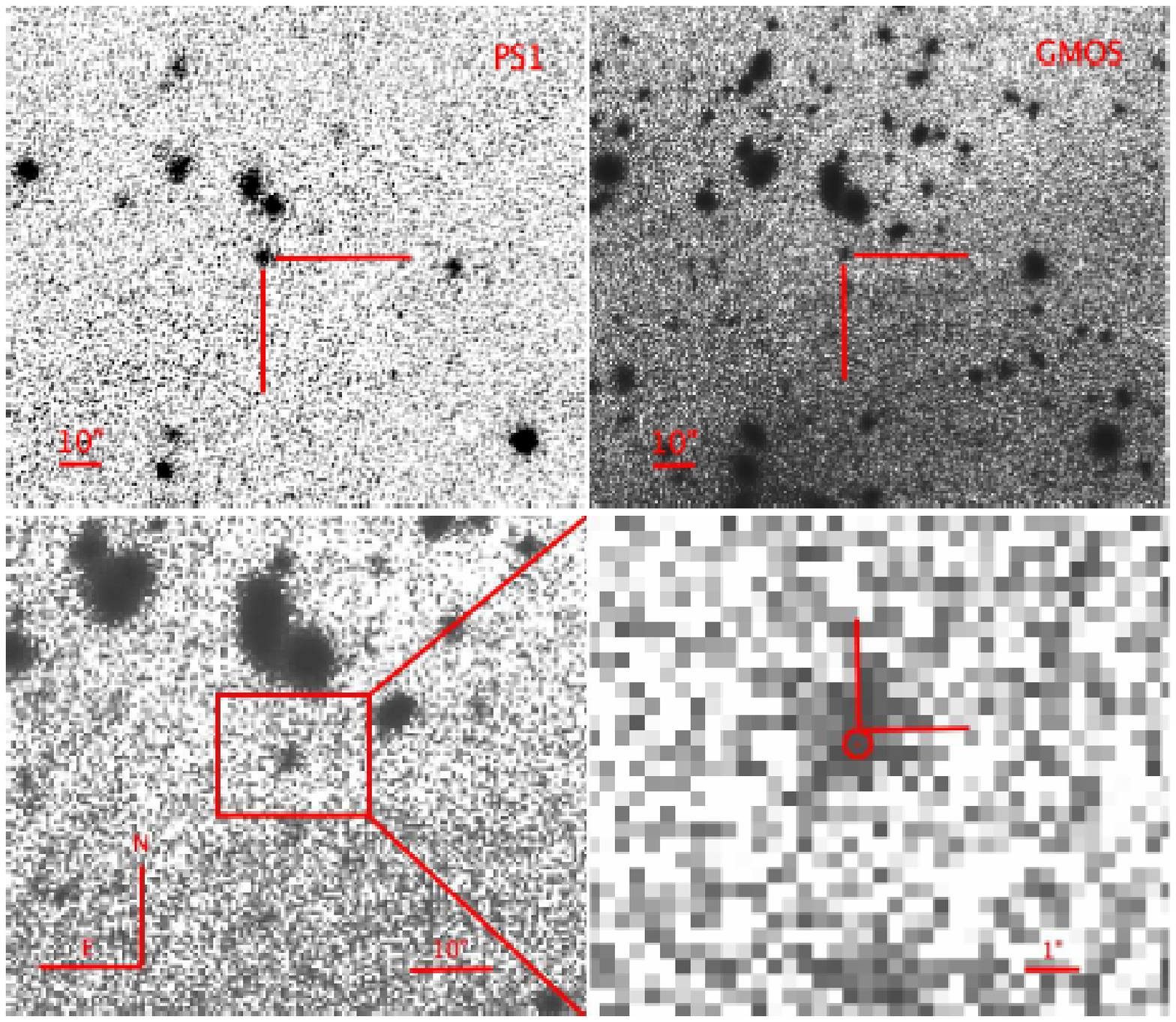}
\end{center}
\caption{PS1 and GN \ips- and \emph{i}-band images of the SLSN-Ic PS1-10pm at peak and after the explosion has faded.  The lower set of images show subsections of the host galaxy of PS1-10pm, 265 days after peak.  The circle in the zoomed, lower right image is centred on the SLSN-Ic position with a radius corresponding to 3$\sigma$.  The perpendicular lines in this image meet at the determined centroid of the galaxy which can be seen to be just inside the 3$\sigma$ boundary of the SLSN-Ic position.}
\label{fig:10pmim}
\end{figure*}

We obtained deep images in \emph{i} and \emph{z} at the position of PS1-10pm with $9\times150$s exposures at GN, $\sim265$d after the \ips-band peak.  The data were reduced as normal by subtracting a bias level gleaned from the overscan region of the Gemini CCD, dividing each image by an appropriate flatfield image and subtracting an appropriately scaled, sourceless fringe frame created using the \emph{gifringe} function in the \emph{gemini} \textsc{iraf}\footnote{\textsc{iraf} is distributed by the
  National Optical Astronomy Observatories, which are operated by the
  Association of Universities for Research in Astronomy, Inc., under
  the cooperative agreement with the National Science Foundation.} package.  Although no host was seen in the PS1 reference templates,
  a faint object can be seen in the deeper Gemini images. 
Aperture photometry was carried out using the aperture photometry procedure available in the Graphical Astronomy and Image Analysis tool software package\footnote{http://astro.dur.ac.uk/$\sim$pdraper/gaia/gaia.html} \citep[\textsc{gaia},][]{naylor97}, giving an
\emph{i}-band magnitude of $\emph{i}=24.99\pm0.42$ and an observed \emph{z}-band magnitude of $\emph{z}=24.86\pm0.31$. These correspond to absolute magnitude values of M$_{3400}\sim-18.7$ and M$_{3900}\sim-18.9$ respectively when corrected to $z=1.206$ and for foreground extinction.

The position of a SN with respect to its host galaxy can provide evidence against it being misclassified as an AGN, provided an offset from the galactic centre is found.  Alignment of the GN $i$-band image with a PS1 \ips-band image of the SLSN-Ic at peak was carried out by first measuring the pixel coordinates of 10 bright stars in a 6.5$'$ x 2.8$'$ field using the \textsc{iraf} \emph{phot} task utilising the centroid centring algorithm on both images. The list of matched coordinates was then used as an input to the \textsc{iraf} \emph{geomap} task to derive a geometric transformation between the two images, allowing for translation, rotation and independent scaling in the \emph{x} and \emph{y} axes.  The root mean square (RMS) of the fit was 0.061$''$ (GMOS pixels are 0.1454$"$ and PS1 GPC pixels are 0.25$"$ after warping). 
 
Aperture photometry was then carried out on the host galaxy (in the Gemini image) and PS1-10pm (in the original \PS\ image). The coordinates of the SLSN-Ic and of the host galaxy were measured in both images with three different centring algorithms provided by the \emph{phot} task; centroid, Gaussian and optimal filtering. This provided a mean position and a standard deviation. The standard deviation of the three measurements was taken as the positional error measurement in $x$ and $y$ of the two objects.

The $x,y$ position of PS1-10pm was then transformed to the coordinate system of the GMOS frame using the transformation defined by the 10 stars in common. This revealed a difference of 2.24 GMOS pixels which, at the 0.1454$''$ resolution of GN (for a 2x2 binned CCD with a pixel scale of 0.0727 arcsec/pixel\footnote{http://www.gemini.edu/sciops/instruments/gmos/imaging/detector-array/gmosn-array-eev}) corresponds to an offset of 0.33$''$ (see Fig.\,\ref{fig:10pmim}).

The total uncertainty in the alignment of the two objects is hence the quadrature sum of the uncertainties in the centroids of the host and PS1-10pm and the RMS of the alignment transformation (see \cite{pos} for example). This was found to be $\sigma=0.806$ pixels (or 0.12$''$) and hence the SLSN-Ic and the galaxy centroid differ by 2.8$\sigma$.  While this is not quite a formal 3$\sigma$ difference, it indicates that the SN is not coincident with the centre of the galaxy, hence supports evidence that PS1-10pm is a SLSN and is not a UV transient event due to any type of AGN variability.

\begin{figure}
\begin{center}
\includegraphics[angle=270,scale=0.35]{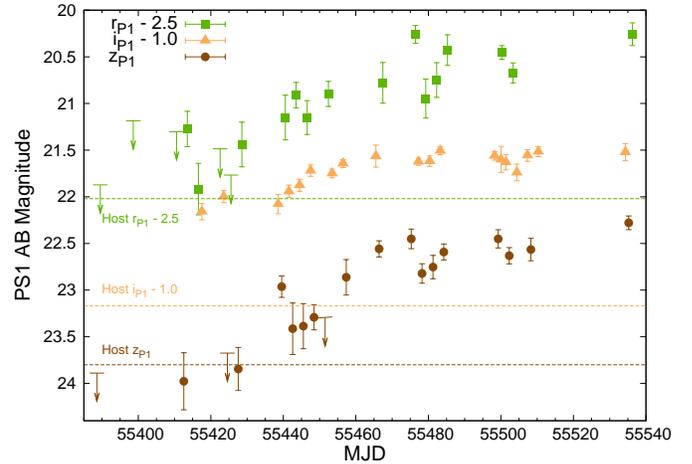}
\caption{Observed \rps-, \ips- and \zps-band light curves of PS1-10ahf.  Measurements of the host galaxy from a deep, WHT  \emph{i}-band observation and deep, GS \emph{r}- and \emph{z}-band images are shown as horizontal lines and any upper limits are indicated with arrows.  During this period, from MJD 55389.54 until 55530.33, 20 non-detections in \gps\ are also recorded down to a photometric limit of 24.59 and a \emph{g}-band limit of $\sim27$ is found for the host from a deep, GS \emph{g}-band observation.}
\label{fig:10ahfoblc}
\end{center}
\end{figure}

 \cite{2013arXiv1311.0026L} has shown that the host of PS1-10pm actually breaks up into 
a resolved source which is significantly extended and has an irregular morphology in Hubble 
Space Telescope images.  They determine  
AB magnitudes of $m_{\rm F606W} = 25.38\pm0.05$ and 
$m_{\rm F110W} = 24.40\pm0.08$. Our $i$ and $z$ band filter mags sit comfortably between
these mags, which would allow four points on the SED to be used for future analysis.

\subsection{PS1-10ahf}
\label{sec:ahf}

PS1-10ahf was first detected on MJD 55414 (6$^{th}$ August 2010) in MD10 at RA\,=\,23$^h$32$^m$28$^s$.3, DEC\,=\,-00$^\circ$21$'$43$''$.6.  The initial light curve of PS1-10ahf showed a faint, slowly rising source clearly detected in the \ips- and \zps-bands ($\ips\sim23.3$). \PS\ followed the field until MJD 55535 (5$^{th}$ December 2010), thereafter the field was dropped from the \PS\ observing cycle due to airmass constraints. The details of these observations can be found in Table \ref{table:10ahf}.  No sign of a host was found in any of the major survey catalogues, nor was it visible in the \PS\ reference stack images.
As a check, manual photometry was also carried out independently of both the automated pipelines described in Section 2.2. The optimal photometry facility within \textsc{gaia} was again used to perform photometry on the target images rather than the difference images. As there was clearly no host contribution in the PS1 reference stack image, there should not be any contribution to the luminosity other than that of the transient.  The manual light curves for the \ips- and \zps-bands were calibrated using 8 SDSS stars in the field of the transient, and in all cases the photometry was consistent with the pipeline measurements.  The epoch of the $i_{\rm P1}$-band maximum was found from a second order polynomial fit and determined to be MJD $55540\pm5$.  Observed \rps-, \ips- and \zps-band light curves for PS1-10ahf can be found in Fig.\,\ref{fig:10ahfoblc}.

\begin{table*}
\tiny
\caption{Observed photometry for PS1-10ahf. No \emph{K}-corrections have been applied.  Phase is in observer frame, not restframe, as more than one possible redshift value is presented in the text.  Note that the \PS\ observations have had any flux from previous reference image removed through image subtraction (although note that no host object can be seen at the location of PS1-10ahf) whereas the late time WHT and GS observations have not.}
\label{table:10ahf}
\begin{tabular}{c c c c c c c c c c}
  \hline
  \hline
{\bf Date} & {\bf MJD} & {\bf Phase (days)} & {\bf r$_{P1}$} & {\bf i$_{P1}$} & {\bf z$_{P1}$} & {\bf \emph{r}} & {\bf \emph{i}} & {\bf \emph{z}} & {\bf Telescope} \\
    \hline
11/07/2010 & 55388.59 & -151.41 & - & - & $>$23.89 & - & - & - & PS1 \\ 
12/07/2010 & 55389.52 & -150.48 & $>$24.37 & - & - & - & - & - & PS1 \\ 
21/07/2010 & 55398.61 & -141.39 & $>$23.69 & - & - & - & - & - & PS1 \\ 
02/08/2010 & 55410.55 & -129.45 & $>$23.80 & - & - & - & - & - & PS1 \\ 
04/08/2010 & 55412.55 & -127.45 & - & - & 23.98 (0.31) & - & - & - & PS1 \\
05/08/2010 & 55413.54 & -126.46 & 23.77 (0.19) & - & -  & - & - & - & PS1 \\
08/08/2010 & 55416.56 & -123.44 & 24.42 (0.28) & - & -  & - & - & - & PS1 \\
09/08/2010 & 55417.56 & -122.44 & - & 23.16 (0.09) & -  & - & - & - & PS1 \\
14/08/2010 & 55422.56 & -117.44 & $>$23.99 & - & - & - & - & - & PS1 \\ 
15/08/2010 & 55423.56 & -116.44 & - & 23.00 (0.06) & - & - & - & - & PS1 \\
16/08/2010 & 55424.57 & -115.43 & - & - & $>$23.68 & - & - & - & PS1 \\ 
17/08/2010 & 55425.56 & -114.44 & $>$24.27 & - & - & - & - & - & PS1 \\ 
19/08/2010 & 55427.57 & -112.43 & - & - & 23.85 (0.23) & - & - & - & PS1 \\
20/08/2010 & 55428.56 & -111.44 & 23.94 (0.24) & - & - & - & - & - & PS1 \\
30/08/2010 & 55438.58 & -101.42 & - & 23.08 (0.10) & - & - & - & - & PS1 \\
31/08/2010 & 55439.54 & -100.46 & - & - & 22.96 (0.12) & - & - & - & PS1 \\
01/09/2010 & 55440.53 & -99.47 & 23.65 (0.24) & - & - & - & - & - & PS1 \\
02/09/2010 & 55441.52 & -98.48 & - & 22.94 (0.07) & - & - & - & - & PS1 \\
03/09/2010 & 55442.57 & -97.43 & - & - & 23.41 (0.28) & - & - & - & PS1 \\
04/09/2010 & 55443.50 & -96.5 & 23.41 (0.14) & - & - & - & - & - & PS1 \\
05/09/2010 & 55444.48 & -95.52 & - & 22.88 (0.06) & - & - & - & - & PS1 \\
06/09/2010 & 55445.52 & -94.48 & - & - & 23.39 (0.24) & - & - & - & PS1 \\
07/09/2010 & 55446.55 & -93.45 & 23.65 (0.18) & - & - & - & - & - & PS1 \\
08/09/2010 & 55447.53 & -92.47 & - & 22.72 (0.06) & - & - & - & - & PS1 \\
09/09/2010 & 55448.47 & -91.53 & - & - & 23.29 (0.13) & - & - & - & PS1 \\
10/09/2010 & 55449.51 & -90.49 & - & - & - & - & - & - & PS1 \\ 
12/09/2010 & 55451.51 & -88.49 & - & - & $>$23.29 & - & - & - & PS1 \\ 
13/09/2010 & 55452.43 & -87.57 & 23.40 (0.13) & - & - & - & - & - & PS1 \\
14/09/2010 & 55453.46 & -86.54 & - & 22.75 (0.05) & - & - & - & - & PS1 \\
17/09/2010 & 55456.45 & -83.55 & - & 22.64 (0.04) & - & - & - & - & PS1 \\
18/09/2010 & 55457.35 & -82.65 & - & - & 22.86 (0.19) & - & - & - & PS1\\ 
19/09/2010 & 55458.33 & -81.67 & - & - & - & - & - & - & PS1\\  
26/09/2010 & 55465.52 & -74.48 & - & 22.56 (0.12) & - & - & - & - & PS1\\ 
27/09/2010 & 55466.41 & -73.59 & - & - & 22.56 (0.09) & - & - & - & PS1\\ 
28/09/2010 & 55467.43 & -72.57 & 23.28 (0.22) & - & - & - & - & - & PS1\\ 
06/10/2010 & 55475.27 & -64.73 & - & - & 22.45 (0.10) & - & - & - & PS1\\ 
07/10/2010 & 55476.45 & -63.55 & 22.76 (0.10) & - & - & - & - & - & PS1\\ 
08/10/2010 & 55477.27 & -62.73 & - & 22.62 (0.04) & - & - & - & - & PS1\\ 
09/10/2010 & 55478.26 & -61.74 & - & - & 22.82 (0.10) & - & - & - & PS1\\ 
10/10/2010 & 55479.27 & -60.73 & 23.45 (0.21) & - & - & - & - & - & PS1\\ 
11/10/2010 & 55480.36 & -59.64 & - & 22.62 (0.06) & - & - & - & - & PS1\\ 
12/10/2010 & 55481.29 & -58.71 & - & - & 22.75 (0.13) & - & - & - & PS1\\ 
13/10/2010 & 55482.27 & -57.73 & 23.25 (0.19) & - & - & - & - & - & PS1\\ 
14/10/2010 & 55483.26 & -56.74 & - & 22.51 (0.04) & - & - & - & - & PS1\\ 
15/10/2010 & 55484.24 & -55.76 & - & - & 22.59 (0.09) & - & - & - & PS1\\ 
16/10/2010 & 55485.26 & -54.74 & 22.93 (0.16) & - & - & - & - & - & PS1\\ 
29/10/2010 & 55498.26 & -41.74 & - & 22.56 (0.05) & - & - & - & - & PS1\\ 
30/10/2010 & 55499.26 & -40.74 & - & - & 22.45 (0.10) & - & - & - & PS1\\ 
31/10/2010 & 55500.05 & -39.95 & - & - & - & - & 22.60 (0.14) & - & WHT\\ 
31/10/2010 & 55500.26 & -39.74 & 22.95 (0.07) & - & - & - & - & - & PS1\\ 
01/11/2010 & 55501.39 & -38.61 & - & 22.63 (0.08) & - & - & - & - & PS1\\ 
02/11/2010 & 55502.30 & -37.7 & - & - & 22.63 (0.09) & - & - & - & PS1\\ 
03/11/2010 & 55503.34 & -36.66 & 23.17 (0.11) & - & - & - & - & - & PS1\\ 
04/11/2010 & 55504.41 & -35.59 & - & 22.74 (0.09) & - & - & - & - & PS1\\ 
07/11/2010 & 55507.33 & -32.67 & - & 22.56 (0.06) & - & - & - & - & PS1\\ 
08/11/2010 & 55508.25 & -31.75 & - & - & 22.57 (0.12) & - & - & - & PS1\\ 
10/11/2010 & 55510.31 & -29.69 & - & 22.52 (0.05) & - & - & - & - & PS1\\ 
30/11/2010 & 55530.33 & -9.67 & - & - & - & - & - & - & PS1\\  
04/12/2010 & 55534.30 & -5.7 & - & 22.52 (0.09) & - & - & - & - & PS1\\ 
05/12/2010 & 55535.24 & -4.76 & - & - & 22.28 (0.07) & - & - & - & PS1\\ 
06/12/2010 & 55536.27 & -3.73 & 22.76 (0.12) & - & - & - & - & - & PS1\\ 
\hline
24/07/2011 & 55766 & 226.0 & - & - & - & 24.52 (0.06) & - & - & GS\\ 
26/07/2011 & 55768 & 228.0 & - & - & - & - & - & 23.80 (0.10) & GS\\ 
08/08/2011 & 55781.14 & 241.14 & - & - & - & - & 24.17 (0.07) & - & WHT\\ 
\hline
 \end{tabular}
 \medskip
\end{table*}

\begin{figure*}
\begin{center}$
\begin{array}{cc}
\includegraphics[angle=270,scale=0.45]{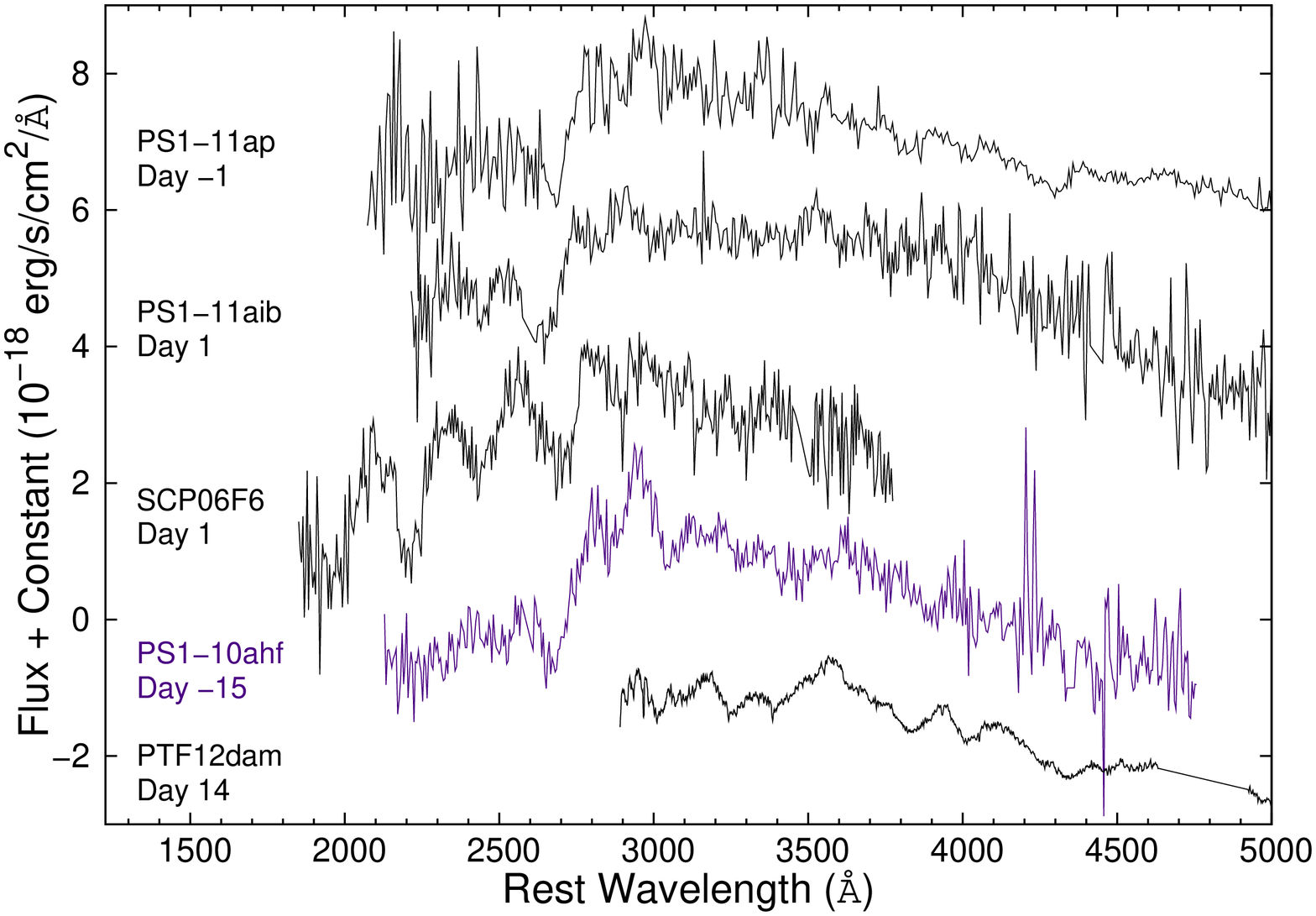} \\
\includegraphics[angle=270,scale=0.45]{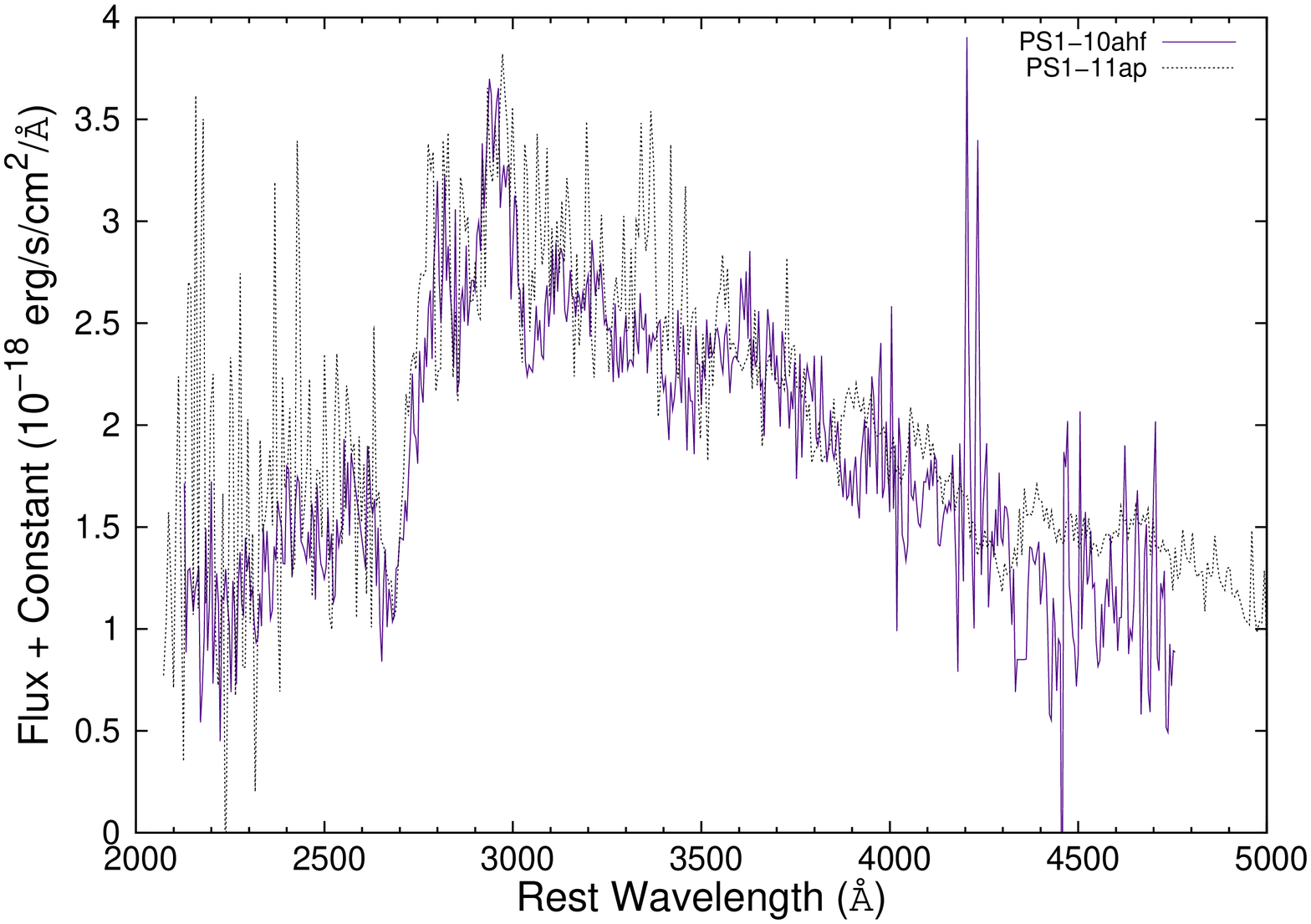}
\end{array}$
\end{center}
\caption{GMOS spectrum of PS1-10ahf at $z=1.1$, taken with GS, compared with PS1-11ap at $z=0.524$, PS1-11aib at $z=0.997$, SCP06F6 at $z=1.189$ and PTF12dam at $z=0.108$.
The lower plot shows an overlay of the PS1-10ahf spectrum with the same PS1-11ap spectrum as in the top panel to emphasise the similarities between the objects.
All of the spectra have been re-binned to 10\AA\ and some chip gaps have been smoothed over.  See references in text.}
\label{fig:10ahfspectra}
\end{figure*}

A spectrum of PS1-10ahf was obtained with GMOS on GS\footnote{Gemini Program ID: GS-2010B-Q-43} on the 6$^{th}$ November 2010 using the R150 grating (G5306) with a 1$''$ slit, giving a useful wavelength range from $\sim4300-8000$\AA.
 A set of $4\times2700$s exposures gave a combined SNR of $\sim19$ in the continuum, when rebinned to 10\AA\ per pixel.
The flux calibrated GMOS spectrum provides a synthetic $r_{\rm P1}$-band magnitude of 
22.8 as calculated in SYNPHOT. This flux (on 20101106) is in reasonable agreement (within $\pm0.2^{m}$)
of the PS1 photometry. The synthetic $g_{\rm P1}$-band magnitude from the same spectrum of 24.7 is 
consistent with the non-detection in the nightly PS1 images.
There are no strong and obvious narrow features either in absorption (e.g. Mg\,{\sc ii} or Ca\,{\sc ii} ISM lines) or in emission (e.g.  nebular lines) to provide an unambiguous redshift. Hence we initially compared it to a range of SNe, including the SLSNe-Ic that we used for the PS1-10pm comparison and the confirmed $z\simeq1$ SLSNe-Ic already from PS1 \citep{10kyawh}.  There are a number of broad absorption or P-Cygni features, and a plausible redshift of $z=1.1$ would put the deepest absorption at a rest wavelength similar to the broad Mg\,{\sc ii} absorption seen in other SLSNe-Ic (Fig.\,\ref{fig:10ahfspectra}). However the spectrum lacks C\,{\sc ii} and Si\,{\sc ii} as detected in previous SLSNe-Ic \citep{bluedeath,10kyawh} and overall is not an entirely convincing match.
Attempts to match the PS1-10ahf spectrum with other features typical of Type Ic SNe, such as Ca\,{\sc ii} H\&K features and some Fe\,{\sc ii} blends, also proved unconvincing.

\begin{figure*}
\begin{center}$
\begin{array}{cc}
\includegraphics[angle=270,scale=0.475]{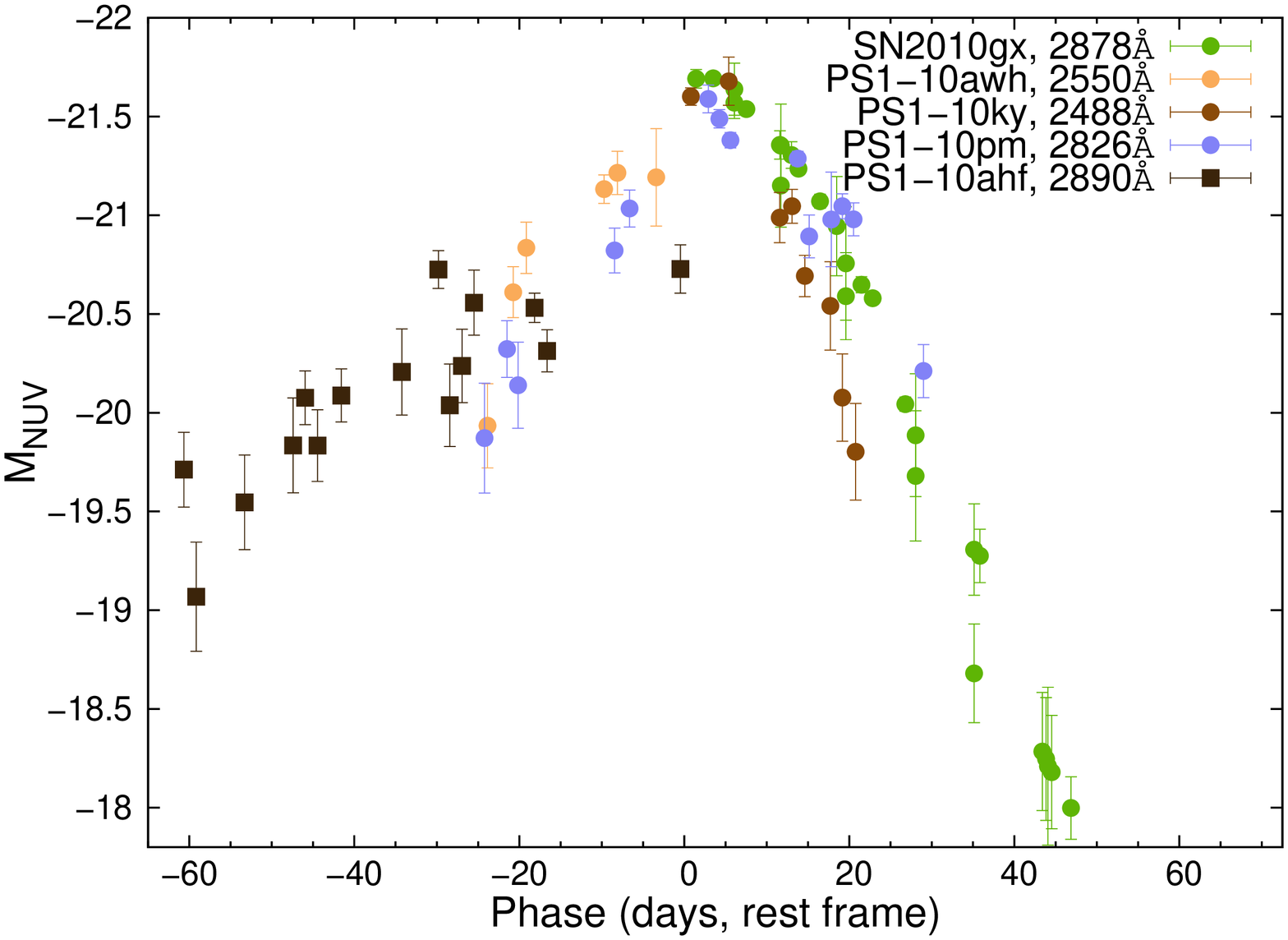} \\
\includegraphics[angle=270,scale=0.475]{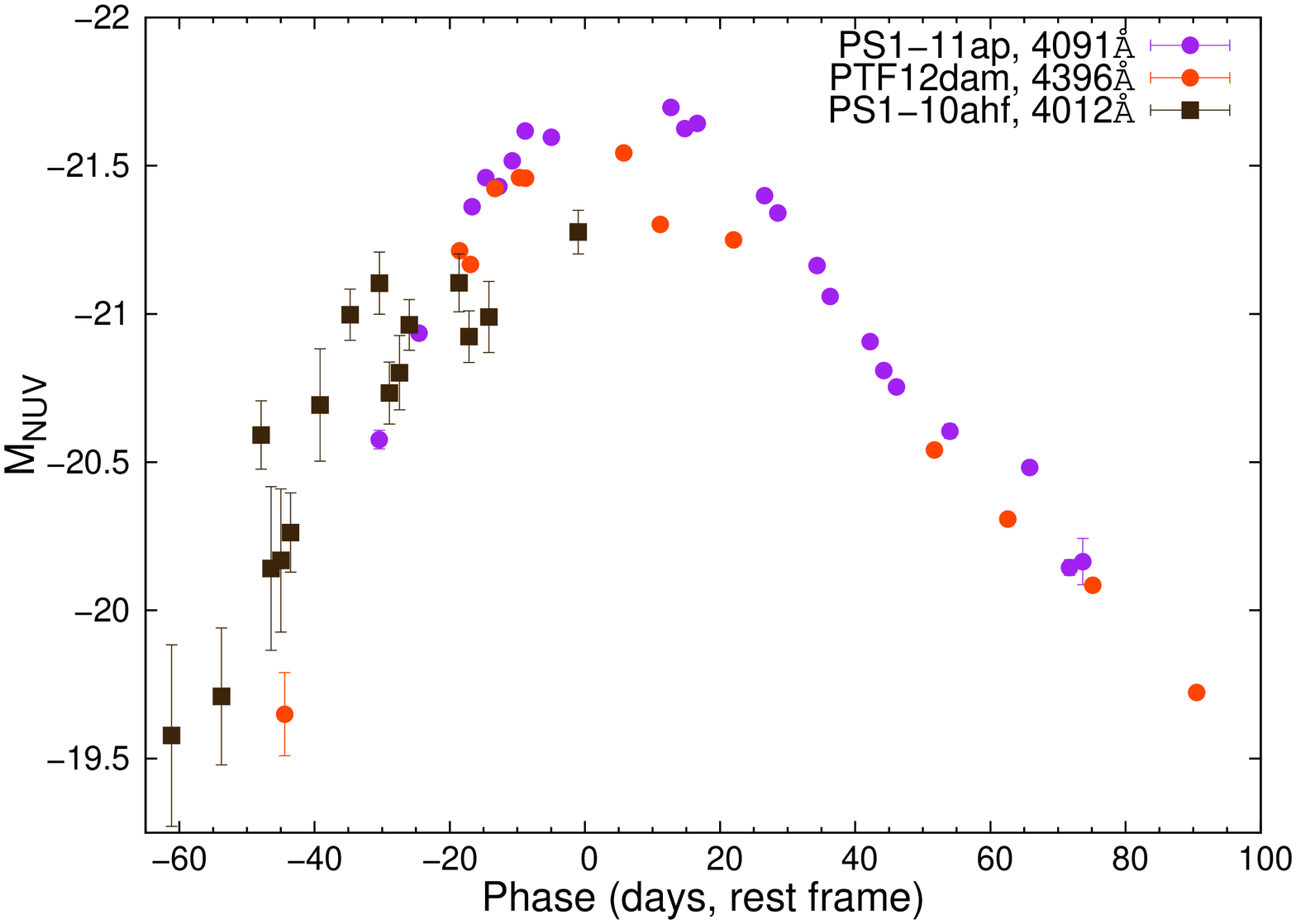}
\end{array}$
\end{center}
\caption{A variation on Fig. \ref{fig:10pmlc}, again showing an absolute \emph{u}-band SN2010gx light curve, absolute \gps-band PS1-10awh and PS1-10ky light curves and an absolute \rps-band PS1-10pm light curve.  An absolute \rps-band light curve of PS1-10ahf is shown here for comparison purposes, with $z=1.1$.  The lower image offers a comparison of a \zps-band light curve of PS1-10ahf with \rps-band PS1-11ap and \emph{g}-band PTF12dam data.  See references within.
}
\label{fig:10ahfablc}
\end{figure*}

Fig.\,\ref{fig:10ahfablc} shows a comparison of an absolute magnitude \rps-band PS1-10ahf light curve with \emph{u}-band data for SN2010gx \citep{10gx}, \gps-band data for PS1-10awh and PS1-10ky \citep{10kyawh} and \rps-band PS1-10pm data (this paper), again created using Eq.\,\ref{eq:abmag}.    The central wavelengths in the rest frame for these filters offer a reliable comparison as shown in Table \ref{table:filters}.  
The measured magnitudes were corrected for cosmological expansion and foreground reddening for Galactic line of sight only \citep{extinct} as again we have no host extinction information.  The foreground extinction and the \cite{exlaw} extinction law implies $A_{i}\simeq0.07$. The long rise time of PS1-10ahf is still prevalent after correcting for time dilation and clearly sets the transient apart from the normal SLSNe-Ic class.

If this redshift of $z=1.1$ from the spectral comparisons is secure then the
transient is a closer match to the slowly evolving SLSNe-Ic PS1-11ap \citep{11ap} and PTF12dam \citep{12dam}, as seen in the lower plot in Fig.\,\ref{fig:10ahfablc}.
PS1-11ap has broad Mg\,{\sc ii} absorption with a line width of $\sim14,500$\,kms$^{-1}$.  The line width of possible Mg\,{\sc ii} absorption was determined to be $\sim12,000$\,kms$^{-1}$ for PS1-10ahf which compares well to PS1-11ap (see lower plot in Fig.\,\ref{fig:10ahfspectra}).
The overall spectral match to PS1-11ap was the closest we could find, after comparing with all known types of SNe for which NUV spectra exist. 
At this redshift the light curve is very broad, even after applying time dilation. The transient has an intrinsically slow rest-frame rise of 60 days in the NUV bands (see Fig.\,\ref{fig:10ahfablc}). 
The rising slope, within the photometric uncertainties is similar to 45-60 day rise time deduced from the modelling of PS1-11ap \citep{11ap}.
Unfortunately we do not sample the decay time after peak for PS1-10ahf, simply due to the length of the PS1 observing season. 

The plots show a peak absolute magnitude M$_{NUV}=-21.39\pm0.07$ for PS1-10ahf however the polynomial fit used to determine the peak MJD suggests that the transient continued to brighten for a short time after the observing period had ended. 
 This illustrates a practical limitation in following the evolution of transients with broad light curves and slow evolution times combined with time dilation at $z > 1$. A typical PS1 observing season for a MD field is 150-180 days, meaning that a transient at $z\sim1$ with a symmetric light curve and rise-time of 40 days (restframe) needs to be discovered close to the start of the observing season for the MD field if one is to sample the full rise and decay time. We were fortunate that this occurred for PS1-11ap \citep[at $z=0.524$,][]{11ap}. In conclusion, we find the most likely match to the light curve and spectrum of PS1-10ahf is with the slowly evolving SLSNe-Ic PS1-11ap.
To further illustrate the connection we show a spectrum of PS1-11aib. The latter transient fell outside the survey window set for this paper (discovered on the 27$^{th}$ July, 2011 in MD09), but it too has a broad, red light curve and a spectrum with very similar absorption features and slope to both PS1-10ahf and PS1-11ap.  PS1-11aib has a convincing detection of Mg\,{\sc ii} $\lambda\lambda$2796,2803 ISM doublet at a redshift of $z=0.997$ and will be discussed in a future PS1 paper (Lunnan et al., in preparation).  Unfortunately due to the high redshift of PS1-10ahf, a direct comparison with other published objects of this class (\bi\ and \dam) is not possible as the rest wavelength ranges do not have a sufficient overlap.
A PTF12dam spectrum is included for completeness but, as can be seen in Fig.\,\ref{fig:10ahfspectra}, the main features of the PS1-10ahf spectrum fall just bluewards of the reach of PTF12dam.
Nevertheless, as is exemplified in the lower plot of Fig.\,\ref{fig:10ahfspectra}, the continuum shape of the PS1-10ahf spectrum and the slowly evolving SLSNe-Ic spectra are very similar.

\subsection{Was PS1-10ahf  a variable BAL QSO?}
\label{ahfqso}

\begin{figure*}
\begin{center}$
\begin{array}{cc}
\includegraphics[angle=270,scale=0.3]{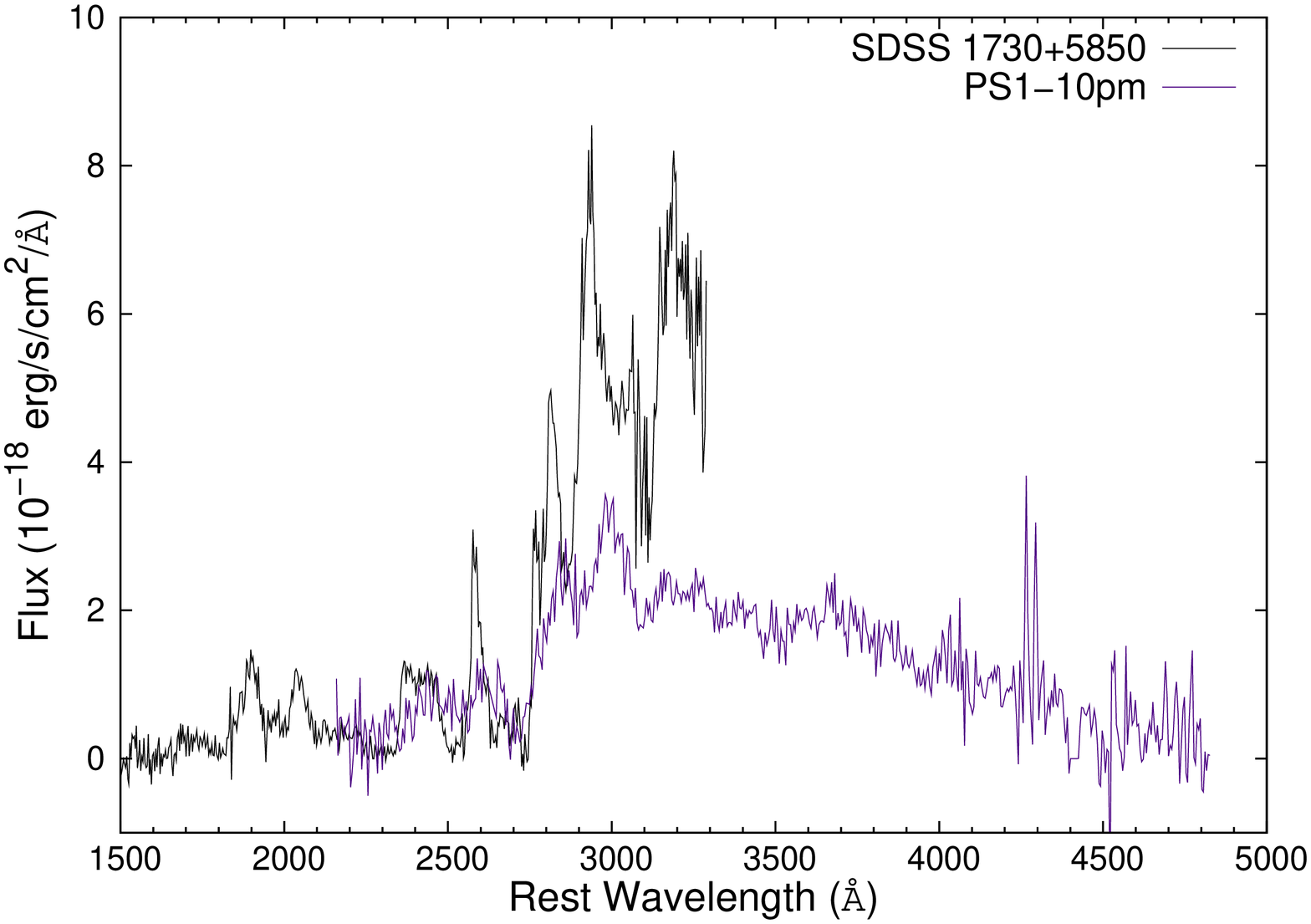} &
\includegraphics[angle=270,scale=0.3]{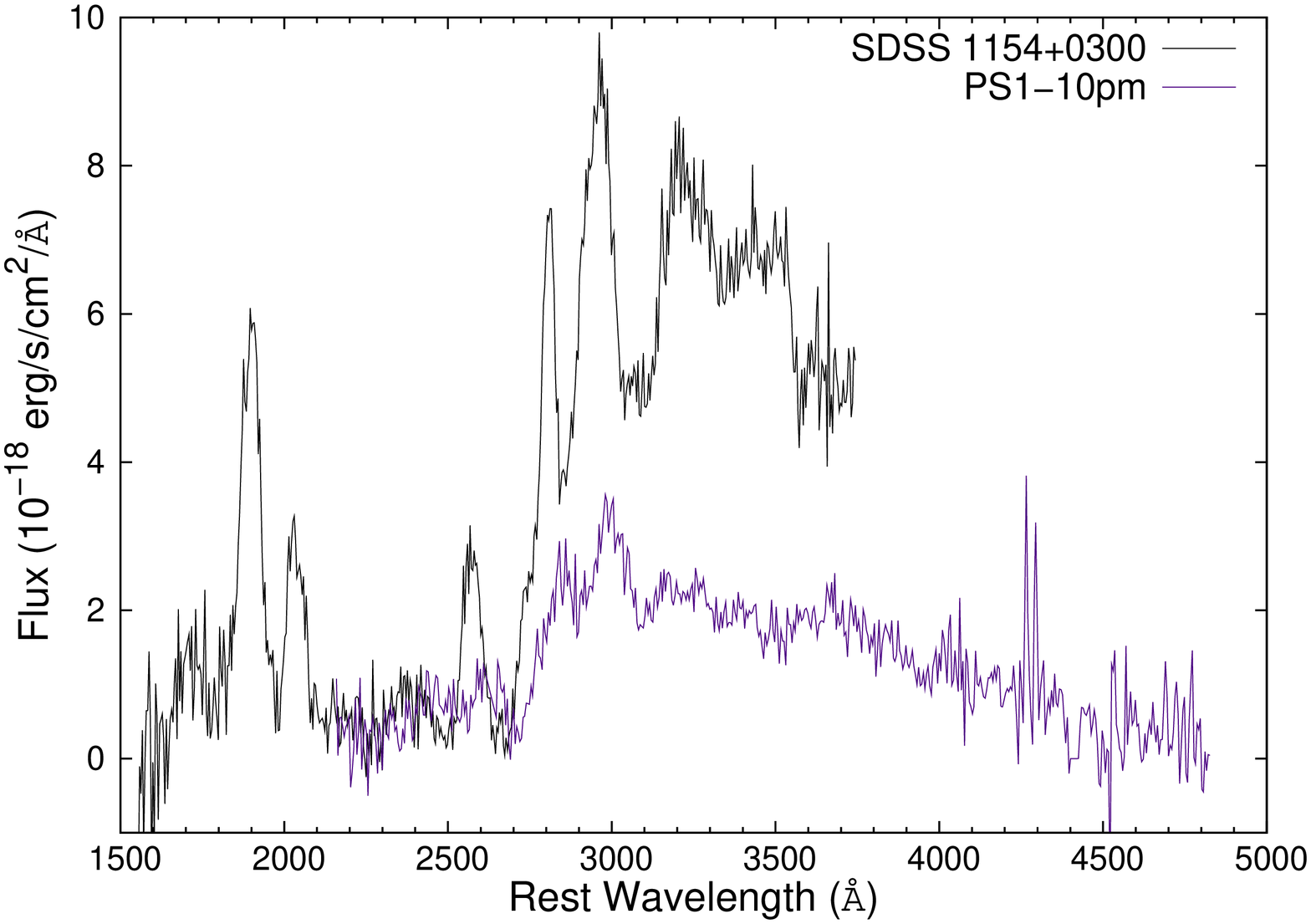} \\
\includegraphics[angle=270,scale=0.3]{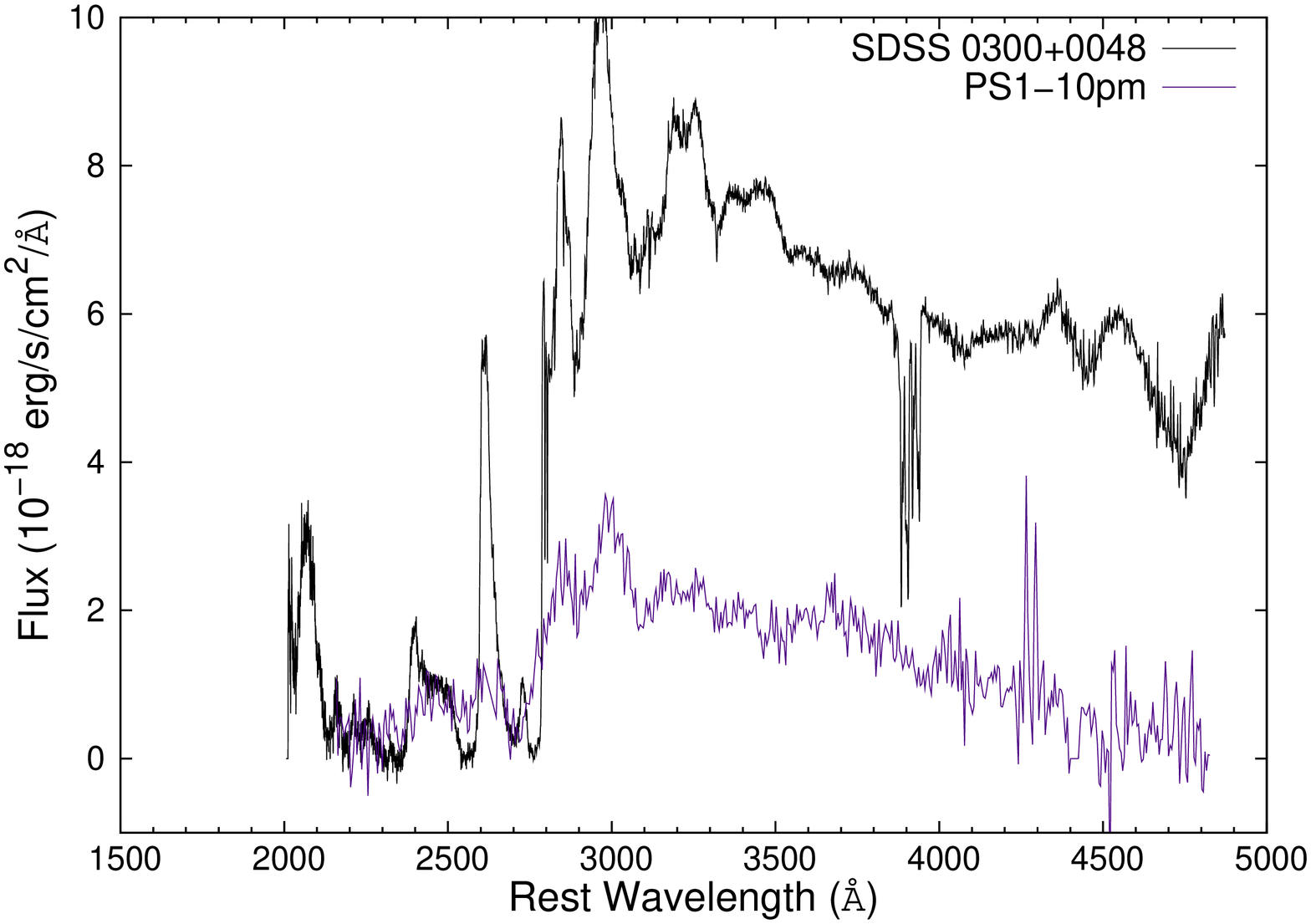} &
\includegraphics[angle=270,scale=0.3]{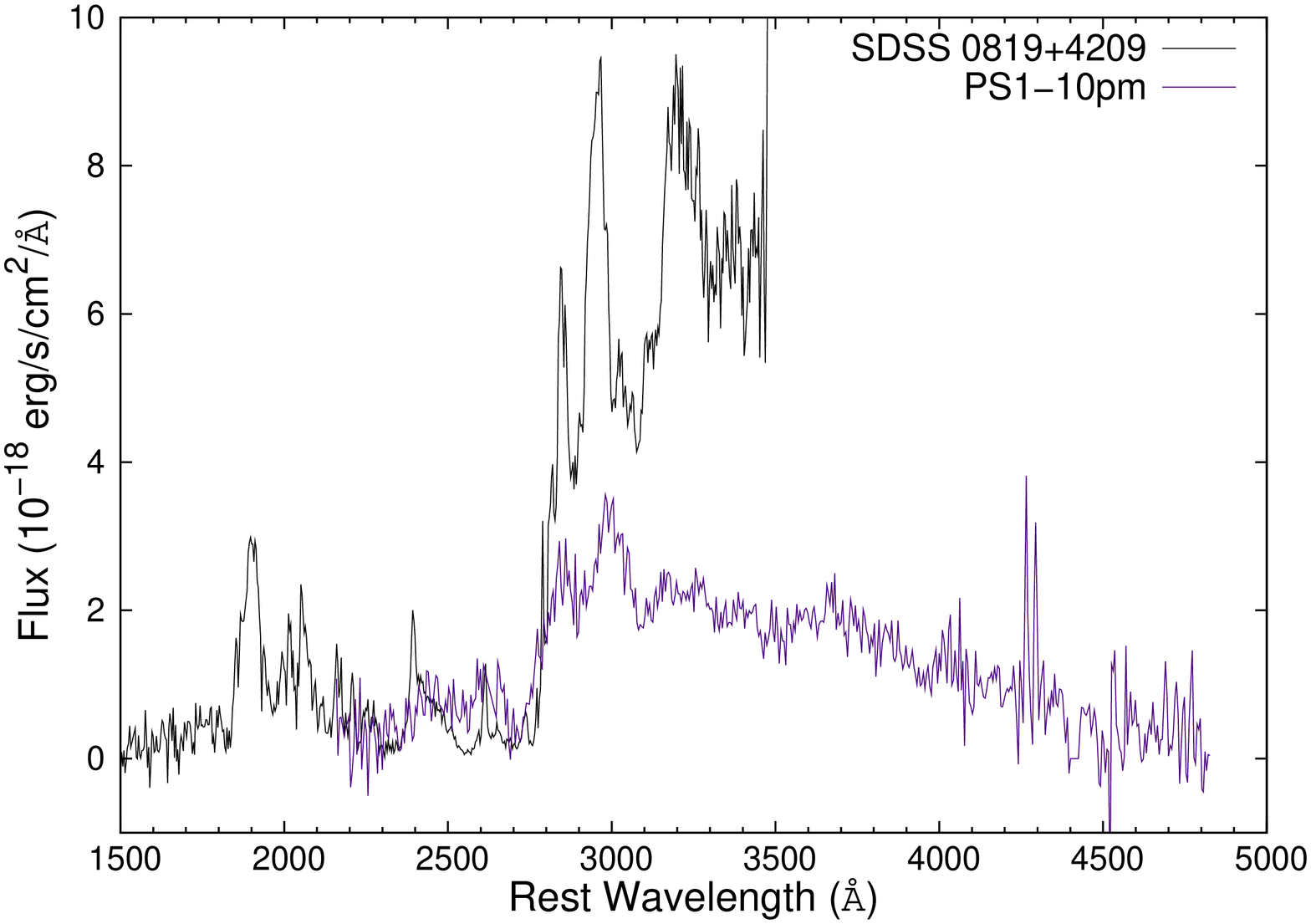} \\
\end{array}$
\end{center}
\caption{A comparison of PS1-10ahf with the SDSS Quasars 1730+5850 ($z=2.035$), 1154+0300 ($z=1.458$), SDSS 0300+0048 ($z=0.892$) and 0819+4209 ($z=1.926$) where $z=1.07$ for PS1-10ahf.  All of the spectra have been re-binned to 10\AA\ and some chip gaps have been smoothed over.  See references in text.}
\label{fig:10ahfspectra2}
\end{figure*}

Initial difficulties in determining any spectral features in the PS1-10ahf GMOS spectrum led us to 
consider if the object was perhaps a broad absorption line quasar (BAL QSO).  Fig.\,\ref{fig:10ahfspectra2} shows a PS1-10ahf spectrum at rest wavelength compared with FeLoBal SDSS 1730+5850, 1154+0300, 0300+0048 and 0819+4209 \citep{Hall} with the redshift of PS1-10ahf set at $z=1.07$.  These BAL QSOs can show absorption troughs $\sim2000 - 20,000$\,kms$^{-1}$ arising from gas with blueshifted velocities up to $66,000$\,kms$^{-1}$ \citep{qso} giving broad spectral features that can be comparable to typical SNe features in low to moderate signal-to-noise spectra.  A number of other features (possibly Fe\,{\sc ii}  and Ca\,{\sc ii}, most apparent in SDSS 0300+0048) in the QSO spectra match shallower absorption features in PS1-10ahf. However there is no convincing match to any of these, particularly as the spectral break in the QSOs due to broad Mg\,{\sc ii}  absorption is much deeper in the QSOs compared to PS1-10ahf.
In Fig.\ref{fig:10ahfspectra2}  we show the QSO spectra compared with PS1-10ahf. We matched the flux level in the restframe continuum region 2200-2700\AA\ of each object to show the contrast in the spectrum break expected if PS1-10ahf was a BAL QSO. In all cases the BAL QSO spectral breaks are significantly larger than that of PS1-10ahf and it seems unlikely that they can be similar objects. While any underlying galaxy flux might be expected to dilute the transient flux of PS1-10ahf, and hence increase the spectral break, the host galaxy is at least 
2 magnitudes fainter than the transient when the spectrum was taken (see Table\,\ref{table:10ahf}). Therefore flux dilution cannot account for the differences. 
In addition, the unusual QSOs of the \cite{Hall} sample are significantly more luminous than PS1-10ahf (if the latter is at $z=1.07$), as they range between $M _{NUV}\sim -23\ {\rm to} -29$, compared to $M_{NUV}\simeq-21$ for 
PS-10ahf.

\begin{figure*}
\begin{center}
\includegraphics[scale=0.6]{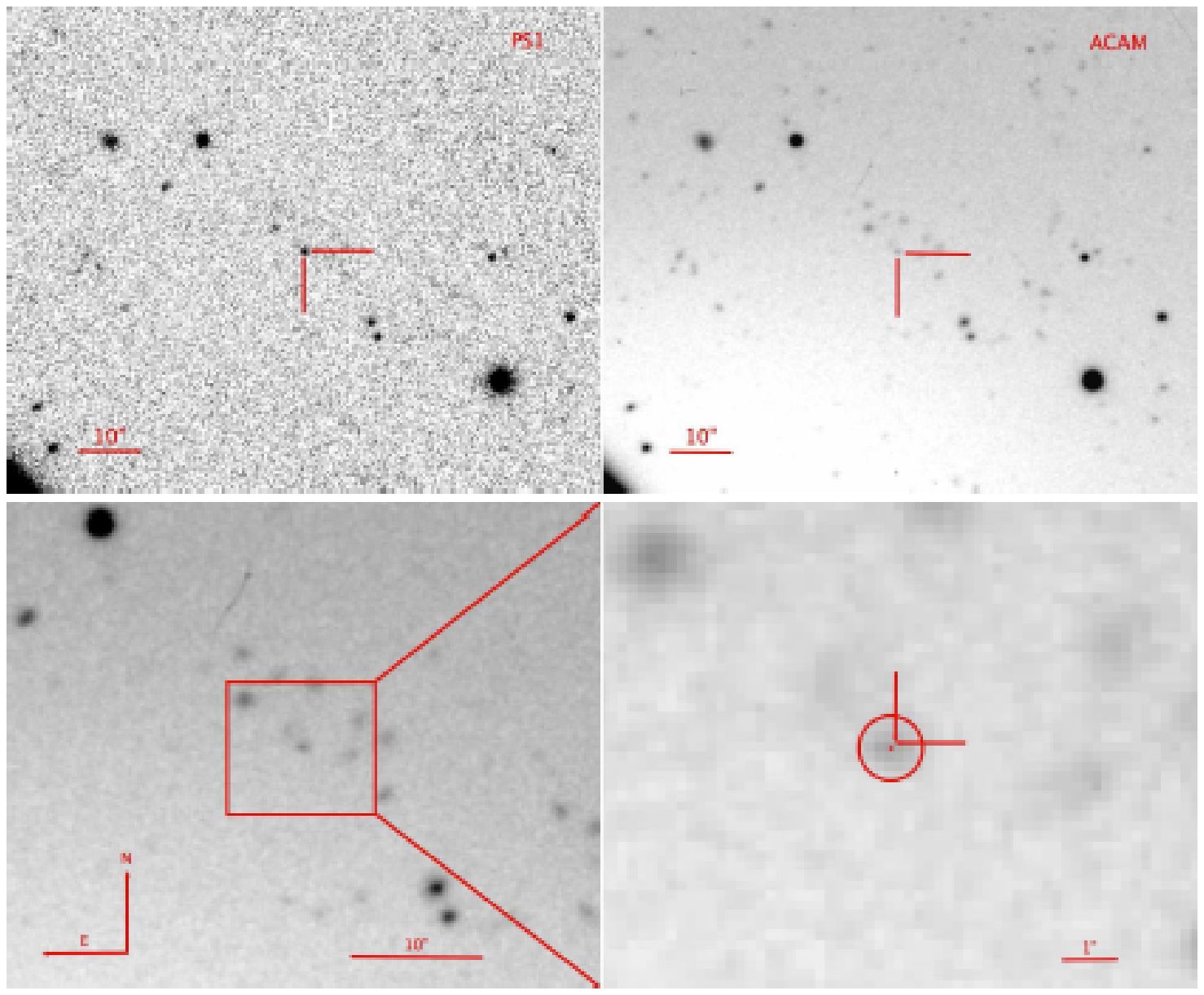}
\end{center}
\caption{PS1 and WHT \ips- and \emph{i}-band images of the possible slowly evolving SLSN-Ic PS1-10ahf at peak and after the explosion has faded.  The circle in the zoomed, lower right image is centred on the galaxy position with a radius corresponding to 3$\sigma$.  The perpendicular lines in this image meet at the determined centroid of the transient position which is well within the determined central region of the galaxy.  Although this does not rule out the object as a slowly evolving SLSN-Ic, this result does not provide conclusive evidence that the transient is not an AGN-like event, unlike the poor spectral comparisons of PS1-10ahf with the Hall et al. (2002) QSOs.}
\label{fig:10ahfim}
\end{figure*}

We carried out a similar search for the host galaxy of PS1-10ahf as for PS1-10pm to align it with the position of the transient.  A deep \emph{i}-image was obtained with a 600s exposure using the ACAM instrument on the WHT, 241 days after the determined peak epoch.  The data was reduced by subtracting a separate bias image and dividing the image by an appropriate flatfield image, taken on the same night.  As can be seen in Fig.\,\ref{fig:10ahfim} an object can clearly be seen at the expected coordinates and an observed \emph{i}-band magnitude of $\emph{i}=24.17\pm0.07$ was determined using the photometry procedures available in the \textsc{gaia} software package \citep{naylor97}. Taking $z=1.1$ and $A_{i}=0.07$ gives an absolute magnitude of M$_{3400}\sim-19.5$.  Note that a point source was again chosen as a reference PSF for the galaxy however, due to high redshift involved, the target appears approximately point-like.  As before, both optimal and aperture photometry procedures were carried out to ensure that this approach was sensible. 

Alignment of the WHT $i$-band image with a PS1 \ips-band image of the transient at peak was carried out using the method described in Section \ref{sec:host}.  The coordinates of 10 bright stars in a 6.5$'$ x 6.5$'$ field were determined using the \textsc{iraf} \emph{phot} task, utilising the centroid centring algorithm on both images. The list of matched coordinates was then used as an input to the \textsc{iraf} \emph{geomap} task to derive a geometric transformation between the two images, allowing for translation, rotation and independent scaling in the \emph{x} and \emph{y} axes.  The RMS of the fit was 0.064$''$.  Aperture photometry was then carried out on the WHT image for a host galaxy measurement and on the original PS1 image for a measurement of PS1-10ahf.  The coordinates of the transient and of the host galaxy were measured in both images with three different centring algorithms provided by the \emph{phot} task; centroid, Gaussian and optimal filtering. This provided a mean position and a standard deviation. The standard deviation of the three measurements was taken as the positional error measurement in $x$ and $y$ of the two objects.  The $x,y$ position of PS1-10ahf was then transformed to the coordinate system of the ACAM frame using the transformation defined by the 10 stars in common.  A separation of 0.148$''$ was determined (using a pixel scale for the ACAM instrument of 0.25 arcsec/pixel\footnote{http://www.ing.iac.es/Astronomy/observing/instruments.html}).  As before, the total uncertainty in the alignment of the two objects is hence the quadrature sum of the uncertainties in the centroids of the host and PS1-10ahf and the RMS of the alignment transformation. This was found to be 0.214$''$ at the pixel scale of ACAM, hence there is no evidence for an offset between the transient and the centroid of its host galaxy.

PS1-10ahf is coincident, within the errors, to the centroid of the host galaxy which doesn't offer the same strong argument of it not being QSO variability when compared with PS1-10pm being offset. However the differences in the spectral features and the relatively low luminosity of PS1-10ahf compared to the \cite{Hall} QSO sample does not favour a QSO origin.  Overall the comparisons above indicate a best match to a slowly evolving SLSN-Ic of the same type as \bi, \dam\ and \ap\ \citep{07bi,DY,12dam,11ap}.


\begin{figure*}
\begin{center}
\vspace{1.25cm}
\includegraphics[angle=270,scale=0.5]{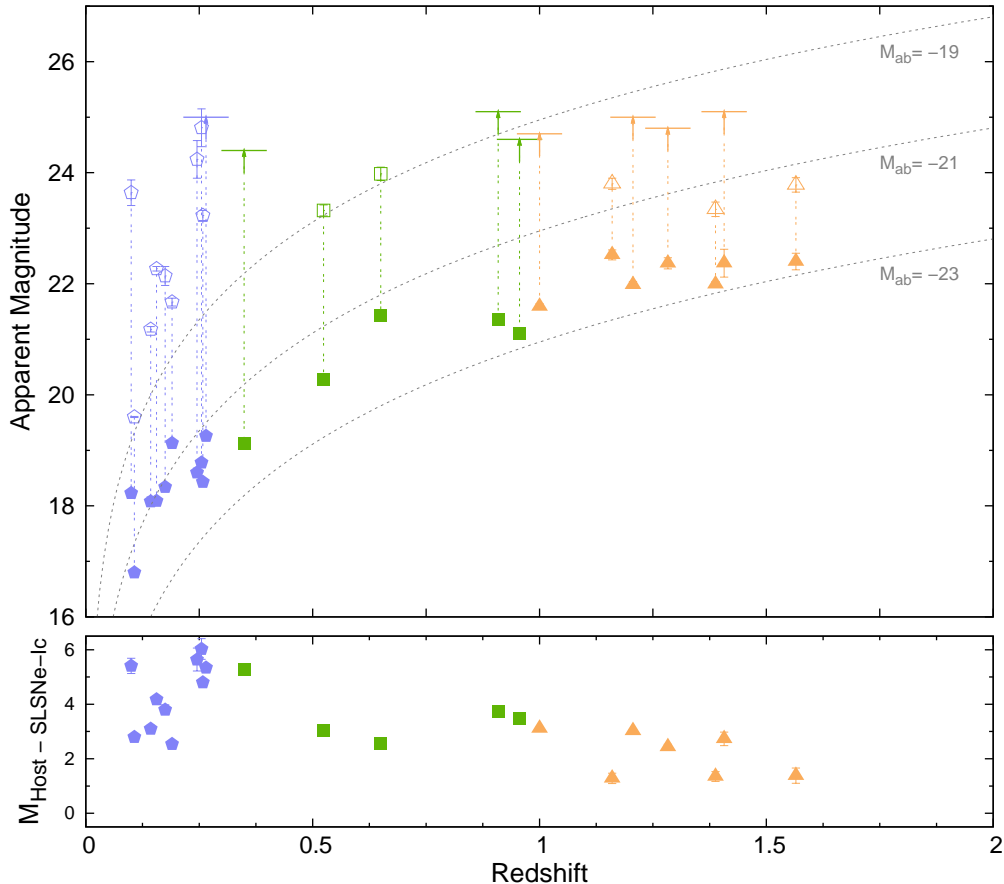}
\end{center}
\caption{Observed SLSNe-Ic apparent magnitudes (opaque shapes) compared with their host galaxy magnitudes (hollow shapes) in the same filter to emphasis the large differences between the two and thus support the method of finding SLSNe-Ic given in the paper.  The blue hexagons show V, \emph{g} and \gps-band data, the green squares show \emph{r} and \rps-band data and the yellow triangles show \emph{i} and \ips-band data.  Any host galaxies with only limiting magnitudes  are shown as arrows.  Host galaxy data were taken from Lunnan et al. (2014) and Nicholl et al. (2014) and the SLSNe-Ic data from various sources (see text for more details).  The three, grey dotted lines represent the trends given by constant absolute magnitudes of -23, -21 and -19 mag.  The lower panel shows the magnitude difference between the SLSNe-Ic and their hosts, plotted against the same redshift range as the upper plot.}
\label{fig:snhost}
\end{figure*}

\section{DISCUSSION}
\label{sec:discussion}

\subsection{SLSNe-Ic in PS1}

By a combination of 2 semi-automated transient alert pipelines and vigorous human screening, the catalogue of all the hostless transients from approximately the first year of the \PS\ survey presented within can be considered to be comprehensive.
Of the 12 spectroscopically confirmed, core-collapse \PS\ SNe with faint host galaxies (see Table \ref{table:CC!}), 5 have been classified as SLSNe-Ic independently of this paper; PS1-10ky and PS1-10awh \citep{10kyawh} and PS1-11ap \citep{11ap}.
PS1-11tt and PS1-11afv fall into the same category with high redshifts obtained from the identification of Mg\,{\sc ii} in their spectrum giving them comparable absolute magnitudes to the superluminous dataset \citep{2013arXiv1311.0026L}.
Evidence from the spectrum and light curves of PS1-10pm provide convincing arguments that it is, in fact, another such event as shown in Section \ref{sec:pm} of this paper.  PS1-10ahf also seems to share characteristics with these objects, although possible reasons for classifying it as an AGN are given in Section \ref{sec:ahf}.
PS1-10afx was originally thought to be a  superluminous event \citep{10afx} 
but more recently has been shown to be  a lensed SN\,Ia \citep{2013ApJ...768L..20Q,quimblens}.
Despite this possible contamination, probing hostless transient objects with all SNe\,Ia removed appears to prove an efficient method of finding  SLSNe-Ic.

Fig.\,\ref{fig:snhost} uses literature data  of published SLSNe-Ic to illustrate the trend of SLSNe-Ic having luminosities consistently brighter than their host galaxies and probes how this might evolve with redshift. 
We have used host galaxy and SLSNe-Ic data from a range of sources for confirmed SLSNe-Ic. 
The data  are taken from this paper; 
 \cite{10gx,bluedeath,10kyawh,berger,10afx,11xk,2013arXiv1311.0026L,11ap,2014arXiv1405.1325N}. 
The peak apparent magnitudes of these SLSNe-Ic are plotted along with the apparent magnitudes of their host galaxies, in the same observed filter.  

We should  highlight possible biases in interpreting this figure. At low redshift ($z < 0.5$) these SLSNe have been  found in a range of surveys which searched for SNe in and outside bright galaxies, such 
as the Palomar Transient Factory \citep{2009PASP..121.1395L} and  La Silla QUEST \citep{2013PASP..125..683B}
However both the CRTS \citep{2009ApJ...696..870D} and  the 
Pan-STARRS1 ``Faint Galaxy Supernova Search"  \citep{11xk}
used catalogue matching which selected transients with significant flux differences between
host object and transient. While there may be some bias in the CRTS and PS1 discovered objects
no survey (or published study) has found a SLSN in a high mass, high luminosity host (with the 
threshold of roughly $M_g \sim -19$). Neither is there any explanation how this would arise
in a survey selection bias. At redshifts above $z > 0.5$ (mostly PS1 objects in the MDS fields) the trend remains,  but of course many of these objects have been selected for classification in PS1 precisely because
they have no obvious host galaxy as we have described in this paper. 
At redshifts below $z=0.5$ the mean difference between host and peak SLSN-Ic magnitude is at least $4.5\pm1.2$ mag. It appears to decrease to $2.2\pm0.8$ mag between redshifts 1-2
(note that the numbers quoted for the magnitude differences here include the limiting host galaxy magnitudes, meaning that they are lower limits on the magnitude difference).

Although there are caveats with the selection methods it is certainly true that no SLSN-Ic below 
$z \sim 1.5$ has been found in a galaxy anywhere near the luminosity of a typical $L^{\star}$ galaxy. 
If normal and slowly evolving SLSNe-Ic can occur in brighter galaxies at low redshift, the 
combined survey power of all professional and amateur searches have not uncovered any. 
Although this lack of evidence does not prove that these events cannot occur in such environments, the large number of SLSN-Ic discoveries exclusively faint galaxies seems to suggest that such localities are preferential.
Additionally the plot suggests that there may be a trend for the host galaxies of SLSNe-Ic to 
be systematically brighter at higher redshift.
Although we have employed the technique of no visible
host to select high$-z$ candidates, this wouldn't explain why the hosts look intrinsically 
$brighter$ than at low redshift in this plot. We again emphasise that there may well be selection
effects and further discoveries from low$-z$ e.g. iPTF and LSQ + PESSTO \citep{2014arXiv1405.1325N} to high$-z$
(e.g. the Dark Energy Survey) are required to shed further light on the host galaxies and expand on 
the detailed work of \cite{2013arXiv1311.0026L}.

\cite{2011ApJ...727...15N}, \cite{janet} and \cite{2013arXiv1311.0026L} attribute the link between
SLSN-Ic and dwarf galaxies as  being physically due to the low metallicities measured (where possible) for the hosts. If the mass-metallicity relation for galaxies evolves over redshift, one might expect a larger fraction of the massive galaxies (within a factor of two of L$^{\ast}$) at redshifts beyond 1.5 to have significantly 
lower metallicity than their low-redshift counterparts. \cite{2006ApJ...644..813E} show that this is visible at $<z>=2.26\pm0.17$ with star-forming galaxies having metallicities which are typically 0.3\,dex lower than low-\emph{z} galaxies of similar mass. If there is a metallicity threshold beyond which SLSNe-Ic are generally not formed due to progenitor evolution \citep{janet,2013arXiv1311.0026L} 
then one night expect the the mass-metallicity evolution of galaxies to cause SLSN-Ic to 
appear in more massive, more luminous galaxies at higher redshift. We may be seeing this effect in 
the trend in  Fig.\,\ref{fig:snhost}, although the scatter is quite large and there are numerous non-detections at the higher redshifts. In addition,  \cite{cooke} present the discoveries of 
two transients with lightcurves that match SLSN-Ic in galaxies with redshifts of 
$z=2.05$ and $z=3.9$. These transients were selected by virtue of being in galaxies with 
estimated photometric redshifts beyond $z>2$ and the method has has an obvious selection bias as their discovery required detection of a host galaxy.  However the difference between 
host galaxy and transient magnitudes are only $+0.3^{m}$ and $-0.18^{m}$. 
This would suggest the trend in Fig.\,\ref{fig:snhost} could continue 
up to higher redshift, but this requires further work in selecting SLSN at redshifts beyond 
$z=2$.

Of interest is the lack of SLSNe-II discoveries within this dataset.  This could suggest an intrinsically lower rate for this particular brand of SLSNe or it could be evidence of the observational bias apparent in this investigation. The small number of SLSNe-II discovered so far have been associated with brighter hosts, which could explain the lack of discoveries here.
It is noted here for completeness that some SLSNe-II have been found in dwarf galaxies \citep[for example SN2006tf,][]{brightII,2011ApJ...727...15N}.

\subsection{Monte-Carlo simulations and estimate of SLSNe-Ic rates}
\label{sec:MC-rates}

\begin{figure*}
\begin{center}
\includegraphics[angle=270,scale=0.5]{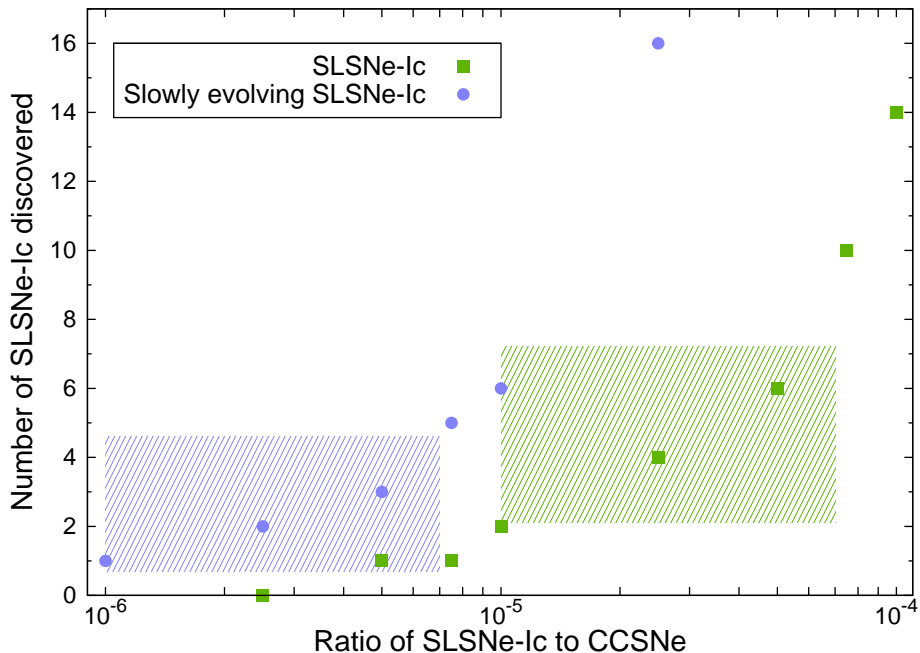}
\end{center}
\caption{The results of the Monte-Carlo simulations.  The two shaded boxes represent the number of probable SLSNe-Ic (green) and slowly evolving SLSNe-Ic (blue) discovered during the search for orphans presented within the paper.  The points represent the ratios of SLSNe-Ic to CCSNe against the number of SLSNe-Ic discovered within the simulation during the same time period.  Thus we can deduce the approximate rate of SLSNe-Ic, of both the normal and the more rare, slowly evolving types.}
\label{fig:snrate}
\end{figure*}

Previous works have carried out rough estimates of the rates of SLSN-Ic to provide
an initial guide to the relative frequency of these transients compared to the 
normal supernova population. 
\cite{bluedeath} estimated their relative rate to be around 1 in every 10,000 core-collapse supernova. 
From the detection of just one event in the Texas Supernova Search (SN2005ap), 
\cite{quimbrate} estimated the SLSN-Ic rate to be 32$^{+77}_{-26}$ events Gpc$^{-1}$\,yr$^{-1}$\,h$^{3}_{71}$. An estimation of the rate with respect to the core-collapse SN rate (within the same
volume) is a useful parameter as it can constrain theories of the progenitor sources, as has
been done with GRB and broad-lined Ic SNe \citep[or hypernova, e.g.][]{2004ApJ...607L..17P}. 
It appears that long duration GRBs rates (LGRBs) are around 1 in every 1000 core-collapse SNe
\citep{2013MNRAS.428.1410G} 
hence the initial rates in \cite{bluedeath} would suggest a very low volumetric rate, roughly 10 
times rarer than LGRBs.  \cite{quimbrate} also compared their volumetric rate to the core-collapse
supernova rate in \cite{2008A&A...479...49B}, to estimate a relative rate of SLSN type Ic to core-collapse SN of
1 in 1000-20,000.  In many of these works, the rate of the rare transients (GRBs or SLSN)
with respect to the CCSN population is assumed to be relatively constant with redshift. This is 
certainly an unknown factor and may evolve due to the  metallicity of the bulk of the star formation 
changing. Since we have low numbers of objects we will assume here that the ratio is constant.

\cite{young} used Monte-Carlo simulations to estimate the number of CCSNe expected to be discovered in all sky transient surveys such as \PS. Since the publication of this paper, the study of SLSNe has rapidly evolved  \citep[for a recent review see][]{gal-yam} .  We use the same Monte-Carlo code  from \cite{young} (recoded in python) and have updated the input data to  include light curves and spectra from SLSNe-Ic and their slowly evolving, \bilike\ counterparts. This allows a calculation of the rates of these SLSNe-Ic from the detections in the \PS\ \MDS\ orphan population. We can determine a robust lower limit and a reasonable estimate of the most likely range for the rates within a redshift of approximately $z<1.4$. 

The simulations were tailored specifically for the \PS\ \MDS; taking into account observational cadence, limiting magnitudes and historical records of time lost due to bad weather, technical difficulties or scheduled maintenance.
A  foreground extinction of $E(B-V)=0.023$ is assumed which is typical for the Galactic line of sight for the MDS fields. No internal galaxy extinction is applied, which is a reasonable assumption for the SLSNe-Ic found to date.
The simulations run in two stages.  The first stage simulates 10, 000 SN events with a population demographic mimicking the input SNe light curves, spectral databases, CCSN to star-formation ratio and star-formation history, all within a volume between $0.3 < z < 1.4$.  The second stage determines the fraction of these events the \PS\ survey would have discovered.

To estimate the limiting magnitude of the MDS images and the efficiency of recovering transients we ran multiple fake star
tests. Five nightly stacks of skycell \#39 (which is approximately one of the 60 CCDs in the focal plane array, warped to the sky)
in MD08 were taken from September 2010, with FWHM of point sources between 1\farcs2-1\farcs6 . Fake stars were added at magnitudes of 21, 22, 23, 24 and 25 (including shot noise) at approximately 80 separate positions on the skycell.
When simulating sources of $>21$ mag only sparse detection efficiency curves are produced and so the points were fitted with an
'S-function' \citep[see Eq. 1 in][]{2008ApJ...681..462D}.
This allowed precise determination of the detection efficiencies.
Half of these were on empty sky regions, to simulate the orphan population and the other half were placed inside resolved galaxies. The IPP image subtraction routine (ppSub) was run using a reference image made of a stack which contained 86 individual images. This is a typical static reference sky product that was used during the period of searching of these transients, and had an image quality of 1\farcs1. 
Sources were catalogued in the difference image, and clear visual detections (in a similar way to how the manual screening was done in the real transient search) were picked for photometric measurements with PSF fitting, as described in Section \ref{sec:observations}. The detection efficiency fell below 98\% at the following $r_{\rm P1}$-band magnitudes : 22.8 (in seeing of 1\farcs6),  22.9 (1\farcs4), 23.3 (1\farcs2), 23.6 (1\farcs0). 
Hence we take the $r_{\rm P1}$-band limiting magnitude to be 23 as the median seeing of individual PS1 images are in the 
range 1\farcs3 to 1\farcs1  (\gps through \zps).
We assume similar for $g_{\rm P1}$- and $i_{\rm P1}$-bands as the exposure times for these are set to retrieve the same limiting magnitude and previous estimates of depth have found the same \cite[e.g. see the extensive discussion in][]{JTwds}. For the $z_{\rm P1}$-band we conservatively take the depth to be 22.4. 

For each of the 10, 000 SNe simulated, the simulations use rest frame SNe spectra to calculate the \emph{K}-corrections attributed to the SN at its assigned redshift for each of the \PS\ \emph{griz}$_{P1}$ filters.  To this end, the simulations require a complete spectral series, covering both the full wavelength and temporal ranges required to generate all possible \emph{K}-corrections, for each of the two SLSNe-Ic classes.
As no single object dataset fulfilled these specifications, composite spectral series were made for the normal and slowly evolving SLSNe-Ic classes.
The slowly evolving SLSNe-Ic series uses data from \dam\ and \ap\ \citep{12dam,11ap} and the SLSNe-Ic series uses \gx, PTF09cnd, PS1-10ky, \xk, SN2012il and PS1-10pm data \citep[][this paper]{10gx, bluedeath, 10kyawh, 11xk}.
Even with data from all the aforementioned SLSNe-Ic however, the full wavelength coverage required for the \cite{young} simulations to work was still not reached at all epochs, particularly at late times when the SLSNe-Ic had become much fainter than their peak magnitudes.  To extend the spectra blue-wards, blackbody fits were employed to extrapolate the observed data.  The 38 day, 65 day and 185 day slowly evolving SLSNe-Ic epochs were fitted with 8500, 7500 and 6500 K blackbody SEDs respectively.  16000, 7000 and 6500 K blackbody SEDs best fitted the -21 day, 50 day and 115 day SLSNe-Ic epochs.  The complete spectral series used for both of these SLSNe-Ic classes can be seen in Fig.\,\ref{fig:mc_spec}.

As each spectrum consists of data from multiple sources, the flux of each epoch had to be scaled to accurately represent the class in question.  A \ap\ absolute magnitude \gps-band light curve was used as a template for the slowly evolving class and absolute \emph{r}-band \xk\ data was used for the SLSNe-Ic class. We set the absolute peak magnitude distributions to be $M_{\rm AB} = -21.5 \pm 0.3$ for the normal SLSNe-Ic and $M_{\rm AB} = -21.25 \pm 0.5$ for the slowly evolving SLSNe-Ic, based on the observed spread of absolute magnitudes that can be seen in Fig.\,\ref{fig:10pmlc} and Fig.\,\ref{fig:10ahfablc} of this paper and in published literature such as \cite{11xk}, \cite{11ap}, 
and \cite{insmartt14},  the latter of which 
which provides the largest compilation of absolute magnitudes of SLSNe to date. 

To scale each spectrum, the \emph{calcphot} task of the \emph{synphot}\footnote{http://www.stsci.edu/institute/software\_hardware/stsdas/synphot} package within \textsc{iraf} was utilised to deduce synthetic absolute magnitude values using an appropriate filter.  Each spectrum was simply multiplied by a constant until the synthetic magnitude matched that of the template light curve at the same phase.
The SDSS filters built-in to \emph{calcphot} were used for this rough comparison but the closeness of their filter functions to that of the \PS\ filters of the template light curves is more than adequate.

For a simulated SN event to be classified as `discovered', we required that the object peaked above an AB magnitude of 22 (in any band), and had a light curve which was detectable above the limiting magnitudes listed above for 100 days in the observer frame (in at least one band). In the PS1 survey we spectroscopically detected 7 SLSNe-Ic  as listed in Table \ref{table:CC!}.
PS1-10awh  was detected for 75 days above the set detection limits and did not strictly meet the criteria. Hence we 
will consider that we have 6 SLSNe-Ic detected (and also check this with a separate Monte Carlo calculation with
the criterion for detection set at 75 days).   
While this is almost certainly incomplete it serves as a baseline observational comparison for the simulated rates and allows lower limits to be placed on the volumetric rates and plausible ranges to be discussed. 

We consider the slowly evolving, SLSNe-Ic and the SLSNe-Ic separately. 
Fig.\,\ref{fig:snrate} uses this information to illustrate the range of possible SLSNe-Ic/CCSNe ratios by comparing the observed data with the results of the Monte-Carlo simulations carried out here.  
For the simulation to mimic the 4 standard SLSNe-Ic which were spectroscopically found during the hostless transient search presented (Table \ref{table:CC!}), the ratio of SLSNe-Ic to CCSNe has to be set to $3^{+3}_{-2}\times10^{-5}$ in the Monte-Carlo simulation \citep[with error values corresponding to $1\sigma$ Gaussian limits taken from][rounded to one decimal place]{limits}.
The slowly evolving type, of which only 2 possible events were discovered during the first year of the \PS\ \MDS, are likely less common and their simulated rate was determined to be only $3^{+4}_{-2}\times10^{-6}$ of the CCSNe rate.
As the slowly evolving SLSNe-Ic remain brighter for a longer duration after their peak luminosity,
they should be easier to detect and hence the fact that we have spectroscopically confirmed fewer of these than the faster declining SLSNe-Ic suggests that they are indeed rarer. This is in agreement with previous suggestions of \cite{gal-yam} and \cite{12dam}.  As a comparison, we ran the Monte Carlo calculation with the requirement of 
75 days, and hence included PS1-10awh as a detected event. We found a relative rate of $\sim10^{-5}$, within the 
error bar of our estimated result of $3^{+3}_{-2}\times10^{-5}$.

However we cannot be confident that we are spectroscopically complete and there could well be SLSNe-Ic in Tables \ref{table:CC?} and \ref{table:CC??} which have not been spectroscopically confirmed. There are approximately 10 SNe in these tables which peak above $M_{\rm AB} =22$ and do not have a possible Type Ia light curve classification. If we regard these as potential SLSNe-Ic which we have not managed to classify then the ratio of 
normal CCSNe to SLSNe-Ic in our spectroscopically confirmed sample would suggest that approximately 60\% of them could be SLSNe-Ic. Thus we consider a plausible upper limit to the number of SLSNe-Ic in our total  detected PS1 MDS sample to be $10^{+4}_{-3}$ \citep[with errors again estimated from][]{limits}, which would imply an upper limit to the rate of $8^{+2}_{-1}\times10^{-5}$ SLSNe-Ic per CCSNe. 

In summary, we have estimated a  range for the rate of SLSNe-Ic compared to the 
rate of CCSNe within a redshift of $0.3\leq z\leq1.4$ of between $3^{+3}_{-2}\times10^{-5}$ and $8^{+2}_{-1}\times10^{-5}$.
The rate of the slowly evolving, SN2007bi-like SLSNe-Ic appear to be a factor of $\sim10$ lower and likely to be 
around $3\times10^{-6}$, although this number is uncertain by about a factor two given the small numbers 
detected.  To put this in context and compare with the only other quantitative rate calculation (albeit at lower
redshift), we find there is about one SLSN-Ic per 12000$-$30000 core-collapse supernovae, whereas \cite{quimbrate} finds
one per 1000-20000. This compares to the rate of LGRBs of around 1 in every 1000 core-collapse SNe
\citep{2013MNRAS.428.1410G}.

\section{CONCLUSIONS}
\label{sec:conclusion}

We have catalogued all the SN-like, hostless transients from the PS1 Medium Deep Survey and, by using multiple spectroscopic programmes and photometric classifiers, filtered out all the SNe\,Ia events. This leaves 
 a promising percentage of remaining objects that seem to fall into the category of SLSNe-Ic.

\begin{itemize}
\item 249 hostless transients were discovered within the first 1.37 yr of the \PS\ \MDS, 133 of which have SN-like features.
\item 40 are spectroscopically confirmed SNe, $\sim17.5\%$ of which are possible SLSNe-Ic.
\item 12 are spectroscopically confirmed, non-type Ia SNe, $\sim60\%$ of which are possible SLSNe-Ic.
\end{itemize}

PS1-10pm and PS1-10ahf were discovered in this way.
Photometric and spectroscopic comparisons place PS1-10pm comfortably in the SLSNe-Ic class.  The classification of PS1-10ahf is not as robust, but reasonably solid photometric and spectroscopic comparisons give it a probable association with the slowly evolving class of SLSNe-Ic such as SN2007bi, PS1-11ap and PTF12dam. 
We highlight that this was a combination of spectroscopic classification (when the SNe were close to peak) and 
photometric classification after the lightcurves had been gathered. A challenge remains to carry out 
accurate photometric classification in real time. 

Using the SLSNe-Ic statistics gathered during the search for orphans and comparing them with Monte-Carlo simulations of SLSNe-Ic, we determined the rate of SLSNe-Ic within a redshift of $0.3\leq z\leq1.4$ to be between  $3^{+3}_{-2}\times10^{-5}$ and $8^{+2}_{-1}\times10^{-5}$ that of the CCSNe rate. 
The ratio of slowly evolving SLSNe-Ic to CCSNe seems to be much lower, at around $3^{+4}_{-2}\times10^{-6}$.

Using a combination of careful photometric analysis and thorough spectroscopic follow-up and the search method of exploiting the common characteristic of the $>2$ mag difference between discovered SLSNe-Ic peak magnitudes and their host galaxies, an ever increasing number of SLSNe-Ic should be found in the next few years from current and future wide-field surveys (Pan-STARRS2, The Dark Energy Survey, The Zwicky Transient Factory, La Silla-QUEST + the Public ESO Spectroscopic Survey of Transient Objects, Large Synoptic Survey Telescope).
\\ \\
{\it Facilities:} Pan-STARRS1, Gemini, William Herschel Telescope

{\small{ \textit{Acknowledgements}. 

    The Pan-STARRS1 Surveys (PS1) have been made possible through
    contributions of the Institute for Astronomy, the University of
    Hawaii, the Pan- STARRS Project Office, the Max-Planck Society and
    its participating institutes, the Max Planck Institute for
    Astronomy, Heidelberg and the Max Planck Institute for
    Extraterrestrial Physics, Garching, The Johns Hopkins University,
    Durham University, the University of Edinburgh, Queen's University
    Belfast, the Harvard- Smithsonian Center for Astrophysics, the Las
    Cumbres Observatory Global Telescope Network Incorporated, the
    National Central University of Taiwan, the Space Telescope Science
    Institute, the National Aeronautics and Space Administration under
    Grant No. NNX08AR22G issued through the Planetary Science Division
    of the NASA Science Mission Directorate, the National Science
    Foundation under Grant No. AST-1238877, and the University of
    Maryland. S.J.S. acknowledges funding from the European Research
    Council under the European Union's Seventh Framework Programme
    (FP7/2007-2013)/ERC Grant agreement no [291222] (PI: S.J. Smartt).
This work is based on observations made with the following telescopes: William Herschel
    Telescope (operated by the Isaac Netwon Group), in the Spanish Observatorio
    del Roque de los Muchachos of the Instituto de Astrofísica de
    Canarias, in the island of La Palma; the Gemini Observatory, which is operated by the 
    Association of Universities for Research in Astronomy, Inc., under a cooperative agreement 
    with the NSF on behalf of the Gemini partnership: the National Science Foundation 
    (United States), the National Research Council (Canada), CONICYT (Chile), the Australian 
    Research Council (Australia), Minist\'{e}rio da Ci\^{e}ncia, Tecnologia e Inova\c{c}\~{a}o 
    (Brazil) and Ministerio de Ciencia, Tecnolog\'{i}a e Innovaci\'{o}n Productiva (Argentina).
    Some observations reported here were obtained at the MMT Observatory, a joint facility of the Smithsonian Institution and the University of Arizona.  Support for SR was provided by NASA through Hubble Fellowship grant \#HST-HF-51312.01 awarded by the Space Telescope Science Institute, which is operated by the Association of Universities for Research in Astronomy, Inc., for NASA, under contract NAS 5-26555.  JT acknowledges support for this work provided by National Science Foundation grant AST-1009749.  SM acknowledges financial support from the Academy of Finland (project: 8120503). We thank A. Gal-Yam and P. Nugent for providing the classification of PS1-11acn, which is the PTF object PTF11dws.}

\appendix

\section{Object tables and example \PS\ photometry}
\label{app:A}

\begin{table}
\begin{center}
\caption{Subaru host detections for some of the orphans presented within the paper and their \emph{z}-band magnitudes.  The last four objects had no associated host and so only a limiting magnitude, again in the \emph{z}-band, is available.}
\label{table:subaru}
\begin{tabular}{c c | c c}
  \hline
  \hline
{\bf PS1 ID} & {\bf M$_\emph{z}$ (\emph{dz})} & {\bf PS1 ID} & {\bf M$_\emph{z}$ (\emph{dz})} \\
    \hline
PS1-10ahq & 22.818 (0.053) &PS1-10abf & $>24.933$ \\
PS1-10aht & 23.864 (0.201) & PS1-10awh & $>24.625$ \\
PS1-10afx & 22.512 (0.075) & PS1-10ky  & $>25.049$ \\
PS1-10dq  & 25.380 (0.470) & PS1-11er  & $>23.510$ \\
PS1-11ad  & 23.626 (0.248) && \\
\hline
 \end{tabular}
 \medskip
\end{center}
\end{table}

\begin{table*}
\begin{center}
\caption{28 SN-like orphans.  The numbers in the \textsc{soft} and \textsc{psnid} columns represent the probability that the algorithms classified an object as the SN-types given in brackets.  Although none of these objects met with the confidence restrictions placed upon the photometric classifiers for identifying SNe (see Tables 4 and 6 for objects that do), it can be seen here that a large number of these objects were given some sort of tentative, photometric SNe classification.}
\label{table:CC??}
\begin{tabular}{lcccccc}
  \hline
  \hline
{\bf Field} & {\bf PS1 ID} & {\bf RA (deg, J2000)} & {\bf Dec (deg, J2000)} & {\bf SOFT} & {\bf PSNID} & {\bf Peak \emph{r}$_{\bf P1}$}\\
    \hline
MD01 & PS1-10bku & 35.5355 & -5.1926 & 0.721 (Ia) & 0.937 (Ia) & 22.389 (0.089) \\
MD01 & PS1-10ags & 34.8215 & -4.0770 & 1 (Ibc) & 1 (II) & 21.894 (0.076) \\
MD02 & PS1-10bkj & 54.0300 & -27.6676 & 0.964 (Ia) & 0.84 (Ibc) & 21.773 (0.084) \\
MD03 & PS1-10bkc & 130.4508 & 43.2094 & 1 (IIL) & 0.999 (Ibc) & 21.968 (0.081) \\
MD03 & PS1-10bmj & 130.7928 & 43.9931 & 0.777 (Ia) & 0.995 (Ibc) & 22.640 (0.186) \\  
MD03 & PS1-10bll & 131.2202 & 42.9031 & 0.616 (Ia) & 0.939 (Ibc) & 22.580 (0.120) \\
MD04 & PS1-11ac & 149.6572 & 1.3447 & 0.73 (Ia) & - & 22.992 (0.163) \\
MD04 & PS1-11cq & 149.7889 & 1.6694 & 0.63 (Ia) & 0.536 (Ibc) & 22.902 (0.161) \\
MD04 & PS1-11ma & 150.7541 & 2.9919 & - & 0.976 (Ia) & 22.387 (0.091) \\
MD04 & PS1-11vp & 150.8929 & 1.1808 & - & - & 22.453 (0.177) \\
MD04 & PS1-11ax & 151.0974 & 2.6413 & 0.512 (Ia) & 0.924 (Ibc) & 22.657 (0.140) \\
MD05 & PS1-11bei & 159.8550 & 59.2007 & - & - & 21.740 (0.150) \\
MD05 & PS1-10vj & 159.6722 & 58.8748 & 0.759 (Ia) & 0.953 (Ia) & 22.255 (0.110) \\
MD06 & PS1-11jl & 183.4510 & 48.0142 & 0.493 (Ia) & 0.994 (Ibc) & 22.702 (0.134) \\
MD06 & PS1-10tk & 185.4339 & 46.2052 & 0.996 (Ibc) & 0.984 (Ia) & 22.468 (0.162) \\  
MD06 & PS1-11zh & 185.7641 & 46.2110 & - & - & 22.407 (0.080) \\
MD06 & PS1-11abu & 183.0532 & 47.7119 & - & - & 22.151 (0.137) \\
MD06 & PS1-11yd & 184.9034 & 46.0195 & - & - & 22.234 (0.088) \\
MD06 & PS1-11za & 183.6234 & 47.8537 & - & - & 21.924 (0.055) \\
MD06 & PS1-11xq & 184.2742 & 46.0311 & - & - & 22.367 (0.105) \\
MD07 & PS1-11pe & 212.0687 & 53.4873 & - & 0.784 (Ia) & 21.916 (0.092) \\
MD07 & PS1-11vq & 213.6972 & 52.3841 & - & - & 22.508 (0.184) \\
MD08 & PS1-10mc & 244.9956 & 54.4228 & 0.527 (Ibc) & 1 (Ia) & 22.374 (0.120) \\
MD08 & PS1-11agk & 242.0433 & 55.4815 & - & - & 23.021 (0.335) \\
MD08 & PS1-11ada & 243.8752 & 53.8923 & - & - & 21.186 (0.100) \\
MD09 & PS1-10byu & 333.6330 & 1.1939 & 0.585 (Ia) & 0.62 (Ia) & 21.923 (0.145) \\
MD10 & PS1-10ceu & 351.6930 & 0.3734 & - & - & 21.480 (0.030) \\
MD10 & PS1-10bkh & 352.5808 & -0.4692 & 0.748 (Ia) & 0.941 (Ia) & 23.126 (0.195) \\
\hline
 \end{tabular}
 \medskip
\end{center}
\end{table*}

\begin{figure*}
\begin{center}$
\begin{array}{cc}
\includegraphics[scale=0.24,angle=270]{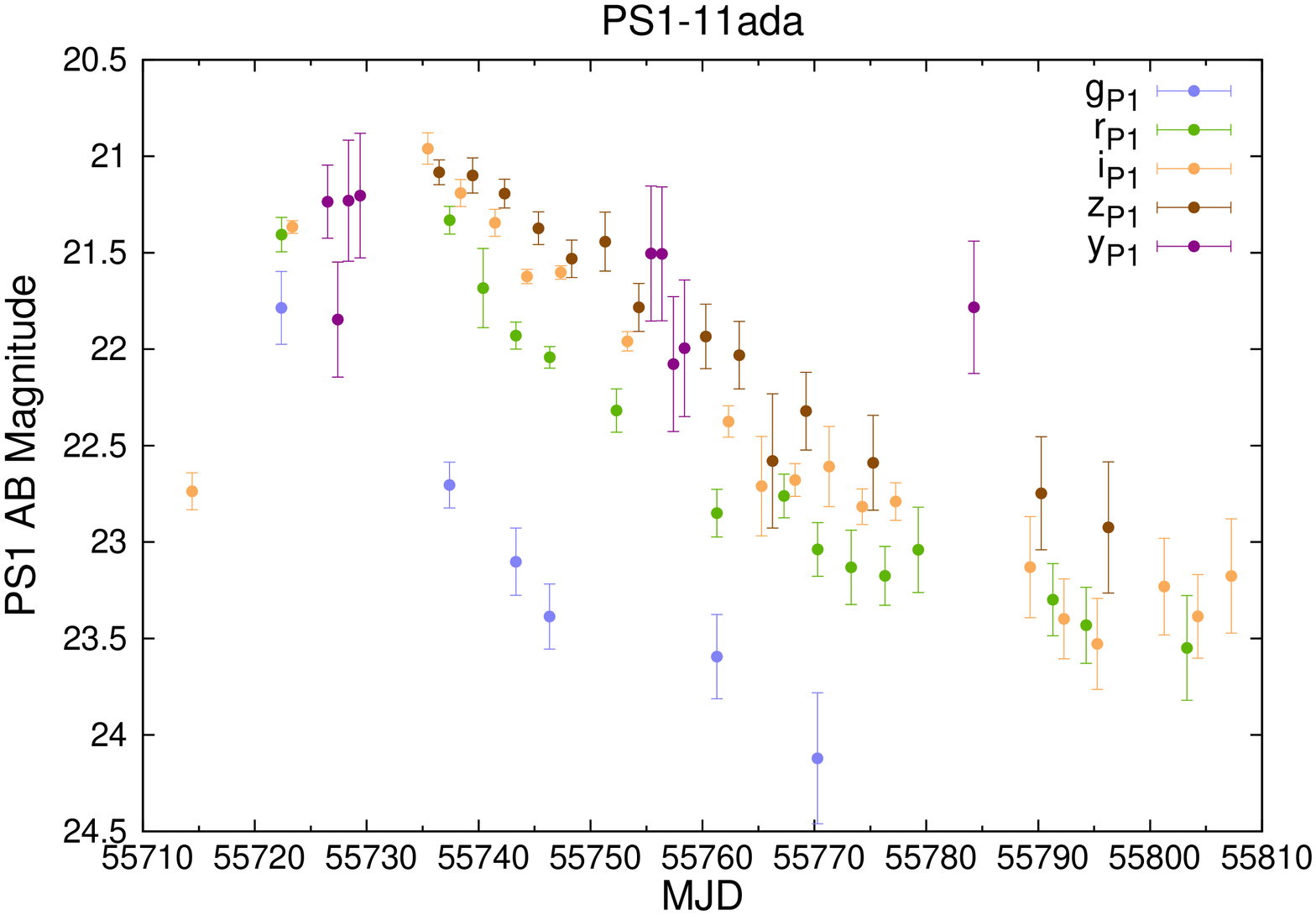} &
\includegraphics[scale=0.24,angle=270]{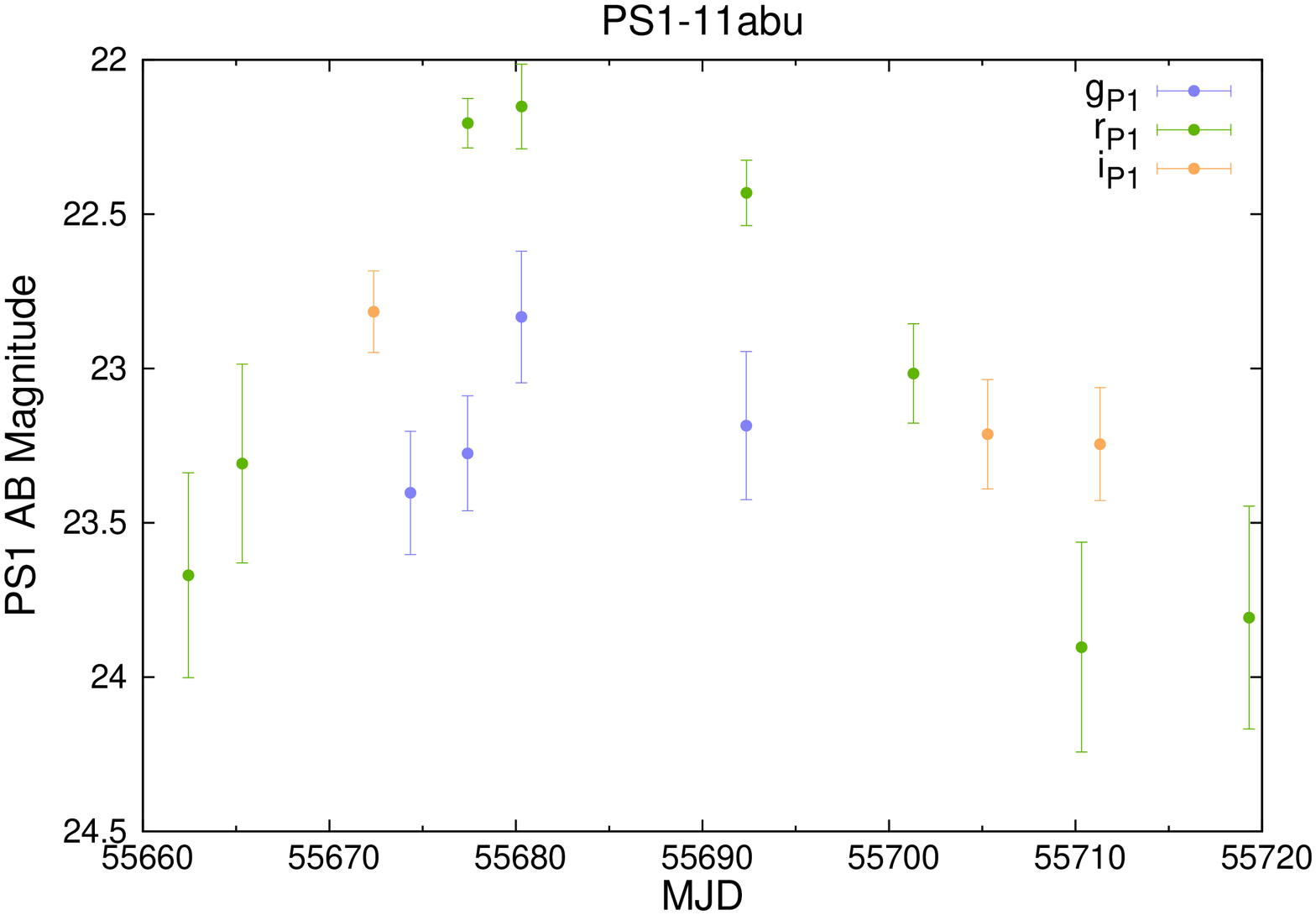} \\
\includegraphics[scale=0.24,angle=270]{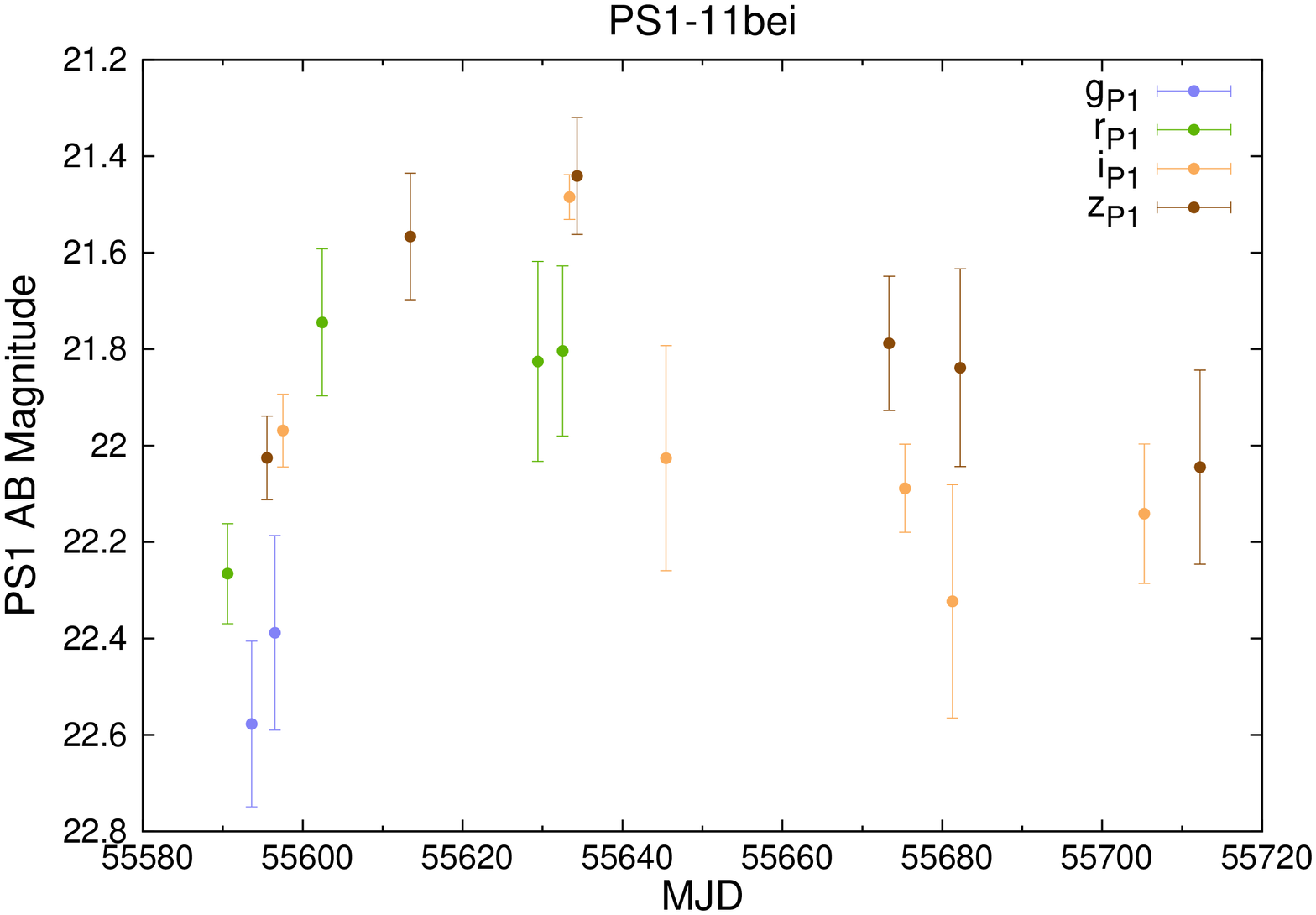} &
\includegraphics[scale=0.24,angle=270]{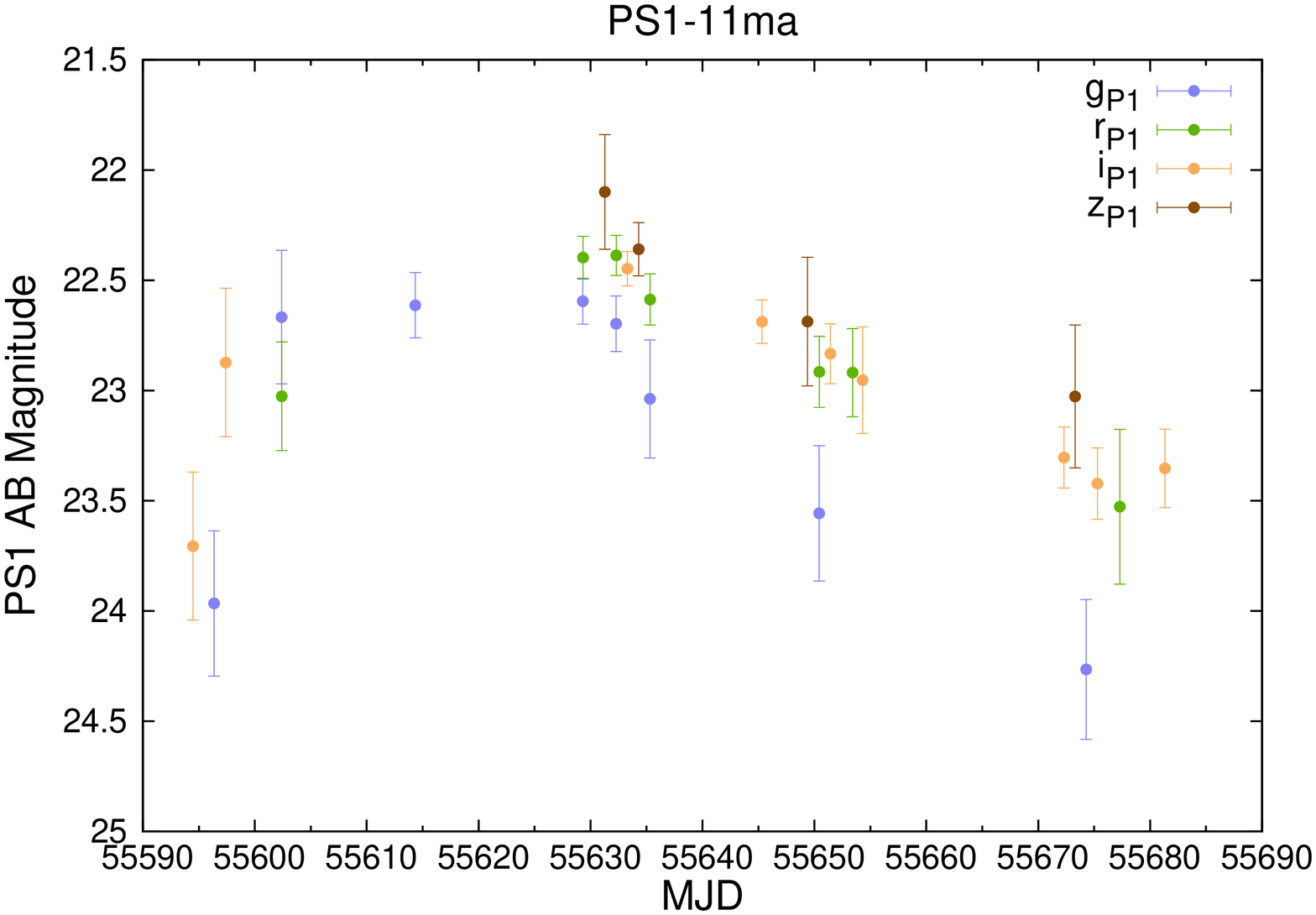}
\end{array}$
\end{center}
\caption{Example light curves from a selection of orphans from Table \ref{table:CC??}.  These transients were hand picked for having typical SN-like light curves exemplified by single, asymmetric peaks.}
\label{fig:CCLC}
\end{figure*}

\begin{table*}
\tiny
\begin{center}
\caption{The 116 unknown orphans.  The acronyms in the `LC Status' column refer to the reasons why an attempt at classification has not been made.  `INC' objects have incomplete light curves, `VAR' objects have variable light curves with no obvious trend and `FAINT' objects fall out of the magnitude range for spectroscopic or reliable photometric classification.  Objects with `RED' status had no detections in the \gps- or \rps-band filters but showed a faint peak in the \ips- and \zps-band filters, possibly as a result of a single, energetic outburst such as a SN at high redshift.}
\label{table:UNK}
\begin{tabular}{lccccc|cccccc}
  \hline
  \hline
{\bf Field} & {\bf PS1 ID} & {\bf RA (deg, J2000)} & {\bf Dec (deg, J2000)} & {\bf Peak  \emph{r}$_{\bf \tiny P1}$} & {\bf LC Status} & {\bf Field} & {\bf PS1 ID} & {\bf RA (deg, J2000)} & {\bf Dec (deg, J2000)} & {\bf Peak \emph{r}$_{\bf \tiny P1}$} & {\bf LC Status} \\
    \hline
MD01 & PS1-10zw & 35.0206 & -4.0609 & 22.813 (0.151) & INC &	MD05 & PS1-11aaq & 161.1660 & 59.4033 & 21.885 (0.073) & INC \\
MD01 & PS1-10zt & 35.9310 & -4.1083 & 22.573 (0.111) & INC &	MD06 & PS1-10sr & 186.7873 & 46.7462 & 22.320 (0.143) & INC \\
MD01 & PS1-10adm & 36.0926 & -5.6325 & 23.096 (0.274) & VAR &	MD06 & PS1-10sl & 185.5830 & 45.9826 & 22.513 (0.118) & INC \\
MD01 & PS1-11db & 37.0777 & -4.2684 & 21.560 (0.087) & INC &	MD06 & PS1-10yq & 183.4461 & 47.1259 & 22.776 (0.133) & FAINT \\
MD01 & PS1-11eu & 35.2756 & -3.1914 & 23.445 (0.244) & FAINT &	MD06 & PS1-10xs & 185.2658 & 46.5539 & 22.884 (0.140) & INC \\
MD01 & PS1-10cal & 35.5671 & -4.5893 & 23.711 (0.326) & FAINT &	MD06 & PS1-10yl & 185.9299 & 46.2831 & 22.462 (0.133) & FAINT \\
MD01 & PS1-10ceo & 35.0395 & -5.0884 & 23.010 (0.110) & VAR &	MD06 & PS1-10ys & 184.5287 & 47.7259 & 22.600 (0.149) & FAINT \\
MD01 & PS1-10cep & 35.1032 & -3.5826 & 23.010 (0.110) & FAINT &	MD06 & PS1-10xt & 184.2873 & 46.1409 & 23.265 (0.164) & INC \\
MD01 & PS1-10bkt & 36.0372 & -2.7882 & 21.505 (0.066) & - & MD06 & PS1-10jm & 184.4595 & 47.4467 & 21.104 (0.097) & - \\
MD02 & PS1-10agq & 52.8398 & -28.5254 & 23.008 (0.184) & VAR &	MD06 & PS1-10cef & 185.2454 & 46.0920 & 23.150 (0.210) & FAINT \\
MD02 & PS1-10ces & 54.4704 & -28.1703 & 22.780 (0.130) & VAR &	MD06 & PS1-11bet & 184.4993 & 46.1507 & 22.530 (0.220) & FAINT \\
MD02 & PS1-11beh & 52.0673 & -27.8645 & 22.780 (0.150) & INC &	MD06 & PS1-11jq & 186.0334 & 47.9493 & 22.347 (0.080) & INC \\
MD02 & PS1-11bej & 52.5304 & -29.0150 & 23.150 (0.230) & INC &	MD06 & PS1-11jt & 186.3847 & 48.1535 & 22.457 (0.120) & INC \\
MD02 & PS1-10aff & 52.8698 & -27.0436 & - & RED  &	MD06 & PS1-11qc & 184.7770 & 47.6353 & 23.258 (0.279) & FAINT \\
MD02 & PS1-10cer & 53.9535 & -28.9196 & 22.440 (0.090) & FAINT &	MD06 & PS1-11wm & 183.6975 & 48.1353 & 22.586 (0.228) & FAINT \\
MD02 & PS1-11bes & 54.2109 & -27.2108 & 23.260 (0.150) & INC &	MD06 & PS1-11vh & 185.6786 & 47.0719 & 22.389 (0.105) & FAINT \\
MD02 & PS1-10ceq & 54.3060 & -28.5226 & - & RED &	MD07 & PS1-10vm & 213.3563 & 52.3595 & 21.891 (0.109) & VAR \\
MD03 & PS1-10cet & 131.8517 & 44.6298 & 23.450 (0.200) & VAR &	MD07 & PS1-11fr & 212.3379 & 53.6687 & 22.347 (0.282) & INC \\
MD03 & PS1-11bem & 130.0729 & 44.9492 & 23.360 (0.230) & INC &	MD07 & PS1-11fo & 211.3801 & 53.4676 & 22.821 (0.347) & INC \\
MD03 & PS1-10awt & 132.1807 & 44.8813 & 22.125 (0.076) & VAR &	MD07 & PS1-11ahj & 213.8155 & 54.3847 & 22.690 (0.150) & VAR \\
MD03 & PS1-10cat & 129.0721 & 45.1472 & 22.915 (0.157) & VAR &	MD07 & PS1-10ceg & 213.4535 & 52.7237 & 23.090 (0.200) & INC \\
MD03 & PS1-10ayj & 129.5958 & 44.7852 & - & RED &	MD07 & PS1-10ceh & 214.5734 & 54.1141 & 22.800 (0.190) & FAINT \\
MD03 & PS1-10can & 129.8584 & 43.7600 & 22.677 (0.217) & VAR &	MD07 & PS1-11bel & 214.3710 & 52.3339 & - & INC \\
MD03 & PS1-10cae & 129.1859 & 44.5547 & 22.584 (0.101) & - & MD07 & PS1-10jn & 212.3330 & 54.2624 & 22.705 (0.141) & - \\
MD03 & PS1-10blu & 130.2135 & 45.2700 & 22.714 (0.208) & FAINT &	MD08 & PS1-10cei & 240.1075 & 54.8165 & - & INC \\
MD03 & PS1-10cak & 130.5093 & 43.9321 & - & RED &	MD08 & PS1-11beu & 240.7690 & 55.2721 & 23.440 (0.180) & INC \\
MD03 & PS1-11gq & 131.8921 & 43.7795 & 22.978 (0.250) & FAINT &	MD08 & PS1-10cej & 241.7209 & 53.8874 & 23.530 (0.190) & VAR \\
MD03 & PS1-11eh & 131.4092 & 45.3290 & 23.177 (0.282) & FAINT &	MD08 & PS1-10cek & 243.0319 & 55.1189 & 22.800 (0.150) & VAR \\
MD03 & PS1-11qj & 131.9785 & 45.2305 & 22.280 (0.166) & FAINT &	MD08 & PS1-11ben & 244.4578 & 55.2028 & 23.960 (0.350) & FAINT \\
MD03 & PS1-11hi & 128.8228 & 44.5885 & 23.659 (0.269) & FAINT &	MD08 & PS1-10mn & 241.1689 & 55.5560 & 22.673 (0.123) & INC \\
MD03 & PS1-11rg & 132.5940 & 43.9940 & 21.874 (0.100) & INC &	MD08 & PS1-10ne & 244.1189 & 53.7972 & 21.097 (0.053) & VAR \\
MD03 & PS1-11ol & 130.8790 & 45.6951 & 22.822 (0.135) & FAINT &	MD08 & PS1-10aeo & 242.0829 & 54.3316 & 22.732 (0.246) & FAINT \\
MD03 & PS1-11ok & 130.6385 & 45.7282 & 23.374 (0.257) & FAINT &	MD08 & PS1-10abn & 240.3384 & 54.7644 & 22.739 (0.140) & FAINT \\
MD03 & PS1-10ayc & 132.3020 & 44.1192 & 23.315 (0.196) & INC &	MD08 & PS1-10aby & 242.0255 & 53.7137 & 23.438 (0.189) & FAINT \\
MD03 & PS1-10aww & 130.3612 & 45.4012 & 22.829 (0.150) & INC &	MD08 & PS1-10abk & 240.8492 & 54.0940 & 23.014 (0.236) & FAINT \\
MD03 & PS1-11se & 129.0951 & 43.6360 & 21.247 (0.053) & INC &	MD08 & PS1-11ug & 241.9200 & 55.4232 & 22.538 (0.237) & FAINT \\
MD03 & PS1-11sd & 131.5120 & 43.4041 & 21.860 (0.151) & INC &	MD08 & PS1-11uc & 240.5109 & 55.0289 & 22.700 (0.101) & FAINT \\
MD04 & PS1-11ber & 148.7119 & 2.5314 & - & RED &	MD08 & PS1-11uf & 244.1813 & 55.1709 & 22.571 (0.154) & FAINT \\
MD04 & PS1-10tv & 150.1942 & 1.3438 & 22.099 (0.258) & INC &	MD08 & PS1-10acg & 245.0462 & 54.3438 & - & RED \\
MD04 & PS1-11cs & 151.0165 & 1.0870 & 21.952 (0.095) & INC &	MD09 & PS1-10aak & 333.4872 & -0.7134 & 21.719 (0.129) & INC \\
MD04 & PS1-11nv & 150.4967 & 2.7301 & 22.541 (0.123) & FAINT &	MD09 & PS1-10aab & 335.0863 & 0.5239 & 22.090 (0.129) & INC \\
MD04 & PS1-11mv & 150.6274 & 1.9906 & 22.076 (0.249) & FAINT &	MD09 & PS1-10bgx & 333.8112 & 0.1249 & 23.490 (0.253) & FAINT \\
MD04 & PS1-11mb & 149.4433 & 0.8251 & 22.278 (0.166) & INC &	MD09 & PS1-10bma & 333.1478 & 0.9874 & 22.657 (0.189) & FAINT \\
MD04 & PS1-11vf & 151.1440 & 2.1694 & 22.355 (0.139) & INC &	MD09 & PS1-10awi & 333.9666 & 0.5064 & 22.731 (0.151) & FAINT \\
MD04 & PS1-11tf & 149.7141 & 0.6388 & 22.116 (0.131) & INC &	MD09 & PS1-10afy & 333.7640 & 0.5902 & 23.187 (0.197) & FAINT \\
MD04 & PS1-11bek & 149.1747 & 2.7484 & 23.150 (0.180) & INC &	MD09 & PS1-11beo & 334.6328 & -0.0450 & - & RED \\
MD04 & PS1-11tl & 149.6978 & 1.1330 & 22.519 (0.113) & INC &	MD09 & PS1-10cel & 334.4344 & -0.6454 & 23.070 (0.160) & FAINT \\
MD04 & PS1-11tr & 150.2001 & 1.3042 & 21.678 (0.083) & INC &	MD09 & PS1-11bep & 332.7250 & 0.6546 & 22.920 (0.160) & FAINT \\
MD05 & PS1-10ac & 160.0895 & 57.9299 & 22.690 (0.154) & INC &	MD10 & PS1-10cem & 352.6395 & -0.0084 & 23.190 (0.140) & INC \\
MD05 & PS1-11ck & 160.2717 & 59.2027 & 22.379 (0.106) & FAINT &	MD10 & PS1-11beq & 353.3164 & 0.0658 & 23.530 (0.350) & FAINT \\
MD05 & PS1-10cec & 164.2774 & 57.9606 & 23.120 (0.120) & INC &	MD10 & PS1-10cen & 351.8198 & -1.4149 & 22.760 (0.110) & FAINT \\
MD05 & PS1-10ced & 162.6750 & 57.4088 & 22.650 (0.170) & FAINT &	MD10 & PS1-10bja & 351.3430 & -0.7578 & 22.597 (0.127) & FAINT \\
MD05 & PS1-10cee & 163.5573 & 59.2589 & 22.800 (0.160) & INC &	MD10 & PS1-10bjy & 351.1140 & -1.1171 & 22.741 (0.192) & FAINT \\
MD05 & PS1-11cb & 162.0916 & 57.3682 & 22.525 (0.185) & FAINT &	MD10 & PS1-10bkg & 352.9823 & -0.8480 & 23.262 (0.200) & FAINT \\
MD05 & PS1-11av & 163.0888 & 57.3474 & 22.718 (0.256) & FAINT &	MD10 & PS1-10biw & 350.9567 & -0.0839 & 22.198 (0.141) & FAINT \\
MD05 & PS1-11bm & 162.6764 & 56.8736 & 22.541 (0.264) & FAINT &	MD10 & PS1-10aaz & 351.3668 & -0.3134 & 22.978 (0.248) & FAINT \\
MD05 & PS1-11re & 159.7510 & 58.7787 & 22.480 (0.139) & FAINT &	MD10 & PS1-10bnj & 351.6292 & -1.1867 & 23.330 (0.315) & FAINT \\
MD05 & PS1-11ta & 160.2264 & 58.4274 & 22.210 (0.144) & FAINT &	MD10 & PS1-10adf & 353.3260 & -0.7616 & 23.413 (0.264) & FAINT \\
\hline
 \end{tabular}
 \medskip
\end{center}
\end{table*}

\begin{figure*}
\begin{center}$
\begin{array}{c c}
\includegraphics[angle=270, scale=0.35]{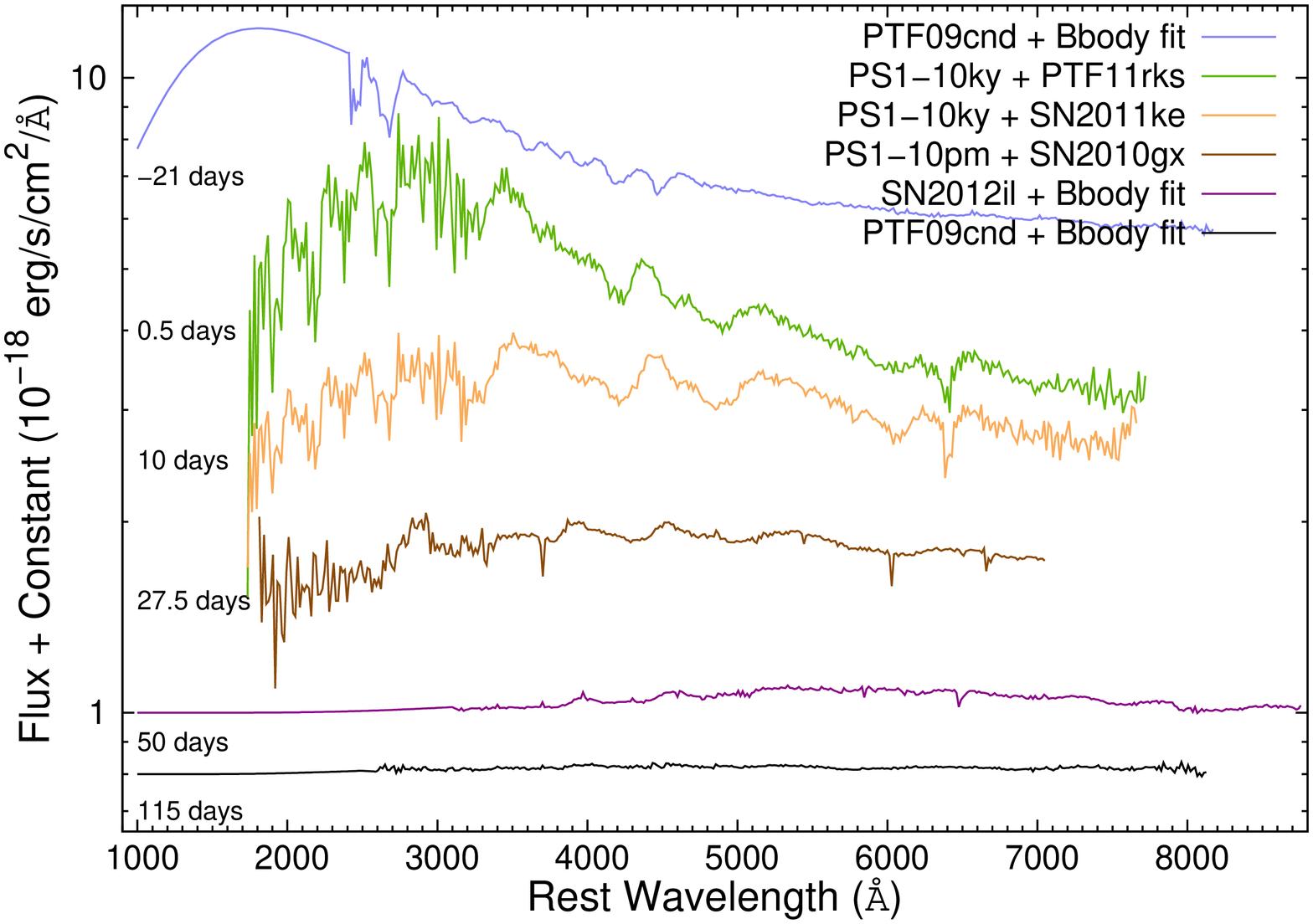} &
\includegraphics[angle=270, scale=0.35]{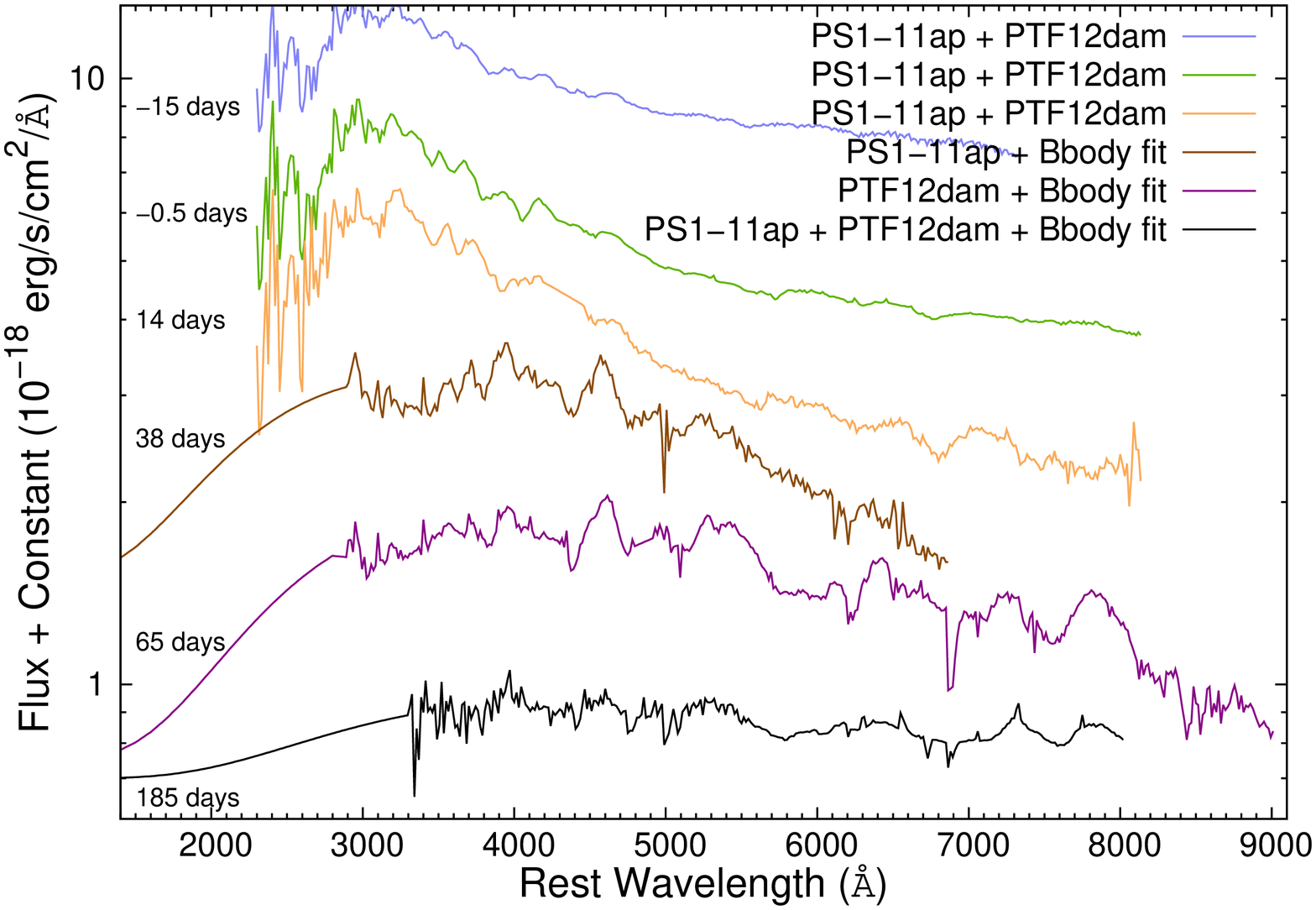}
\end{array}$
\caption{\textrm{The composite spectral series' for the normal ({\it left-hand figure}) and slowly evolving ({\it right-hand figure}) SLSNe-Ic classes.  Each spectrum is composed of a mix of objects of the same class at similar epochs and appropriate blackbody curves, as indicated in the figure keys.} \label{fig:mc_spec}}
\end{center}
\end{figure*}

\end{document}